\documentclass[journal]{IEEEtran}
\usepackage{setspace}
\usepackage{cite,graphicx,amsmath,amssymb}
\usepackage{hyperref}
\usepackage{subfigure}
\usepackage{fancyhdr}
\usepackage{mdwmath}
\usepackage{mdwtab}
\usepackage{balance}
\usepackage{xcolor}
\usepackage{bm}
\usepackage{amsthm}
\usepackage{threeparttable}
\usepackage{multirow}
\usepackage{flafter}
\usepackage{makecell}
\usepackage{diagbox}
\usepackage{tcolorbox}
\usepackage{physics}
\usepackage{bbding}
\usepackage{pifont}
\usepackage{enumerate}

\theoremstyle{definition}

\newtheorem{theorem}{Theorem}

\newtheorem{lemma}{Lemma}

\newtheorem{corollary}{Corollary}

\providecommand{\url}[1]{#1}
\allowdisplaybreaks
\usepackage{lineno}
\usepackage{cite}
\usepackage{hyperref}
\usepackage{amsmath,amssymb,amsfonts}
\usepackage{amsmath}
\usepackage{amsthm}

\usepackage{graphicx}
\usepackage{textcomp}
\usepackage{xcolor}
\usepackage{graphicx}
\usepackage{float}
\usepackage{subfigure}
\usepackage{amsmath}
\usepackage{amsfonts,amssymb}
\usepackage{mathrsfs}
\usepackage{mathtools}
\usepackage{algorithm}
\usepackage{algorithmicx}
\usepackage{algpseudocode}
\usepackage{bm}
\usepackage{multirow}
\usepackage{array}
\usepackage{amssymb}
\usepackage{amsmath}
\usepackage{cite}
\usepackage{url}
\usepackage{xcolor}
\usepackage{cite,graphicx,amsmath,amssymb}
\usepackage{subfigure}
\usepackage{fancyhdr}
\usepackage{mdwmath}
\usepackage{mdwtab}
\usepackage{caption}
\usepackage{amsthm}
\usepackage{setspace}
\usepackage{bm}
\usepackage{mathtools}
\usepackage{dsfont}
\usepackage{bbm}
\usepackage{yhmath}
\usepackage{enumerate}
\usepackage{tcolorbox}
\usepackage{array}
\usepackage{bbding}
\usepackage{pifont}
\theoremstyle{definition}
\theoremstyle{definition}
\usepackage{xcolor}
\newcommand\ytl[2]{
\parbox[b]{6em}{\hfill{\color{cyan}\bfseries #1}~$\cdots\cdots$~}\makebox[0pt][c]{$\bullet$}\vrule\quad \parbox[c]{14cm}{\vspace{4pt}\color{red!40!black!80}\raggedright #2.\\[7pt]}\\[-3pt]}
\makeatletter
\begin{document}

\title{Near-Field Communications:\\ A Comprehensive Survey}
\author{Yuanwei Liu, \IEEEmembership{Fellow, IEEE}, Chongjun Ouyang, \IEEEmembership{Member, IEEE}, Zhaolin Wang, \IEEEmembership{Member, IEEE},\\ Jiaqi Xu, \IEEEmembership{Member, IEEE}, Xidong Mu, \IEEEmembership{Member, IEEE}, and A. Lee Swindlehurst, \IEEEmembership{Fellow, IEEE}\\
\thanks{Yuanwei Liu is with the Department of Electrical and Electronic Engineering, The University of Hong Kong, Hong Kong (email: yuanwei@hku.hk).}
\thanks{Chongjun Ouyang is with the School of Electrical and Electronic Engineering, College of Engineering and Architecture, University College Dublin, Dublin, D04 V1W8 Ireland, and also with the School of Electronic Engineering and Computer Science, Queen Mary University of London, E1 4NS London, U.K. (e-mail: chongjun.ouyang@ucd.ie).}
\thanks{Zhaolin Wang is with the School of Electronic Engineering and Computer Science, Queen Mary University of London, E1 4NS London, U.K. (e-mail: zhaolin.wang@qmul.ac.uk).}
\thanks{Jiaqi Xu and A. Lee Swindlehurst are with the Department of Electrical Engineering and Computer Science, University of California, Irvine, CA 92697, USA (email: \{xu.jiaqi, swindle\}@uci.edu).}
\thanks{Xidong Mu is with the Centre for Wireless Innovation (CWI), Queen's University Belfast, BT3 9DT Belfast, U.K. (e-mail: x.mu@qub.ac.uk).}}
\maketitle
\begin{abstract}
Multiple-antenna technologies are evolving towards larger aperture sizes, extremely high frequencies, and innovative antenna types. This evolution is fostering the emergence of near-field communications (NFC) in future wireless systems. Considerable attention has been directed towards this cutting-edge technology due to its potential to enhance the capacity of wireless networks by introducing increased spatial degrees of freedom (DoFs) in the range domain. Within this context, a comprehensive review of the state of the art on NFC is presented, with a specific focus on its \romannumeral1) fundamental operating principles, \romannumeral2) channel modeling, \romannumeral3) performance analysis, \romannumeral4) signal processing techniques, and \romannumeral5) integration with other emerging applications. Specifically, \romannumeral1) the basic principles of NFC are characterized from both physics and communications perspectives, unveiling its unique properties in contrast to far-field communications. \romannumeral2) Building on these principles, deterministic and stochastic near-field channel models are explored for spatially-discrete (SPD) and continuous-aperture (CAP) arrays. \romannumeral3) Based on these models, existing contributions to near-field performance analysis are reviewed in terms of DoFs/effective DoFs (EDoFs), the power scaling law, and transmission rate. \romannumeral4) Existing signal processing techniques for NFC are systematically surveyed, which include channel estimation, beamforming design, and low-complexity beam training. \romannumeral5) Major issues and research opportunities in incorporating near-field models into other promising technologies are identified to advance NFC's deployment in next-generation networks. Throughout this paper, promising directions are highlighted to inspire future research endeavors in the realm of NFC, underscoring its significance in the advancement of wireless communication technologies. 
\end{abstract}

\section{Introduction}
The advent of fifth-generation (5G) wireless networks has profoundly transformed various aspects of daily life and catalyzed significant industrial advancements. According to Qualcomm, 5G is projected to generate up to US\$ 13.2 trillion in global sales activity by 2035 \cite{campbell20195g}. Building on 5G's success, both academic and industrial sectors are now paving the way for the development of next-generation (NG) wireless networks. The envisioned NG applications---including the metaverse, digital twin technology, ultra-high-definition streaming, and extended reality (XR)---require exceptionally high data rates, utilization of ultra-wide frequency bands, and support for massive connectivity \cite{connecting20226g}.

These ambitious objectives necessitate the development of revolutionary wireless technologies to advance NG networks. A pivotal element in this NG wireless landscape is the adoption of extremely large aperture arrays (ELAAs) and higher frequencies \cite{cui2022near}. The transition to ELAAs in high-frequency bands represents a significant paradigm shift from traditional far-field communications (FFC) to near-field communications (NFC), which involves more than just quantitative increases in antenna size and carrier frequency; it marks a qualitative leap in how wireless systems are conceived and implemented \cite{liu2023near}.

\begin{figure}[!t]
    \centering
    \subfigure[EM field boundary.]
    {
        \includegraphics[width=0.4\textwidth]{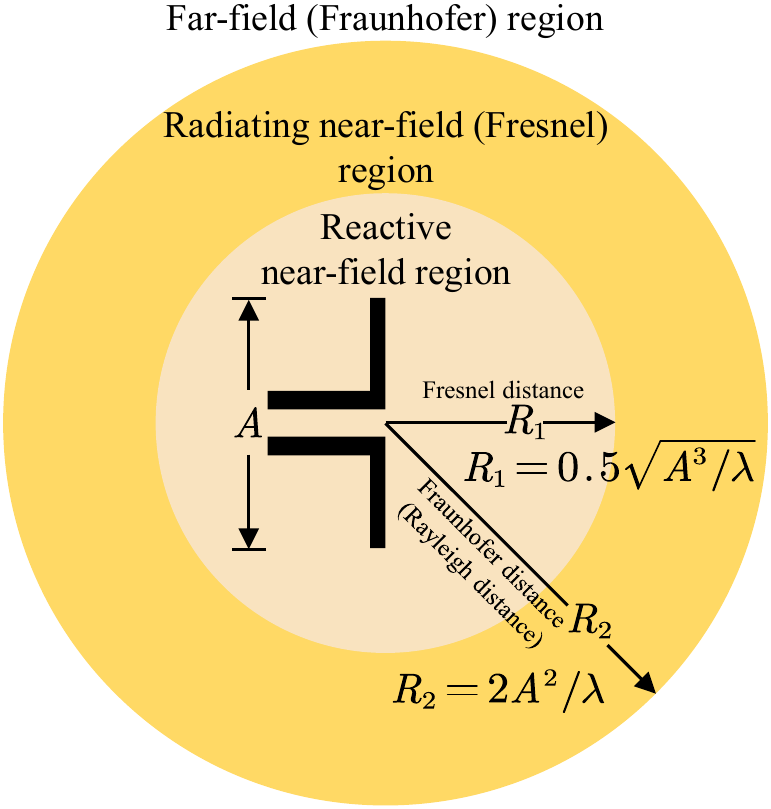}
	   \label{Figure: Field_Boundary1}	
    }
   \subfigure[Typical changes of antenna amplitude pattern shape from reactive near field toward the far field.]
    {
        \includegraphics[width=0.4\textwidth]{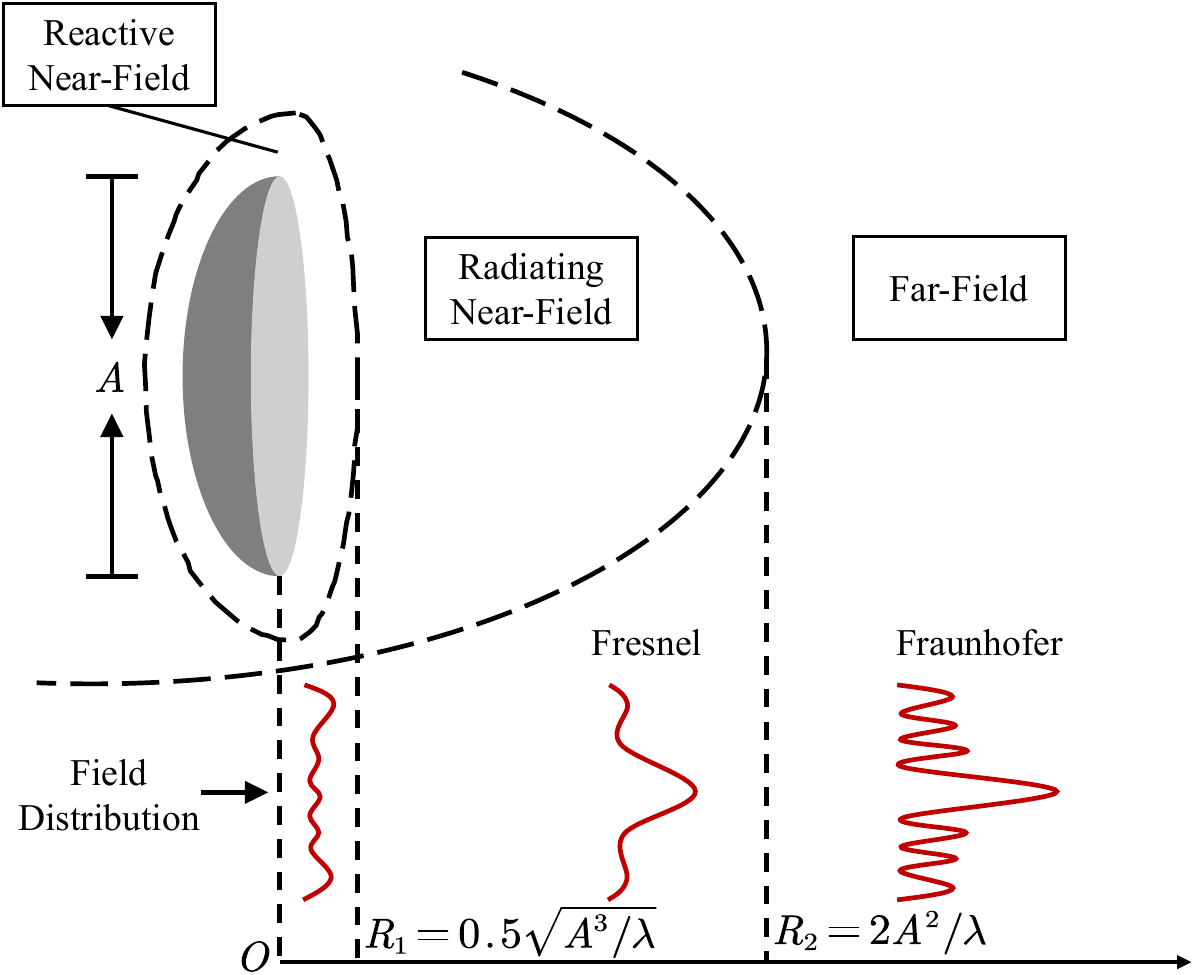}
	   \label{Figure: Field_Boundary2}	
    }
\caption{EM field regions of an antenna, where $A$ and $\lambda$ denote the physical dimension of the antenna array and signal wavelength, respectively.}
    \label{Figure: Field_Boundary}
\end{figure}

\subsection{From Plane Wave to Spherical Wave}
\begin{figure}[!t]
 \centering
\includegraphics[width=0.45\textwidth]{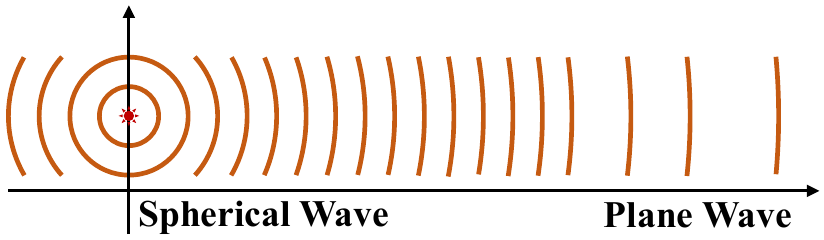}
\caption{The flattening of spherical waves with distance.}
\label{Figure: Point_Source}
\end{figure}

The electromagnetic (EM) radiation field emitted by antennas is conventionally categorized into two distinct regions: the far-field (Fraunhofer) region and the radiating near-field (Fresnel) region \cite{balanis2016antenna}. This division is defined by the Fraunhofer distance (or Rayleigh distance) $\frac{2A^2}{\lambda}$, where $A$ represents the physical dimension of the antenna array and $\lambda$ is the signal wavelength \cite{balanis2016antenna}. This boundary condition is illustrated in {\figurename} {\ref{Figure: Field_Boundary}}. Beyond the Fraunhofer distance, in the far-field region, EM waves exhibit propagation characteristics that differ significantly from those observed in the near-field region. Far-field EM propagation is typically approximated using plane waves, whereas near-field EM propagation requires precise modeling using spherical waves \cite{liu2023nearArxiv,ouyang2024primer}, as depicted in {\figurename} {\ref{Figure: Point_Source}}.

Limited by the dimensions of current antenna arrays and the frequency bands in which they operate, the Fraunhofer distance in existing cellular systems typically extends only a few meters, making near-field effects negligible. As a result, prevalent cellular communications heavily rely on theories and techniques from FFC. However, driven by rapid advancements in wireless technology, NG wireless communications are embracing ELAAs and higher frequencies to meet the escalating demand for communication services. The deployment of massive antenna arrays and the use of high-frequency bands allow NFC to be effective at distances of hundreds of meters, presenting new opportunities for the development of NFC theories and techniques.

As an illustrative example, consider an antenna array with dimensions $A=4$ m---a size plausible for future conformal arrays, such as those deployed on building facades. For such an array, the Fraunhofer distance is calculated to be 373.3 meters for signals at 3.5 GHz (Frequency Range 1) \cite{dahlman20205g}. This distance increases to 2986.7 meters at 28 GHz (Frequency Range 2). These calculations indicate that the enlarged antenna aperture combined with shortened wavelengths in ELAAs significantly extends the near-field region, a facet previously overlooked in traditional wireless transmissions. In a nutshell, the exploration of NFC represents an exciting and relatively unexplored area that is of significant interest to NG wireless systems.
\subsection{A Brief History of Near-Field Communications}
Before delving deeper, we present a brief overview of the historical development of NFC. The history of NFC is marked by significant milestones spanning several centuries. This journey begins with foundational contributions to the understanding of wave propagation, initially focused on light waves, and gradually evolves to include the exploration of EM radio waves.

In 1678, Dutch scientist Christiaan Huygens presented his ``Wave Theory of Light'' to the Académie des Sciences in Paris. This theory posited that each point along a moving wavefront acts as a point source emitting a spherical wave, an idea now recognized as Huygens' Principle. Originally a preliminary chapter in his work ``Dioptrica'', Huygens's theory was officially published in 1690 under the title ``Traité de la Lumière'' (Treatise on Light) \cite{huygens1885traite}. This seminal work provided the first fully mathematized, mechanistic explanation of an unobservable physical phenomenon---the propagation of light. Initially deduced through mathematical inference, this principle was later experimentally supported by Thomas Young's double-slit diffraction experiment in 1801 \cite{young1832bakerian}.

Huygens' ``Wave Theory of Light'', along with Young's experiments, challenged Isaac Newton's ``Corpuscular Theory of Light''. Subsequently, the wave theory of light gained widespread acceptance, paving the way for significant research into the diffraction patterns of light waves. In optics, diffraction phenomena are categorized into two typical scenarios based on the distance between the diffracting object and the observed pattern. Near-field diffraction occurs when the pattern is observed close to the diffracting object, whereas far-field diffraction is observed at significantly longer distances.

In 1821, German physicist Joseph von Fraunhofer constructed the first diffraction grating, consisting of 260 closely spaced parallel wires. Adept in the mathematical wave theory of light, Fraunhofer used his diffraction grating to measure the wavelengths of specific colors and dark lines in the solar spectrum \cite{von1824neue}. This remarkable achievement earned him personal nobility, bestowing him with the title ``Ritter von'', signifying knighthood.

From 1815 to 1822, French engineer and physicist Augustin-Jean Fresnel submitted a series of seminal memoirs to the French Academy of Sciences, advancing his understanding of diffraction. Fresnel quantified Huygens's principle of secondary spherical waves and Young's principle of diffraction, which provided the first comprehensive explanation of diffraction by straight edges. This included the initial wave-based elucidation of rectilinear propagation. Importantly, Fresnel also established criteria to differentiate between far-field and near-field diffraction zones. During this period, he developed several foundational concepts, including the Fresnel approximation, and defined the near-field and far-field regions.

In honor of the seminal contributions of Fraunhofer and Fresnel, the terms ``Fraunhofer region'' and ``Fresnel region'' are used to designate the far-field and radiating near-field, respectively. The boundaries between the far-field and radiating near-field, as well as between the radiating near-field and reactive near-field, are termed the Fraunhofer distance and Fresnel distance, respectively. Moreover, the criteria set by Fresnel for determining the diffraction region are commonly referred to as the Fraunhofer condition.

Until 1891, British scientist John William Strutt, also known as Baron Rayleigh or Lord Rayleigh, characterized the distance over which a diverging or converging beam of light remains approximately collimated. This distance, which marks the boundary between the near and far field, is calculated using the formula $\frac{A^2}{2\lambda}$ for an antenna or array with aperture $A$ operating at wavelength $\lambda$. It represents the axial distance from a radiating aperture to a point where the phase difference between the axial ray and an edge ray reaches $\frac{\pi}{2}$ \cite{rayleigh1891x}. This distance is referred to as the Rayleigh distance in honor of Baron Rayleigh, though this specific expression is not commonly used today. More recently, the Rayleigh distance is often expressed as $\frac{2A^2}{\lambda}$.

It is essential to note that these contributions to wave propagation were primarily focused on light waves. However, in 1887, German physicist Heinrich Hertz demonstrated the existence of radio waves using what is now recognized as a dipole antenna. This discovery shifted significant research attention towards the near-field propagation of EM radio waves. In 1947, Cutler \emph{et al.} \cite{cutler1947microwave} suggested defining the field boundary as the axial distance from a radiating aperture to a point where the phase difference between the axial ray and an edge ray is $\frac{\pi}{8}$, calculated as $\frac{2A^2}{\lambda}$. Subsequently, in 1956, Polk \cite{polk1956optical} derived an exact expression for the Fresnel distance.

In 1983, the first patent related to NFC was issued, based on radio-frequency identification (RFID) technology. This innovation enabled communications between two electronic devices over short distances, typically just a few centimeters \cite{walton1983portable}. It is important to clarify that this technology primarily utilized the reactive near-field region for communications, which differs from the focus of this paper\footnote{In our paper, the term ``NFC'' specifically refers to wireless communications influenced by near-field EM effects resulting from the use of large-aperture arrays and high-frequency bands. The resulting near-field region is dominated by the radiating near field rather than the reactive near field \cite{ouyang2024impact}.}. For further details on research progress in the reactive near field, please refer to \cite{coskun2015survey,kim2017review}.

\begin{table*}[!t]\footnotesize
\caption{Timeline of NFC Milestones}
\centering
\begin{minipage}[t]{0.84\linewidth}
\color{gray}
\rule{\linewidth}{1pt}
\ytl{1678}{Huygens presented his ``Wave Theory of Light''}
\ytl{1801}{Young presented the double-slit diffraction experiment}
\ytl{1815}{Fresnel presented a series of memoirs about his understanding of diffraction}
\ytl{1821}{Fraunhofer constructed the first diffraction grating}
\ytl{1887}{Hertz demonstrated the existence of radio waves}
\ytl{1891}{Lord Rayleigh calculated the Rayleigh distance $\frac{A^2}{2\lambda}$}
\ytl{1947}{Cutler \emph{et al.} reformulated the Rayleigh distance as $\frac{2A^2}{\lambda}$}
\ytl{1956}{Polk calculated the Fresnel distance}
\ytl{1983}{The first patent on RFID-based NFC was granted}
\ytl{1984}{Winters formulated the initial theory of MIMO}
\ytl{1994}{The first patent on MIMO was granted}
\ytl{1996}{Foschini laid down crucial theoretical foundations for MIMO}
\ytl{1999}{Driessen and Foschini utilized a spherical wave-based model to characterize LoS MIMO}
\ytl{2003}{Jiang \emph{et al.} proposed to use spherical wave-based models to  describe short-range MIMO}
\ytl{2010}{Marzetta proposed the concept of massive MIMO}
\ytl{2015}{Channel measurement results on massive MIMO necessitated the use of a spherical wave-based channel model}
\ytl{2016}{Prather proposed the concept of holographic MIMO}
\ytl{2017}{Hu \emph{et al.} re-showed the potential of large intelligent surfaces in enhancing wireless transmissions}
\ytl{2018}{Amiri \emph{et al.} proposed the concept of ELAA}
\ytl{2023}{The first tutorial review of NFC was presented}
\rule{\linewidth}{1pt}%
\end{minipage}%
\label{Table: timeline}
\end{table*}

The RFID-based NFC systems developed initially in 1983 generally employed single-antenna transceivers. In 1984, Winters introduced the preliminary theory of multiple-input multiple-output (MIMO) communications \cite{winters1984optimum}. A decade later, in 1994, Paulraj and Kailath were awarded a patent that proposed the use of MIMO arrays to increase the number of distinct spatial channels, thereby enhancing communication capacity or overall system performance \cite{paulraj1994increasing}. Building on this innovation, in 1996, Foschini at Bell Labs established fundamental theoretical groundwork for MIMO systems \cite{foschini1996layered}.

Since 1996, extensive research has focused on the practical measurements and testing of MIMO channels to validate their performance. In 1999, Driessen and Foschini \cite{driessen1999capacity} introduced a spherical wave-based model to describe line-of-sight (LoS) propagation in indoor MIMO systems. Researchers during this era observed that the capacity of short-range MIMO channels, reconstructed from measured path parameters, always fell short of that measured directly \cite{molisch2002capacity,jiang2002path}. In 2003, Jiang and Ingram identified incorrect LoS path modeling as the cause of these mismatches, advocating the use of spherical wave models for more accurate channel response reconstructions \cite{jiang2003distributed}. By 2005, Jiang further endorsed spherical-wave models for short-range MIMO, underlining their significance to NFC \cite{jiang2005spherical}.

In 2010, Marzetta proposed the concept of massive MIMO, which involves deploying numerous antennas at both the base station (BS) and receivers to enhance cell coverage and facilitate beamforming \cite{marzetta2010noncooperative}. Subsequent channel measurements for massive MIMO systems suggested that large antenna arrays significantly expand their near-field regions, necessitating spherical wave-based channel models for large-aperture arrays \cite{zhou2015spherical}. The period from 2016 to 2018 saw the introduction of groundbreaking concepts such as holographic radio-frequency (RF) systems \cite{prather20235g, prather2017optically}, large intelligent surfaces \cite{hu2017potential}, and ELAAs \cite{amiri2018extremely}. These innovations, characterized by exceptionally large antenna dimensions, marked a shift from FFC to NFC, highlighting the importance of spherical-wave propagation. This shift attracted substantial research interest within the community. Fast forward to 2023, Liu \emph{et al.} \cite{liu2023near} published the first tutorial overview paper on NFC, providing a comprehensive guide for researchers from various fields about the evolving landscape of NFC technologies.

For reference, the key milestones of NFC are concisely outlined in the timeline depicted in Table {\ref{Table: timeline}}. This timeline synthesizes the historical developments related to near-field propagation and spherical waves discussed previously. Although NFC is not a new concept within the field of wireless communications, its renewed prominence is linked to the advent of innovative antenna configurations and emerging applications. These developments are further elaborated upon in Sections \ref{sec:NFA} and \ref{sec:Incorporating 6G Wireless}, respectively.
\subsection{Prior Works and Motivations}
Although several short magazine papers, tutorials, and surveys have explored the landscape of NFC, our work differs from these existing studies as follows.
\subsubsection{Magazine Papers}
Several magazine papers, such as \cite{cui2022near,zhang20236g,liu2023nearArxiv,an2023toward}, have provided high-level introductions to the unique properties of NFC compared to FFC. These articles discussed topics like channel modeling, beamforming design, and applications, highlighting the benefits, challenges, and future research opportunities in NFC-based wireless networks. Other magazine papers have focused on specific aspects of NFC, including wireless power transfer (WPT) \cite{zhang2022nearwpt}, beam management (beam training, tracking, and scheduling) \cite{you2023near}, degrees of freedom (DoFs) analysis \cite{ouyang2023near}, and integration with reconfigurable intelligent surfaces (RISs) \cite{mu2024reconfigurable}. These papers identified open research opportunities, challenges, and potential solutions. Near-field sensing (NISE) has also been a hot topic, with discussions in works focusing on separate sensing and communications \cite{wang2023terahertz} and integrated sensing and communications (ISAC) \cite{qu2023near,cong2023near,wang2023rethinking}.
\subsubsection{Tutorials}
Several tutorials have provided in-depth, step-by-step explanations of NFC concepts, methodologies, and techniques. One previous work \cite{liu2023near} covered deterministic channel modeling, hybrid beamfocusing architectures, and performance analysis for NFC. This tutorial was extended in \cite{wang2023tutorial,lu2023tutorial,bjornson2024towards}, with additional emphasis on stochastic channel modeling \cite{wang2023tutorial}, performance analysis \cite{lu2023tutorial}, physical aspects along with EM properties \cite{bjornson2024towards}, and practical design considerations \cite{wang2023tutorial,lu2023tutorial,bjornson2024towards}. Another tutorial focused on channel modeling and low-complexity signal processing schemes \cite{han2023towards}. Additionally, a tutorial on EM information theory (EMIT) for holographic MIMO included NFC as a case study \cite{wei2024electromagnetic}.
\subsubsection{Surveys}
Surveys on recent advances related to NFC include an overview of holographic MIMO with NFC supported by continuous arrays as a case study \cite{gong2022holographic}. Another survey reviewed recent advances and open challenges in ISAC, including near-field ISAC \cite{lu2024integrated}. However, these works primarily focused on holographic MIMO and ISAC, respectively, rather than NFC, as both can support NFC and FFC.

The arguments above suggest that the current literature primarily provides high-level overviews (i.e., magazine papers) and tutorials (step-by-step teaching materials) on general concepts or specific aspects of NFC. Moreover, the available surveys, such as \cite{gong2022holographic} and \cite{lu2024integrated}, treat NFC only as a selected case study. Consequently, a comprehensive survey of the historical developments and recent advances in the key properties of NFC is still lacking. Such a survey is crucial for researchers to understand the milestones and state-of-the-art developments in this field, potentially inspiring further contributions. Additionally, the fundamental performance limits of NFC and some potential applications in wireless networks are not fully explored. There is also an absence of comparative analysis of the mathematical tools commonly used for performance evaluations and optimizations in NFC. This paper aims to fill these gaps by offering a comprehensive literature survey of the principles governing NFC. Table \ref{tab:Comparision with Other Overview Papers}\footnote{The ``Novel Antenna Types'' category in Table \ref{tab:Comparision with Other Overview Papers} refers to both novel array architectures and novel array geometries, which are further detailed in Section \ref{sec:NFA} and Section \ref{Section: Array Geometries and Control Techniques}, respectively.} compares this treatise with existing magazine papers, surveys, and tutorials on NFC.

\begin{table*}[!t]
\caption{Comparison of this work with existing magazine/survey/tutorial papers on NFC. Here, $\bigstar$ and $\blacksquare$ refer to ``discussed in detail'' and ``mentioned but not considered in detail'', respectively. Moreover, ``EDoF'' refers to ``effective DoFs'', ``NLoS'' refers to ``non-LoS'', ``SPD'' refers to ``spatially-discrete'', and ``CAP'' refers to ``continuous-aperture''. The existing papers are listed in chronological order.}
\label{tab:Comparision with Other Overview Papers}
\centering
\resizebox{0.99\textwidth}{!}
{\begin{tabular}{|ll|llllllllllll|llllll|lll|}
\hline
\multicolumn{2}{|l|}{\multirow{2}{*}{\textbf{Existing Works}}}                                             & \multicolumn{12}{l|}{\textbf{Magazine Papers (High-Level Overviews)}}                                                                                                                                                                                                                                                                                                                                                                                                                                                                                       & \multicolumn{6}{l|}{\textbf{Tutorials (Step-by-Step Teaching Materials)}}                                                                                                                                                                                                                            & \multicolumn{3}{l|}{\textbf{Surveys (Literature Survey)}}                                                                      \\ \cline{3-23} 
\multicolumn{2}{|l|}{}                                                                                     & \multicolumn{1}{l|}{\cite{zhang2022nearwpt}} & \multicolumn{1}{l|}{\cite{cui2022near}} & \multicolumn{1}{l|}{\cite{liu2023nearArxiv}} & \multicolumn{1}{l|}{\cite{zhang20236g}} & \multicolumn{1}{l|}{\cite{wang2023terahertz}} & \multicolumn{1}{l|}{\cite{you2023near}} & \multicolumn{1}{l|}{\cite{ouyang2023near}} & \multicolumn{1}{l|}{\cite{qu2023near}} & \multicolumn{1}{l|}{\cite{an2023toward}} & \multicolumn{1}{l|}{\cite{cong2023near}} & \multicolumn{1}{l|}{\cite{wang2023rethinking}} & \cite{mu2024reconfigurable} & \multicolumn{1}{l|}{\cite{liu2023near}} & \multicolumn{1}{l|}{\cite{han2023towards}} & \multicolumn{1}{l|}{\cite{wang2023tutorial}} & \multicolumn{1}{l|}{\cite{lu2023tutorial}} & \multicolumn{1}{l|}{\cite{bjornson2024towards}} & \cite{wei2024electromagnetic} & \multicolumn{1}{l|}{\cite{gong2022holographic}} & \multicolumn{1}{l|}{\cite{lu2024integrated}} & This Work \\ \hline
\multicolumn{2}{|l|}{\textbf{Year}}                                                                        & \multicolumn{1}{l|}{2022}                    & \multicolumn{1}{l|}{2023}               & \multicolumn{1}{l|}{2023}                    & \multicolumn{1}{l|}{2023}               & \multicolumn{1}{l|}{2023}                     & \multicolumn{1}{l|}{2023}               & \multicolumn{1}{l|}{2023}                  & \multicolumn{1}{l|}{2023}              & \multicolumn{1}{l|}{2023}                & \multicolumn{1}{l|}{2023}                & \multicolumn{1}{l|}{2023}                      & 2024                        & \multicolumn{1}{l|}{2023}               & \multicolumn{1}{l|}{2023}                  & \multicolumn{1}{l|}{2024}                    & \multicolumn{1}{l|}{2024}                  & \multicolumn{1}{l|}{2024}                       & 2024                          & \multicolumn{1}{l|}{2023}                       & \multicolumn{1}{l|}{2024}                    & 2024      \\ \hline
\multicolumn{2}{|l|}{\textbf{Timeline}}                                                                    & \multicolumn{1}{l|}{}                        & \multicolumn{1}{l|}{}                   & \multicolumn{1}{l|}{}                        & \multicolumn{1}{l|}{}                   & \multicolumn{1}{l|}{}                         & \multicolumn{1}{l|}{}                   & \multicolumn{1}{l|}{}                      & \multicolumn{1}{l|}{}                  & \multicolumn{1}{l|}{}                    & \multicolumn{1}{l|}{}                    & \multicolumn{1}{l|}{}                          &                             & \multicolumn{1}{l|}{}                   & \multicolumn{1}{l|}{}                      & \multicolumn{1}{l|}{}                        & \multicolumn{1}{l|}{}                      & \multicolumn{1}{l|}{}                           &                               & \multicolumn{1}{l|}{}                           & \multicolumn{1}{l|}{}                        & $\bigstar$ \\ \hline
\multicolumn{1}{|l|}{\multirow{4}{*}{\textbf{NFC Principles}}}       & \textbf{Physics-based}              & \multicolumn{1}{l|}{}                        & \multicolumn{1}{l|}{}                   & \multicolumn{1}{l|}{}               & \multicolumn{1}{l|}{}                   & \multicolumn{1}{l|}{}                         & \multicolumn{1}{l|}{}                   & \multicolumn{1}{l|}{}                      & \multicolumn{1}{l|}{}                  & \multicolumn{1}{l|}{}                    & \multicolumn{1}{l|}{}                    & \multicolumn{1}{l|}{}                          &                             & \multicolumn{1}{l|}{$\bigstar$}          & \multicolumn{1}{l|}{}                      & \multicolumn{1}{l|}{$\blacksquare$}               & \multicolumn{1}{l|}{}                      & \multicolumn{1}{l|}{$\bigstar$}                  & $\bigstar$                     & \multicolumn{1}{l|}{}                           & \multicolumn{1}{l|}{}                        & $\bigstar$ \\ \cline{2-23} 
\multicolumn{1}{|l|}{}                                               & \textbf{Communications-based}       & \multicolumn{1}{l|}{$\bigstar$}               & \multicolumn{1}{l|}{$\bigstar$}          & \multicolumn{1}{l|}{$\bigstar$}               & \multicolumn{1}{l|}{$\bigstar$}          & \multicolumn{1}{l|}{$\bigstar$}                & \multicolumn{1}{l|}{$\bigstar$}          & \multicolumn{1}{l|}{$\bigstar$}             & \multicolumn{1}{l|}{$\bigstar$}         & \multicolumn{1}{l|}{$\bigstar$}           & \multicolumn{1}{l|}{$\bigstar$}           & \multicolumn{1}{l|}{$\bigstar$}                 & $\bigstar$                   & \multicolumn{1}{l|}{$\bigstar$}          & \multicolumn{1}{l|}{$\bigstar$}             & \multicolumn{1}{l|}{$\bigstar$}               & \multicolumn{1}{l|}{$\bigstar$}             & \multicolumn{1}{l|}{$\bigstar$}                  & $\bigstar$                     & \multicolumn{1}{l|}{}                           & \multicolumn{1}{l|}{}                        & $\bigstar$ \\ \cline{2-23} 
\multicolumn{1}{|l|}{}                                               & \textbf{Novel Antenna Types}        & \multicolumn{1}{l|}{}                        & \multicolumn{1}{l|}{}                   & \multicolumn{1}{l|}{}                        & \multicolumn{1}{l|}{}                   & \multicolumn{1}{l|}{}                         & \multicolumn{1}{l|}{}                   & \multicolumn{1}{l|}{}                      & \multicolumn{1}{l|}{}                  & \multicolumn{1}{l|}{}                    & \multicolumn{1}{l|}{}                    & \multicolumn{1}{l|}{}                          & $\bigstar$                   & \multicolumn{1}{l|}{}                   & \multicolumn{1}{l|}{$\bigstar$}             & \multicolumn{1}{l|}{}                        & \multicolumn{1}{l|}{$\bigstar$}                      & \multicolumn{1}{l|}{$\bigstar$}                  &                               & \multicolumn{1}{l|}{}                           & \multicolumn{1}{l|}{}                        & $\bigstar$ \\ \cline{2-23} 
\multicolumn{1}{|l|}{}                                               & \textbf{EM Information Theory}         & \multicolumn{1}{l|}{}                        & \multicolumn{1}{l|}{}                   & \multicolumn{1}{l|}{}                        & \multicolumn{1}{l|}{}                   & \multicolumn{1}{l|}{}                         & \multicolumn{1}{l|}{}                   & \multicolumn{1}{l|}{$\blacksquare$}             & \multicolumn{1}{l|}{}                  & \multicolumn{1}{l|}{}                    & \multicolumn{1}{l|}{}                    & \multicolumn{1}{l|}{}                          &                             & \multicolumn{1}{l|}{}                   & \multicolumn{1}{l|}{}                      & \multicolumn{1}{l|}{}                        & \multicolumn{1}{l|}{$\blacksquare$}             & \multicolumn{1}{l|}{$\bigstar$}                  & $\bigstar$                     & \multicolumn{1}{l|}{}                           & \multicolumn{1}{l|}{}                        & $\blacksquare$ \\ \hline
\multicolumn{1}{|l|}{\multirow{5}{*}{\textbf{Channel Modeling}}}     & \textbf{LoS (SPD)}                  & \multicolumn{1}{l|}{}                        & \multicolumn{1}{l|}{}                   & \multicolumn{1}{l|}{$\blacksquare$}               & \multicolumn{1}{l|}{$\blacksquare$}          & \multicolumn{1}{l|}{$\blacksquare$}                         & \multicolumn{1}{l|}{}                   & \multicolumn{1}{l|}{$\blacksquare$}             & \multicolumn{1}{l|}{}                  & \multicolumn{1}{l|}{$\blacksquare$}           & \multicolumn{1}{l|}{}                    & \multicolumn{1}{l|}{$\blacksquare$}                 & $\blacksquare$                   & \multicolumn{1}{l|}{$\bigstar$}          & \multicolumn{1}{l|}{$\bigstar$}             & \multicolumn{1}{l|}{$\bigstar$}               & \multicolumn{1}{l|}{$\bigstar$}             & \multicolumn{1}{l|}{$\bigstar$}                  &                               & \multicolumn{1}{l|}{}                           & \multicolumn{1}{l|}{}                        & $\bigstar$ \\ \cline{2-23} 
\multicolumn{1}{|l|}{}                                               & \textbf{LoS (CAP)}                  & \multicolumn{1}{l|}{}                        & \multicolumn{1}{l|}{}                   & \multicolumn{1}{l|}{$\blacksquare$}               & \multicolumn{1}{l|}{}                   & \multicolumn{1}{l|}{}                         & \multicolumn{1}{l|}{}                   & \multicolumn{1}{l|}{$\blacksquare$}             & \multicolumn{1}{l|}{}                  & \multicolumn{1}{l|}{}                    & \multicolumn{1}{l|}{}                    & \multicolumn{1}{l|}{}                          & $\blacksquare$                   & \multicolumn{1}{l|}{$\bigstar$}          & \multicolumn{1}{l|}{}                      & \multicolumn{1}{l|}{$\bigstar$}               & \multicolumn{1}{l|}{$\blacksquare$}             & \multicolumn{1}{l|}{$\bigstar$}                  &                               & \multicolumn{1}{l|}{$\bigstar$}                  & \multicolumn{1}{l|}{}                        & $\bigstar$ \\ \cline{2-23} 
\multicolumn{1}{|l|}{}                                               & \textbf{NLoS (SPD)}                 & \multicolumn{1}{l|}{}                        & \multicolumn{1}{l|}{}                   & \multicolumn{1}{l|}{}                        & \multicolumn{1}{l|}{}                   & \multicolumn{1}{l|}{}                         & \multicolumn{1}{l|}{}                   & \multicolumn{1}{l|}{}                      & \multicolumn{1}{l|}{}                  & \multicolumn{1}{l|}{}                    & \multicolumn{1}{l|}{}                    & \multicolumn{1}{l|}{}                          &                             & \multicolumn{1}{l|}{$\blacksquare$}          & \multicolumn{1}{l|}{$\bigstar$}             & \multicolumn{1}{l|}{$\bigstar$}               & \multicolumn{1}{l|}{$\bigstar$}             & \multicolumn{1}{l|}{$\blacksquare$}                  &                               & \multicolumn{1}{l|}{}                           & \multicolumn{1}{l|}{}                        & $\bigstar$ \\ \cline{2-23} 
\multicolumn{1}{|l|}{}                                               & \textbf{NLoS (CAP)}                 & \multicolumn{1}{l|}{}                        & \multicolumn{1}{l|}{}                   & \multicolumn{1}{l|}{}                        & \multicolumn{1}{l|}{}                   & \multicolumn{1}{l|}{}                         & \multicolumn{1}{l|}{}                   & \multicolumn{1}{l|}{}                      & \multicolumn{1}{l|}{}                  & \multicolumn{1}{l|}{}                    & \multicolumn{1}{l|}{}                    & \multicolumn{1}{l|}{}                          &                             & \multicolumn{1}{l|}{}                   & \multicolumn{1}{l|}{}                      & \multicolumn{1}{l|}{$\bigstar$}               & \multicolumn{1}{l|}{}                      & \multicolumn{1}{l|}{}                           &                               & \multicolumn{1}{l|}{$\bigstar$}                  & \multicolumn{1}{l|}{}                        & $\bigstar$ \\ \cline{2-23} 
\multicolumn{1}{|l|}{}                                               & \textbf{Hybrid LoS/NLoS}            & \multicolumn{1}{l|}{}                        & \multicolumn{1}{l|}{}                   & \multicolumn{1}{l|}{}                        & \multicolumn{1}{l|}{}                   & \multicolumn{1}{l|}{}                         & \multicolumn{1}{l|}{}                   & \multicolumn{1}{l|}{}                      & \multicolumn{1}{l|}{}                  & \multicolumn{1}{l|}{}                    & \multicolumn{1}{l|}{}                    & \multicolumn{1}{l|}{}                          & $\blacksquare$                   & \multicolumn{1}{l|}{}                   & \multicolumn{1}{l|}{}                      & \multicolumn{1}{l|}{$\blacksquare$ }               & \multicolumn{1}{l|}{$\blacksquare$}             & \multicolumn{1}{l|}{}                           &                               & \multicolumn{1}{l|}{}                           & \multicolumn{1}{l|}{}                        & $\blacksquare$  \\ \hline
\multicolumn{1}{|l|}{\multirow{6}{*}{\textbf{Performance Analysis}}} & \textbf{DoFs/EDoFs (SPD)}           & \multicolumn{1}{l|}{}                        & \multicolumn{1}{l|}{}                   & \multicolumn{1}{l|}{$\blacksquare$}               & \multicolumn{1}{l|}{}                   & \multicolumn{1}{l|}{}                         & \multicolumn{1}{l|}{}                   & \multicolumn{1}{l|}{$\bigstar$}             & \multicolumn{1}{l|}{$\blacksquare$}         & \multicolumn{1}{l|}{}                    & \multicolumn{1}{l|}{}                    & \multicolumn{1}{l|}{}                          & $\blacksquare$                   & \multicolumn{1}{l|}{$\blacksquare$}          & \multicolumn{1}{l|}{}                      & \multicolumn{1}{l|}{$\blacksquare$}                        & \multicolumn{1}{l|}{$\bigstar$}             & \multicolumn{1}{l|}{$\blacksquare$}                  &                               & \multicolumn{1}{l|}{}                           & \multicolumn{1}{l|}{}                        & $\bigstar$ \\ \cline{2-23} 
\multicolumn{1}{|l|}{}                                               & \textbf{DoFs/EDoFs (CAP)}           & \multicolumn{1}{l|}{}                        & \multicolumn{1}{l|}{}                   & \multicolumn{1}{l|}{$\blacksquare$}               & \multicolumn{1}{l|}{}                   & \multicolumn{1}{l|}{}                         & \multicolumn{1}{l|}{}                   & \multicolumn{1}{l|}{$\bigstar$}             & \multicolumn{1}{l|}{}                  & \multicolumn{1}{l|}{}                    & \multicolumn{1}{l|}{}                    & \multicolumn{1}{l|}{}                          & $\blacksquare$                   & \multicolumn{1}{l|}{$\blacksquare$}          & \multicolumn{1}{l|}{}                      & \multicolumn{1}{l|}{}                        & \multicolumn{1}{l|}{$\bigstar$}             & \multicolumn{1}{l|}{$\blacksquare$}                  &                               & \multicolumn{1}{l|}{$\bigstar$}                  & \multicolumn{1}{l|}{}                        & $\bigstar$ \\ \cline{2-23} 
\multicolumn{1}{|l|}{}                                               & \textbf{Power Scaling Law (SPD)}    & \multicolumn{1}{l|}{}                        & \multicolumn{1}{l|}{}                   & \multicolumn{1}{l|}{$\blacksquare$}               & \multicolumn{1}{l|}{}                   & \multicolumn{1}{l|}{}                         & \multicolumn{1}{l|}{}                   & \multicolumn{1}{l|}{}                      & \multicolumn{1}{l|}{}                  & \multicolumn{1}{l|}{}                    & \multicolumn{1}{l|}{}                    & \multicolumn{1}{l|}{}                          & $\blacksquare$                   & \multicolumn{1}{l|}{$\bigstar$}          & \multicolumn{1}{l|}{}                      & \multicolumn{1}{l|}{}                        & \multicolumn{1}{l|}{$\bigstar$}             & \multicolumn{1}{l|}{}                           &                               & \multicolumn{1}{l|}{}                           & \multicolumn{1}{l|}{}                        & $\bigstar$ \\ \cline{2-23} 
\multicolumn{1}{|l|}{}                                               & \textbf{Power Scaling Law (CAP)}    & \multicolumn{1}{l|}{}                        & \multicolumn{1}{l|}{}                   & \multicolumn{1}{l|}{$\blacksquare$}               & \multicolumn{1}{l|}{}                   & \multicolumn{1}{l|}{}                         & \multicolumn{1}{l|}{}                   & \multicolumn{1}{l|}{}                      & \multicolumn{1}{l|}{}                  & \multicolumn{1}{l|}{}                    & \multicolumn{1}{l|}{}                    & \multicolumn{1}{l|}{}                          &                             & \multicolumn{1}{l|}{$\bigstar$}          & \multicolumn{1}{l|}{}                      & \multicolumn{1}{l|}{}                        & \multicolumn{1}{l|}{}                      & \multicolumn{1}{l|}{}                           &                               & \multicolumn{1}{l|}{}                           & \multicolumn{1}{l|}{}                        & $\bigstar$ \\ \cline{2-23} 
\multicolumn{1}{|l|}{}                                               & \textbf{Transmission Rate (SPD)}    & \multicolumn{1}{l|}{}                        & \multicolumn{1}{l|}{$\bigstar$}          & \multicolumn{1}{l|}{}                        & \multicolumn{1}{l|}{}                   & \multicolumn{1}{l|}{}                         & \multicolumn{1}{l|}{}                   & \multicolumn{1}{l|}{}                      & \multicolumn{1}{l|}{}                  & \multicolumn{1}{l|}{}                    & \multicolumn{1}{l|}{}                    & \multicolumn{1}{l|}{}                          &                             & \multicolumn{1}{l|}{}                   & \multicolumn{1}{l|}{}                      & \multicolumn{1}{l|}{}                        & \multicolumn{1}{l|}{$\bigstar$}             & \multicolumn{1}{l|}{$\bigstar$}                  &                               & \multicolumn{1}{l|}{}                           & \multicolumn{1}{l|}{}                        & $\bigstar$ \\ \cline{2-23} 
\multicolumn{1}{|l|}{}                                               & \textbf{Transmission Rate (CAP)}    & \multicolumn{1}{l|}{}                        & \multicolumn{1}{l|}{}                   & \multicolumn{1}{l|}{}                        & \multicolumn{1}{l|}{}                   & \multicolumn{1}{l|}{}                         & \multicolumn{1}{l|}{}                   & \multicolumn{1}{l|}{}                      & \multicolumn{1}{l|}{}                  & \multicolumn{1}{l|}{}                    & \multicolumn{1}{l|}{}                    & \multicolumn{1}{l|}{}                          &                             & \multicolumn{1}{l|}{}                   & \multicolumn{1}{l|}{}                      & \multicolumn{1}{l|}{}                        & \multicolumn{1}{l|}{}                      & \multicolumn{1}{l|}{}                           &                               & \multicolumn{1}{l|}{$\bigstar$}                  & \multicolumn{1}{l|}{}                        & $\bigstar$ \\ \hline
\multicolumn{1}{|l|}{\multirow{3}{*}{\textbf{Signal Processing}}}    & \textbf{Channel Estimation}         & \multicolumn{1}{l|}{}                        & \multicolumn{1}{l|}{$\bigstar$}          & \multicolumn{1}{l|}{}                        & \multicolumn{1}{l|}{$\blacksquare$}          & \multicolumn{1}{l|}{}                         & \multicolumn{1}{l|}{}                   & \multicolumn{1}{l|}{}                      & \multicolumn{1}{l|}{$\blacksquare$}         & \multicolumn{1}{l|}{$\bigstar$}           & \multicolumn{1}{l|}{}                    & \multicolumn{1}{l|}{}                          & $\blacksquare$                  & \multicolumn{1}{l|}{}                   & \multicolumn{1}{l|}{$\bigstar$}             & \multicolumn{1}{l|}{$\bigstar$}               & \multicolumn{1}{l|}{$\bigstar$}             & \multicolumn{1}{l|}{$\bigstar$}                  &                               & \multicolumn{1}{l|}{$\bigstar$}                  & \multicolumn{1}{l|}{}                        & $\bigstar$ \\ \cline{2-23} 
\multicolumn{1}{|l|}{}                                               & \textbf{Beamforming Design}         & \multicolumn{1}{l|}{}                        & \multicolumn{1}{l|}{$\blacksquare$}          & \multicolumn{1}{l|}{$\bigstar$}               & \multicolumn{1}{l|}{$\bigstar$}          & \multicolumn{1}{l|}{$\bigstar$}                & \multicolumn{1}{l|}{$\bigstar$}          & \multicolumn{1}{l|}{}                      & \multicolumn{1}{l|}{$\blacksquare$}         & \multicolumn{1}{l|}{$\bigstar$}           & \multicolumn{1}{l|}{$\blacksquare$}           & \multicolumn{1}{l|}{$\bigstar$}                 & $\blacksquare$                   & \multicolumn{1}{l|}{$\bigstar$}          & \multicolumn{1}{l|}{$\bigstar$}             & \multicolumn{1}{l|}{$\bigstar$}               & \multicolumn{1}{l|}{$\bigstar$}             & \multicolumn{1}{l|}{}                           &                               & \multicolumn{1}{l|}{$\bigstar$}                  & \multicolumn{1}{l|}{}                        & $\bigstar$ \\ \cline{2-23} 
\multicolumn{1}{|l|}{}                                               & \textbf{Beam Training}              & \multicolumn{1}{l|}{}                        & \multicolumn{1}{l|}{$\blacksquare$}          & \multicolumn{1}{l|}{}                        & \multicolumn{1}{l|}{}                   & \multicolumn{1}{l|}{}                         & \multicolumn{1}{l|}{$\bigstar$}          & \multicolumn{1}{l|}{}                      & \multicolumn{1}{l|}{}                  & \multicolumn{1}{l|}{}                    & \multicolumn{1}{l|}{$\bigstar$}           & \multicolumn{1}{l|}{}                          & $\bigstar$                   & \multicolumn{1}{l|}{$\bigstar$}          & \multicolumn{1}{l|}{}                      & \multicolumn{1}{l|}{}                        & \multicolumn{1}{l|}{$\bigstar$}             & \multicolumn{1}{l|}{}                           &                               & \multicolumn{1}{l|}{$\blacksquare$}                           & \multicolumn{1}{l|}{}                        & $\bigstar$ \\ \hline
\multicolumn{1}{|l|}{\multirow{4}{*}{\textbf{Applications}}}         & \textbf{NISE}                       & \multicolumn{1}{l|}{}                        & \multicolumn{1}{l|}{}                   & \multicolumn{1}{l|}{$\bigstar$}               & \multicolumn{1}{l|}{$\blacksquare$}          & \multicolumn{1}{l|}{$\bigstar$}                & \multicolumn{1}{l|}{$\blacksquare$}          & \multicolumn{1}{l|}{}                      & \multicolumn{1}{l|}{$\bigstar$}         & \multicolumn{1}{l|}{$\bigstar$}           & \multicolumn{1}{l|}{$\bigstar$}           & \multicolumn{1}{l|}{$\bigstar$}                 &                             & \multicolumn{1}{l|}{}                   & \multicolumn{1}{l|}{$\blacksquare$}             & \multicolumn{1}{l|}{}                        & \multicolumn{1}{l|}{$\bigstar$}             & \multicolumn{1}{l|}{}                           &                               & \multicolumn{1}{l|}{$\blacksquare$}                           & \multicolumn{1}{l|}{$\blacksquare$}               & $\bigstar$ \\ \cline{2-23} 
\multicolumn{1}{|l|}{}                                               & \textbf{Massive Connectivity}       & \multicolumn{1}{l|}{}                        & \multicolumn{1}{l|}{}                   & \multicolumn{1}{l|}{$\blacksquare$}               & \multicolumn{1}{l|}{$\blacksquare$}          & \multicolumn{1}{l|}{}                         & \multicolumn{1}{l|}{}                   & \multicolumn{1}{l|}{}                      & \multicolumn{1}{l|}{}                  & \multicolumn{1}{l|}{}                    & \multicolumn{1}{l|}{}                    & \multicolumn{1}{l|}{}                          &                             & \multicolumn{1}{l|}{}                   & \multicolumn{1}{l|}{}                      & \multicolumn{1}{l|}{$\blacksquare$}               & \multicolumn{1}{l|}{}                      & \multicolumn{1}{l|}{}                           &                               & \multicolumn{1}{l|}{}                           & \multicolumn{1}{l|}{}                        & $\blacksquare$ \\ \cline{2-23} 
\multicolumn{1}{|l|}{}                                               & \textbf{Sustainability Enhancement} & \multicolumn{1}{l|}{$\bigstar$}               & \multicolumn{1}{l|}{}                   & \multicolumn{1}{l|}{$\blacksquare$}               & \multicolumn{1}{l|}{$\blacksquare$}          & \multicolumn{1}{l|}{}                         & \multicolumn{1}{l|}{$\blacksquare$}          & \multicolumn{1}{l|}{}                      & \multicolumn{1}{l|}{}                  & \multicolumn{1}{l|}{}                    & \multicolumn{1}{l|}{}                    & \multicolumn{1}{l|}{}                          &                             & \multicolumn{1}{l|}{}                   & \multicolumn{1}{l|}{}                      & \multicolumn{1}{l|}{}                        & \multicolumn{1}{l|}{}                      & \multicolumn{1}{l|}{}                           &                               & \multicolumn{1}{l|}{$\blacksquare$}                           & \multicolumn{1}{l|}{}                        & $\blacksquare$ \\ \cline{2-23} 
\multicolumn{1}{|l|}{}                                               & \textbf{Information Safeguarding}   & \multicolumn{1}{l|}{}                        & \multicolumn{1}{l|}{}                   & \multicolumn{1}{l|}{$\blacksquare$}               & \multicolumn{1}{l|}{}                   & \multicolumn{1}{l|}{}                         & \multicolumn{1}{l|}{}                   & \multicolumn{1}{l|}{}                      & \multicolumn{1}{l|}{}                  & \multicolumn{1}{l|}{}                    & \multicolumn{1}{l|}{}                    & \multicolumn{1}{l|}{}                          &                             & \multicolumn{1}{l|}{}                   & \multicolumn{1}{l|}{}                      & \multicolumn{1}{l|}{$\bigstar$}               & \multicolumn{1}{l|}{}                      & \multicolumn{1}{l|}{}                           &                               & \multicolumn{1}{l|}{}                           & \multicolumn{1}{l|}{}                        & $\blacksquare$ \\ \hline
\end{tabular}}
\end{table*}

\subsection{Contributions}
The key contributions of our work are outlined as follows.
\begin{enumerate}
  \item \textbf{\emph{Fundamental Investigation of NFC Principles:}} We explore NFC principles through the lenses of EM theory in physics and information transmission theory in wireless communications. We categorize near-field channel models into two fundamental types based on EM wave propagation. Our investigation examines the physical properties of the near-field region and transitions to communication-theoretic and information-theoretic aspects.
  \item \textbf{\emph{Overview of Near-Field Channel Models:}} We provide a detailed overview of basic near-field channel models for spatially-discrete (SPD) and continuous-aperture (CAP) arrays. Regarding each array type, we review existing channel models, including deterministic LoS models and statistical multipath models. We focus on the spatial non-stationarity of near-field channels, a key feature distinguishing NFC from FFC.
  \item \textbf{\emph{Performance Evaluation Techniques for NFC:}} We develop performance evaluation techniques for NFC, using the reviewed near-field channel models. We incorporate three representative performance metrics: DoFs/effective DoFs (EDoFs), power scaling law, and transmission rate. Our summary includes current research contributions and outlines their advantages and limitations.
  \item \textbf{\emph{Signal Processing Techniques for NFC:}} We explore signal processing techniques for NFC in terms of channel estimation, beamforming design, and low-complexity beam training. For each optimization objective, we consolidate protocols and approaches, accounting for the unique properties of the near-field region and novel antenna and array architectures dedicated to NFC.
  \item \textbf{\emph{Integration with Other Technologies:}} We identify significant research opportunities by incorporating near-field models into other promising NG applications, such as NISE, massive connectivity, sustainability enhancement, and information safeguarding. We discuss potential solutions and future research avenues.
\end{enumerate}

\begin{figure}[!t]
 \centering
\includegraphics[width=0.4\textwidth]{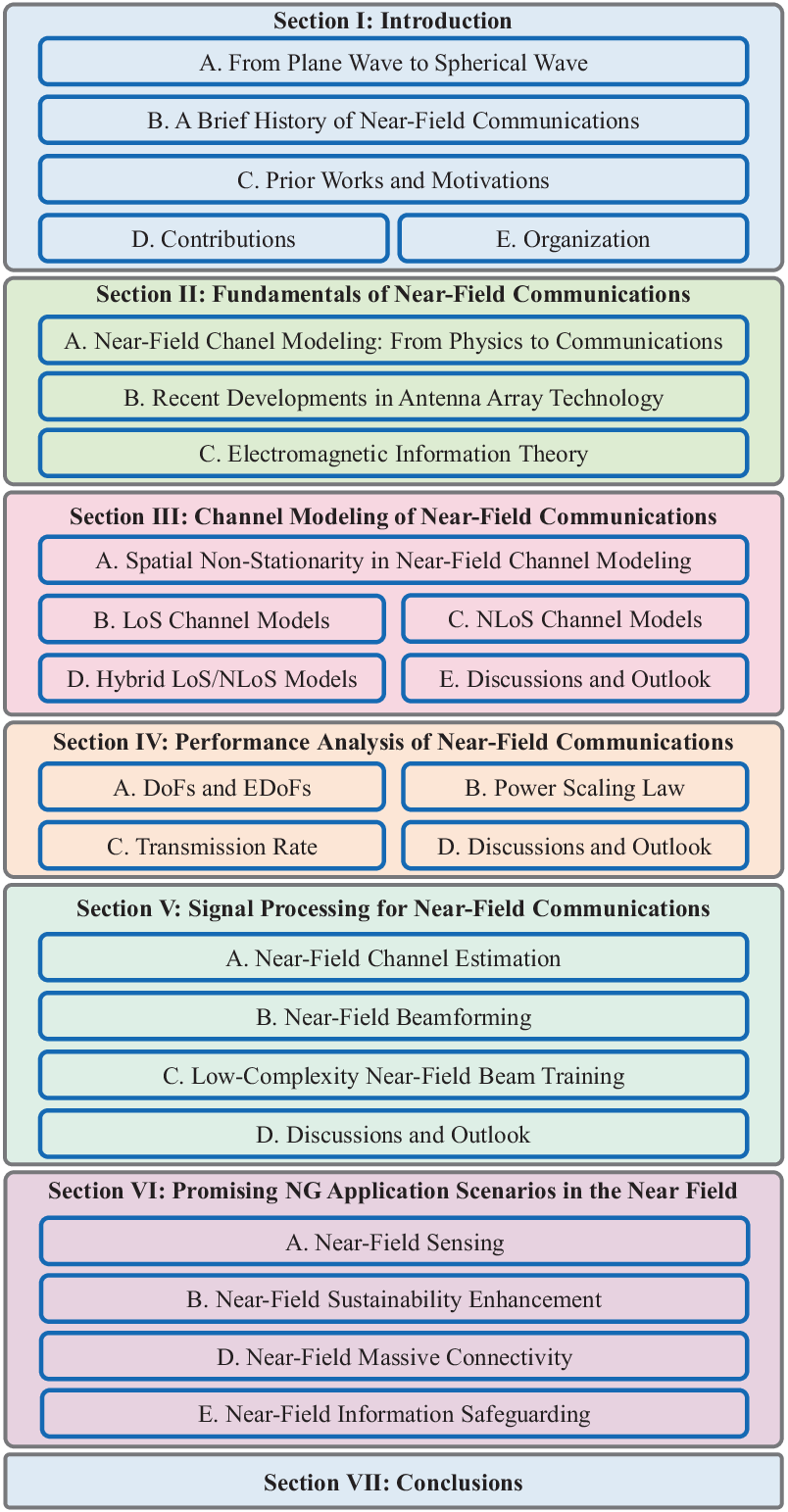}
\caption{Condensed overview of this survey on NFC.}
\label{Figure: NFC_Outline}
\end{figure}

\subsection{Organization}
The remainder of this paper is structured as follows. Section \ref{section:Fundamentals of NFC} explores the fundamental principles of NFC from the perspectives of physics and EM theory. Section \ref{Sec:CMoN} investigates near-field LoS and non-LoS (NLoS) channel models for both SPD and CAP arrays. In Section \ref{Section: Performance Analysis of NFC}, we review current contributions to near-field performance analysis. Signal processing techniques for NFC, including channel estimation, beamforming, and beam training, are summarized in Section \ref{Section: Signal Processing for NFC}. Section \ref{sec:Incorporating 6G Wireless} examines the incorporation of near-field models into other emerging technologies. Finally, Section \ref{sec:conclusion} concludes the paper. {\figurename} {\ref{Figure: NFC_Outline}} illustrates the organizational structure of the paper, and Table \ref{tab:LIST OF ACRONYMS} presents key abbreviations used throughout this treatise.

\begin{table*}[!t]
\caption{List of Acronyms}
\label{tab:LIST OF ACRONYMS}
\centering
\resizebox{0.7\textwidth}{!}{
\begin{tabular}{|l||l|l||l|}
\hline
1D	&	One-Dimensional	&	NISE	&	Near-Field Sensing	\\ \hline
2D	&	Two-Dimensional	&	NLoS	&	Non-Line-of-Sight	\\ \hline
3D	&	Three-Dimensional	&	NOMA	&	Non-Orthogonal Multiple Access	\\ \hline
4D	&	Four-Dimensional	&	NUSW	&	Non-Uniform Spherical-Wave	\\ \hline
5G	&	Fifth-Generation	&	OMP	&	Orthogonal Matching Pursuit	\\ \hline
AoA	&	Angle of Arrival	&	PAS	&	Power Angular Spectrum	\\ \hline
AoD	&	Angle of Departure	&	PL	&	Polarization Loss	\\ \hline
ASR	&	Average Sum Rate	&	PLOS	&	Power Location Spectrum	\\ \hline
ATR	&	Average Transmission Rate	&	PPBSM	&	Physical Propagation-Based Stochastic Model	\\ \hline
BBU	&	Baseband Unit	&	PS	&	Phase Shifter	\\ \hline
BS	&	Base Station	&	PSWF 	&	Prolate Spheroidal Wave Function	\\ \hline
CAP	&	Continuous-Aperture	&	RCS	&	Radar Cross-Section	\\ \hline
CBSM	&	Correlation-Based Stochastic Model	&	RF	&	Radio-Frequency	\\ \hline
CCM	&	Clustered Channel Model	&	RFID	&	Radio-Frequency Identification	\\ \hline
CRB	&	Cram´er-Rao Bound	&	RHS	&	Reconfigurable Holographic Surface	\\ \hline
CSI	&	Channel State Information	&	RIS	&	Reconfigurable Intelligent Surface	\\ \hline
DMA	&	Dynamic Metasurface Antenna	&	RoS	&	Rough Surface	\\ \hline
DoF	&	Degree of Freedom	&	RRS	&	Reconfigurable Refractive Surface	\\ \hline
DSM	&	Double-Scattering Model	&	RZF	&	Regularized Zero-Forcing	\\ \hline
EAL	&	Effective Aperture Loss	&	SBL	&	Sparse Bayesian Learning	\\ \hline
EDoF	&	Effective Degree of Freedom	&	SIC	&	Successive Interference Cancellation	\\ \hline
ELAA	&	Extremely Large Aperture Array	&	SINR	&	Signal-to-Interference-Plus-Noise Ratio	\\ \hline
EM	&	Electromagnetic	&	SISO	&	Single-Input Single-Output	\\ \hline
EMIT	&	Electromagnetic Information Theory	&	SNR	&	Signal-to-Noise Ratio	\\ \hline
ETPA	&	Equal Transmit Power Allocation	&	SPD	&	Spatially-Discrete	\\ \hline
ETR	&	Ergodic Transmission Rate	&	SSM	&	Single-Scattering Model	\\ \hline
FDC	&	Finite-Dimensional Channel	&	STAR	&	Simultaneously Transmitting and Reflecting	\\ \hline
FFC	&	Far-Field Communications	&	S-V	&	Saleh-Valenzuela (S-V)	\\ \hline
FPBSM	&	Fourier Plane-Wave-Based Stochastic Model 	&	SVD	&	Singular Value Decomposition	\\ \hline
FSPL	&	Free-Space Path Loss	&	SWIPT	&	Simultaneous Wireless Information and Power Transfer	\\ \hline
i.i.d.	&	Independent and Identically Distributed	&	THz	&	Terahertz (THz)	\\ \hline
i.n.i.d.	&	Independent but not Identically Distributed	&	TTD	&	True Time Delay	\\ \hline
IoT	&	Internet-of-Things 	&	UCA	&	Uniform Circular Array	\\ \hline
ISAC	&	Integrated Sensing and Communications	&	ULA	&	Uniform Linear Array	\\ \hline
ISI	&	Inter-Symbol Interference	&	UPA	&	Uniform Planar Array	\\ \hline
LMMSE	&	Linear Minimum Mean Square Error	&	UPL	&	Unequal Path Loss	\\ \hline
LoS	&	Line-of-Sight	&	USW	&	Uniform Spherical-Wave	\\ \hline
LS	&	Least-Squares	&	VCR	&	Virtual Channel Representation	\\ \hline
MIMO	&	Multiple-Input Multiple-Output	&	VR	&	Visibility Region	\\ \hline
MISO	&	Multiple-Input Single-Output	&	w.r.t. 	&	With Respect To	\\ \hline
MLA	&	Modular Linear Array 	&	WDM	&	Wavenumber-Division Multiplexing	\\ \hline
MMSE	&	Minimum Mean Square Error	&	WPT	&	Wireless Power Transfer	\\ \hline
mmWave	&	Millimeter-Wave	&	WSMS	&	Widely-Spaced Multi-Subarray	\\ \hline
MRC	&	Maximum Ratio Combining	&	XL-MIMO	&	Extremely Large Multiple-Input Multiple-Output	\\ \hline
MRDN	&	Multiple Residual Dense Network	&	XR	&	Extended Reality	\\ \hline
NFC	&	Near-Field Communications	&	ZF	&	Zero-Forcing	\\ \hline
NG	&	Next-Generation	&		&		\\ \hline
\end{tabular}}
\end{table*}
\section{Fundamentals of Near-Field Communications}\label{section:Fundamentals of NFC}
The discovery of EM induction sparked the wireless revolution, enabling information to be transmitted through the air without the constraints of physical wires. The synergy between physics and EM theory has played a fundamental and indispensable role in wireless communications. Today, we are witnessing a pivotal moment where the integration of advanced antenna development and EM theory continues to elevate wireless technology to new heights. Therefore, it is essential to understand NFC from the perspectives of physics, communication, and information theory. In this section, we will first discuss three categories of near-field channel models from a physics perspective. We will then summarize recent developments in antenna technology for NFC from a communication perspective. Moreover, we will provide a concise introduction to EMIT.

\subsection{Near-Field Chanel Modeling: From Physics to Communications}\label{sec:FPC}
From the perspective of physics, the wireless channel response is governed by the propagation of EM waves in specific radio environments. Although the laws of EM radiation are elegantly described by Maxwell's equations, the complexity of channel modeling arises from the varying geometrical and physical properties of both transceivers and radio environments. As such, calculating the channel response purely from a theoretical standpoint is nearly impossible for any practical communication system. To address this challenge, pioneers in wireless communications developed various techniques and models to approximate the physical channel.

For instance, Friis' formula \cite{goldsmith2005wireless} provides a far-field approximation for the path loss of a wireless channel by considering an omnidirectional transmit antenna, a receive antenna with a certain gain, and a free-space radio environment. In urban and indoor environments, a radio signal transmitted from a fixed source encounters multiple objects. To approximate these complex radio environments, techniques such as ray-tracing \cite{patzold2002mobile}, finite-state Markov models \cite{350282}, and other simple wireless channel paradigms \cite{734678} were developed. In practical scenarios, blockages and multipath propagation give rise to shadowing \cite{778178} and multipath fading, which can be modeled using Rayleigh \cite{601747} or Rician \cite{rice1951reflection} fading models.

However, all the aforementioned techniques assume the far-field approximation derived from Friis’ original formula. In contrast, other EM-based techniques do not rely on far-field approximations and are applicable to the near-field region. Generally speaking, near-field channel models can be loosely classified into two categories: CAP models and SPD models. In most FFC scenarios, CAP models are not necessary because SPD models provide reasonably good accuracy \cite{liu2023near}. However, as the aperture size and spatial density of the transceiver antennas increase, CAP models become essential for accurately characterizing NFC channels.

\subsubsection{Continuous-Aperture Models}
As introduced in Section \ref{sec:NFA}, most newly developed antenna configurations tend to have very high spatial densities. For these types of arrays, only CAP channel models can accurately characterize their spatially-continuous array responses. CAP channel models are a category of physically compliant models based on different physics principles. Two of the most commonly-used principles are the \textbf{EM wave equation} (which stems from \emph{Maxwell's equations}) and the \textbf{Huygens-Fresnel principle} \cite{huygens1885traite} (which originates from \emph{optics}). Both principles describe the propagation of EM waves within a medium or vacuum.

In wireless communication channel modeling, these two principles lead to different approaches: Green's function-based approaches and integral equation approaches. For a monochromatic field varying in time as ${\rm{e}}^{-{\rm{j}}\omega t}$, the EM wave equation reduces to the \emph{Helmholtz equation} \cite{born2013principles}, and the Green's function method is used to solve the Helmholtz equation. Conversely, for EM waves propagating in homogeneous media, the \emph{Huygens-Fresnel principle} states that every point on a wavefront is itself the source of spherical wavelets. Based on this principle, describing the propagation of EM waves can be transformed into an integration problem.
\subsubsection{Spatially-Discrete Models}
\begin{figure}[!t]
    \centering
    \includegraphics[width=0.45\textwidth]{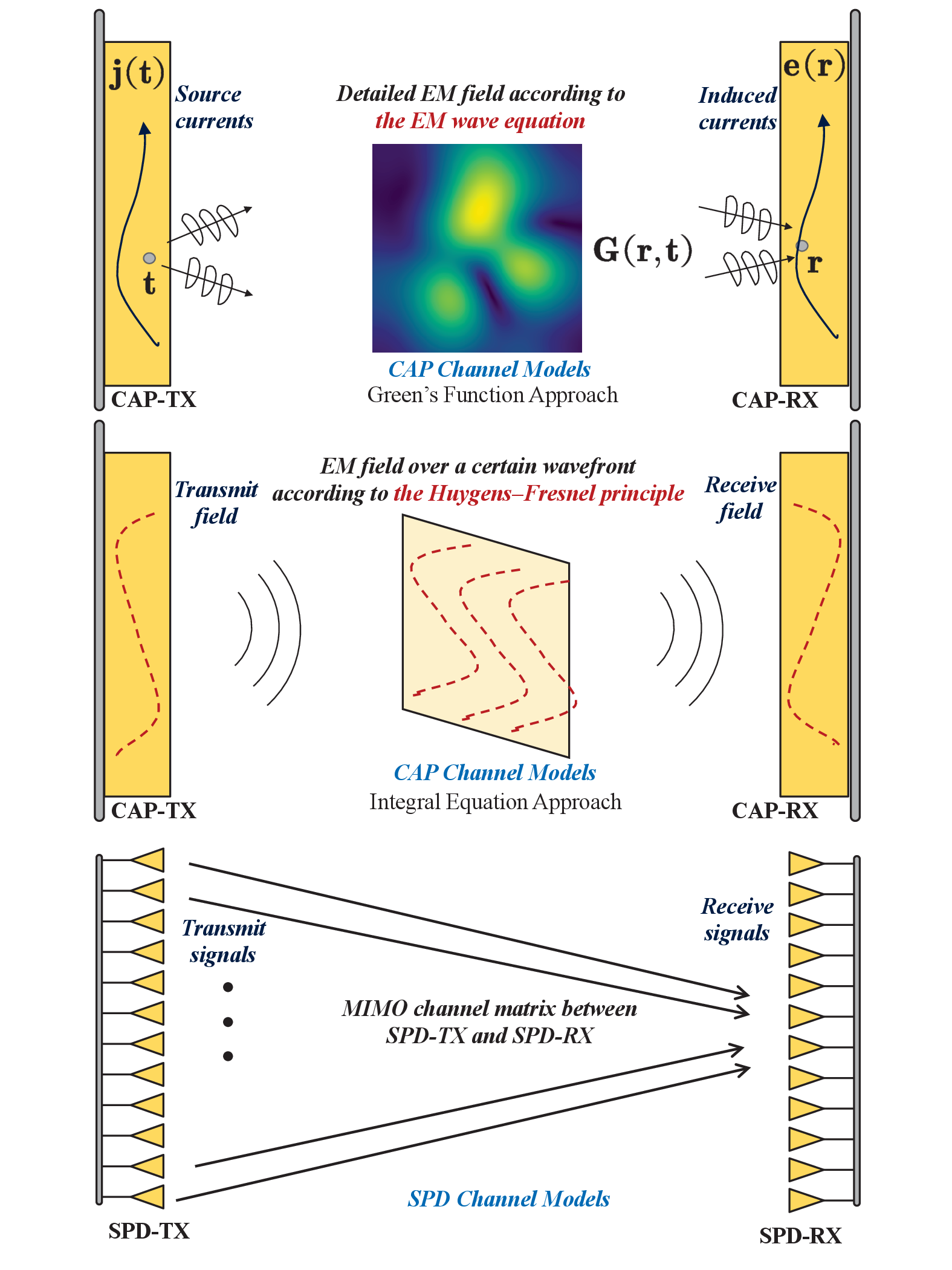}
    \caption{Illustrating CAP and SPD channel models. From top to bottom: Green's function-based approach, integral equation approach, and SPD models.}
    \label{fig:na}
\end{figure}

SPD models are used in channel modeling to describe the behavior of antennas or antenna arrays that are physically separated in space. In these scenarios, the volume or surface integrals arising from Green's functions can be reduced to summations over the antennas \cite{xu_star}. Since the proposal of multi-antenna communication systems more than two decades ago, MIMO channels have been represented by a collection of SPD gains; see Fig. 2 in \cite{1090417} for an example. As illustrated in {\figurename} {\ref{fig:na}}, Green's function-based approaches generally require detailed information about the radiation source, such as the current distribution of the transmitter \cite{xu_vt}. In contrast, integral equation approaches and SPD models only require knowledge of the signal strength distribution on a given wavefront or at each transmit antenna. These channel models reveal the unique properties of NFC. In Section \ref{Sec:CMoN}, we will provide a comprehensive overview of near-field channel modeling. 

\subsection{Recent Developments in Antenna Array Technology}\label{sec:NFA}
With the increasing demand for enhanced quality in wireless service, the role and requirements of antennas and antenna arrays have undergone a profound transformation. In response to these evolving needs, recent years have seen the emergence of various new forms of antennas and array architectures, particularly in challenging near-field environments. Antennas have evolved from being mere conduits for signals to becoming dynamic, adaptable, and intelligent components. Modern antennas actively shape, steer, and manage the flow of data to meet complex service demands.

\begin{figure*}[!t]
    \centering
    \subfigure[Holographic MIMO.]
    {
        \includegraphics[height=0.4\textwidth]{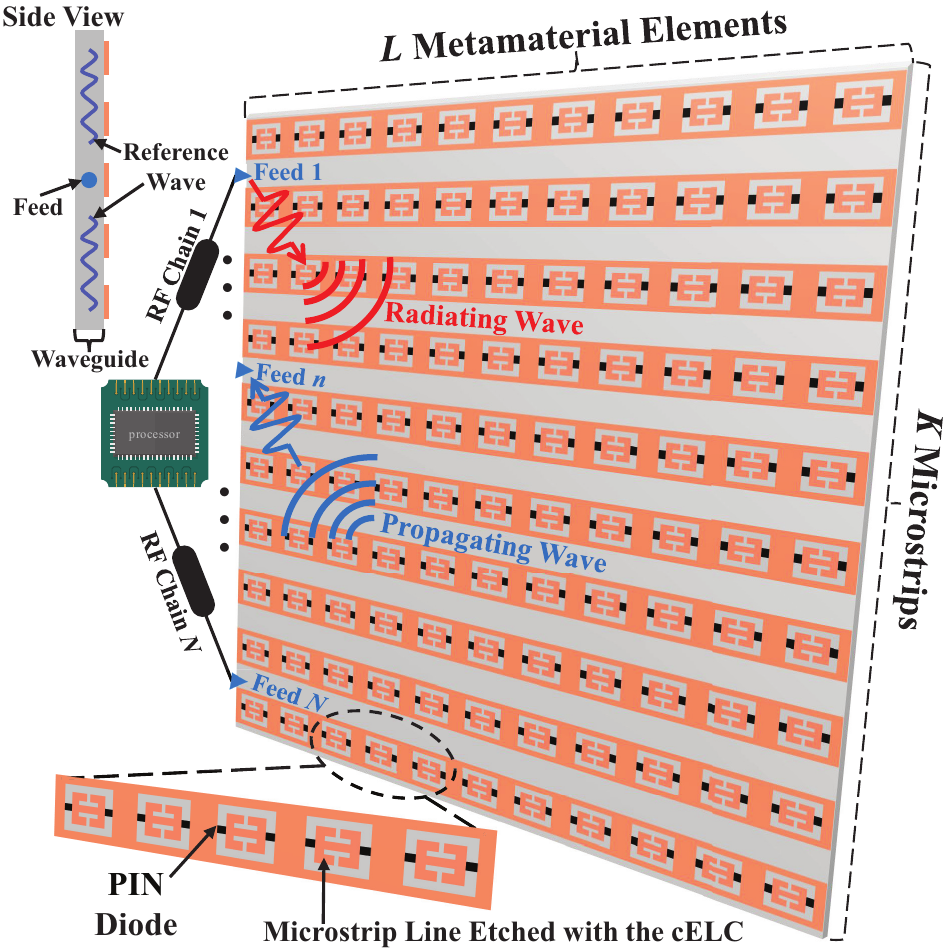}
	   \label{Figure: New_Antenna1}	
    }
    \subfigure[RISs.]
    {
        \includegraphics[height=0.4\textwidth]{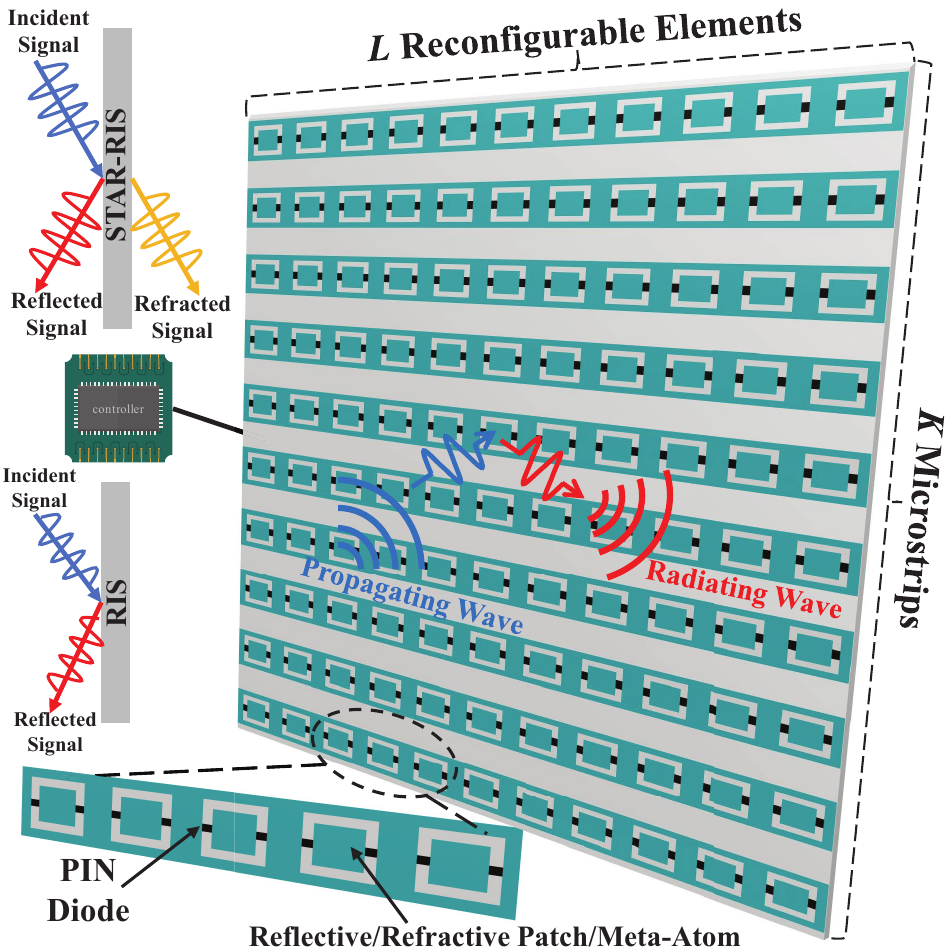}
	   \label{Figure: New_Antenna2}	
    }\hspace{-5pt}
    \subfigure[DMAs.]
    {
        \includegraphics[height=0.4\textwidth]{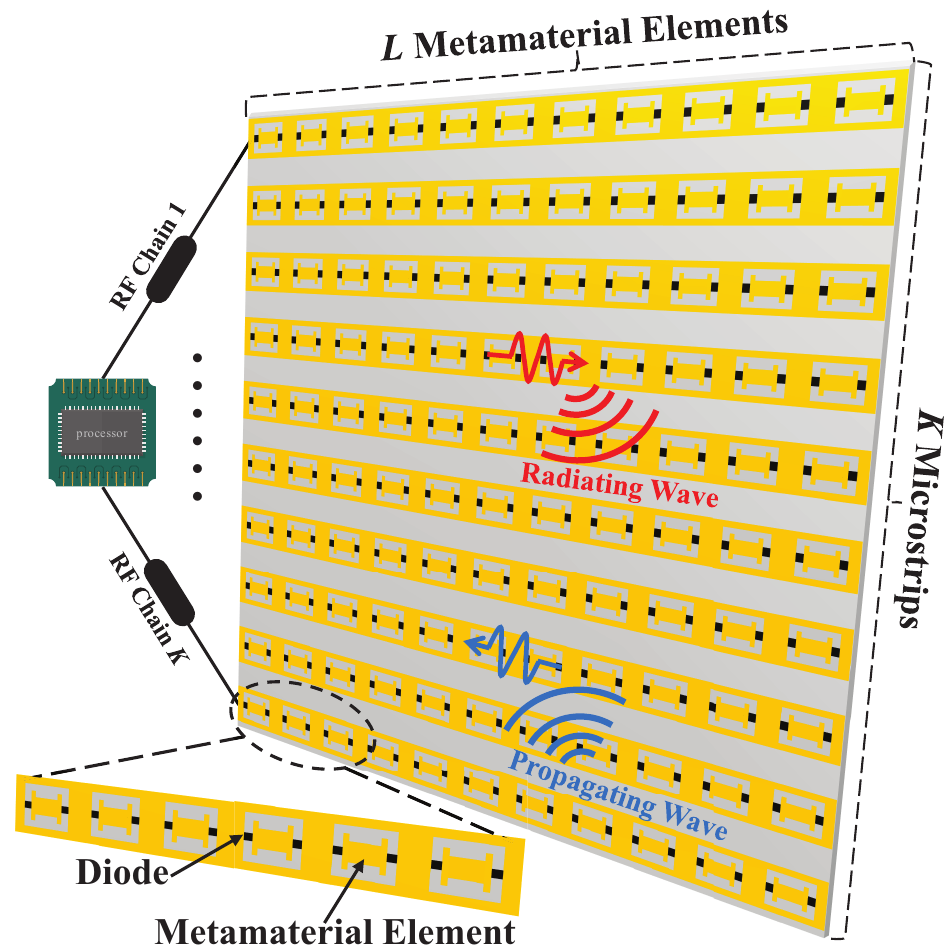}
	   \label{Figure: New_Antenna3}	
    }
   \subfigure[Fluid Antennas.]
    {
        \includegraphics[height=0.4\textwidth]{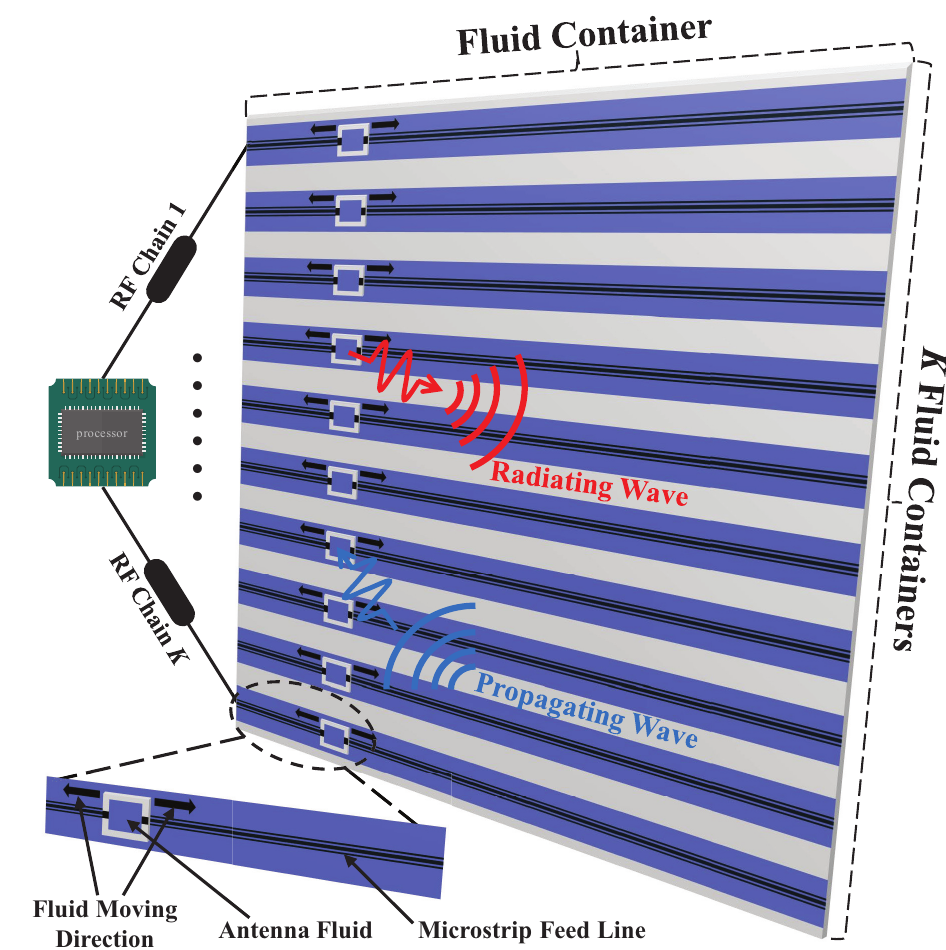}
	   \label{Figure: New_Antenna4}	
    }
\caption{New array architectures.}
    \label{Figure: New_Antenna}
\end{figure*}

{\figurename} {\ref{Figure: New_Antenna}} illustrates four innovative array architectures: \romannumeral1) holographic MIMO \cite{deng2023reconfigurable}, \romannumeral2) RISs and simultaneously transmitting and reflecting (STAR)-RISs \cite{9424177,mu2024simultaneously}, \romannumeral3) dynamic metasurface antennas (DMAs) \cite{9324910}, and \romannumeral4) fluid antennas \cite{9264694}. Each of these architectures can be advantageously designed to address specific challenges posed by the complexities of NFC. These emerging antenna array technologies share common traits such as large aperture sizes and enhanced near-field beamforming capabilities. 

\emph{Extremely large MIMO (XL-MIMO)} is a baseline technology for the four antenna configurations mentioned above, characterized by its substantial aperture size \cite{amiri2018extremely}. XL-MIMO uses a massive number of antennas, often in the hundreds or thousands, to serve multiple users simultaneously with highly directional beams. In the near-field region, XL-MIMO can generate focused beams, enhancing both signal strength and spatial reuse, making it suitable for densely populated areas and indoor environments.

In XL-MIMO, the antenna spacing is typically about half a wavelength. Implementation can be achieved through either a fully digital beamforming structure, where each antenna has an independent RF chain, or a hybrid beamforming structure using a large-scale phased array \cite{ying2023reconfigurable}. Both structures entail high hardware costs \cite{gao2018low} and may incur relatively high energy consumption, especially when many antennas are active simultaneously for beamforming or signal processing \cite{BJORNSON20193}.

In the subsequent sections, we provide a concise survey of the four array architectures shown in {\figurename} {\ref{Figure: New_Antenna}} and compare them with XL-MIMO.
\subsubsection{\textbf{Holographic MIMO}}
Holographic MIMO can be implemented using either an SPD or a CAP array. An SPD array uses multiple antenna elements across a fixed aperture.  Reducing the antenna spacing on this fixed aperture allows for the deployment of more antennas, thus improving system performance. When the spacing approaches zero, an SPD array becomes a CAP array. Unlike SPD arrays, CAP arrays feature an infinite number of dimensionless antennas with infinitesimal spacing. Throughout this paper, we use ``CAP array'' and ``CAP antenna array'' interchangeably to refer to EM devices with a spatially continuous aperture for clarity and consistency. In both SPD and CAP implementations, holographic MIMO can generate programmable three-dimensional (3D) signal patterns by manipulating its radiated EM waves. Each pattern consists of beams directed in different directions, enabling precise control over signal direction and coverage. This pattern resembles a ``3D hologram'' captured by a ``holographic MIMO camera'', hence the name holographic MIMO.

{\figurename} {\ref{Figure: New_Antenna1}} illustrates the SPD implementation, also known as a reconfigurable holographic surface (RHS). The RHS comprises densely packed sub-wavelength metamaterial elements, forming a novel hybrid beamforming structure with analog beamforming implemented through simple diode-based controllers. This structure has been experimentally shown to be more energy-efficient than phased arrays \cite{deng2021reconfigurable,deng2023reconfigurableh}. The CAP counterpart \cite{9136592} can be viewed as an active EM surface that enables significantly higher channel capacity than XL-MIMO through appropriate beamforming design \cite{zhang2023pattern}. These findings suggest that holographic MIMO can be more energy-efficient than XL-MIMO, regardless of its spatial continuity. In NFC, holographic MIMO dynamically shapes signal beams, optimizing connectivity for users and Internet-of-Things (IoT) devices, especially in complex indoor environments.


\subsubsection{\textbf{RISs and STAR-RISs}}
RISs and STAR-RISs are planar surfaces equipped with a grid of closely spaced passive elements capable of reflecting, refracting, or absorbing EM waves \cite{9424177}. These surfaces can be dynamically adjusted to optimize signal strength and phase shifts, reshaping the wireless environment to mitigate interference and enhance signal coverage. The spatial density of the passive elements can be high, especially in applications requiring fine-grained control. RISs are highly energy-efficient since the individual reconfigurable elements do not require active power sources; they manipulate signals passively, consuming minimal energy \cite{ahead}. In the near field, RISs can create localized high-intensity zones for efficient information or energy transfer.
\subsubsection{\textbf{Dynamic Metasurface Antennas}}
Metasurface antennas comprise sub-wavelength structured elements that manipulate the phase, amplitude, and polarization of incoming waves \cite{9324910, Boyarsky}. DMA arrays can achieve very high spatial densities, and their energy consumption depends on their specific design and operation. In general, DMAs offer beam tailoring capabilities and enable the processing of transmitted and received signals in the analog domain with dynamic reconfigurability using simplified transceiver hardware. In the near field, DMAs enable fine-tuning of signals, particularly valuable for applications like wireless power transfer \cite{zhang2024near}. Additionally, DMA-based architectures require much less power and cost compared to conventional XL-MIMO antenna arrays \cite{9324910}.
\subsubsection{\textbf{Fluid Antennas}}
Fluid antennas use electrically reconfigurable metamaterials to change their shape and properties \cite{9264694}. These systems can precisely adjust the antenna's position\footnote{The repositioning of a fluid antenna may not necessarily involve physical movement. This adjustment can be also achieved by switching on or off the units in an array of compact RF pixels. A fluid antenna that relies on physical movement is also referred to as a movable antenna \cite{zhu2023movable}.} within a specified aperture, allowing for numerous possible array configurations. This high resolution in positioning not only facilitates novel interference mitigation \cite{wong2022fluidantenna} but also enables fluid antennas to achieve high transmit/receive spatial diversity through reconfigurable liquids and dynamic changes in antenna positions \cite{9264694}. The diversity gain provided by fluid antennas compensates for significant path losses in NFC. Additionally, the mobility of fluid antennas allows for more energy-efficient radiation, making fluid antenna systems more energy-efficient than XL-MIMO \cite{wong2020fluid}.

The spatial density and energy consumption of these new antenna array technologies can vary widely based on their design, configuration, and application. As antenna arrays become more densely populated and require more precise control, it is beneficial to describe them using spatially continuous models. 
\subsection{Electromagnetic Information Theory (EMIT)}\label{Section: EM Information Theory (EIT) and DoFs}
The previous introduction to new antenna forms suggests that CAP arrays may become the dominant technology in the future antenna market. However, understanding the information transmission capabilities of CAP arrays is challenging, requiring multidisciplinary knowledge from both information theory and EM theory. Traditional SPD models fail to describe the spatial characteristics of CAP arrays, leading to the development of EMIT that specifically addresses information transmission in CAP arrays \cite{migliore2008electromagnetics,migliore2018horse,zhu2022electromagnetic}.

One of the most significant differences between EMIT and traditional information theory is that traditional information theory is based on \emph{matrix theory}, while EMIT is grounded in continuous \emph{operator theory} within Hilbert space. This shift from matrix calculations to operator calculations involves complex integrals and lacks closed-form solutions. Additionally, due to the continuous operators introduced by EM propagation, signals, channels, and noise transform into random fields rather than random variables. This makes designing precoders and equalizers more challenging and complicates the exploration of capacity-achieving transmissions. Moreover, in EMIT, both thermal noise and EM noise must be considered. Thermal noise arises from imprecise measurements, while EM noise is due to EM waves not generated by the input source. EM noise is generally non-white, further complicating analysis. These challenges are just two examples; many other puzzles remain in this field, as detailed in recent tutorials \cite{wei2024electromagnetic}.

Due to these challenges, current research on EMIT is limited to point-to-point networks and linear arrays \cite{gruber2008new,wan2023mutual,wan2023can}. Initial attempts at multiuser precoding design for CAP arrays approximate Hilbert operators with orthogonal discrete vectors, yielding methods that are not clearly different from traditional approaches \cite{zhang2023pattern}. Moreover, research on multiuser EMIT, such as capacity region characterization, remains an open problem. More research efforts are needed to facilitate the deployment of CAP arrays in the future.
\section{Channel Modeling of Near-Field Communications}\label{Sec:CMoN}
Having discussed the fundamentals of NFC, we next investigate near-field channel modeling. In particular, we highlight spatial non-stationarity in near-field channels and provide a comprehensive discussion of non-stationary near-field channel modeling in both LoS and NLoS scenarios.
\subsection{Spatial Non-Stationarity in Near-Field Channel Modeling}\label{Channel modeling of NFC: Spatial Non-Stationarity in NFC}
\begin{figure}[!t]
    \centering
    \subfigure[Stationary MIMO.]
    {
        \includegraphics[height=0.3\textwidth]{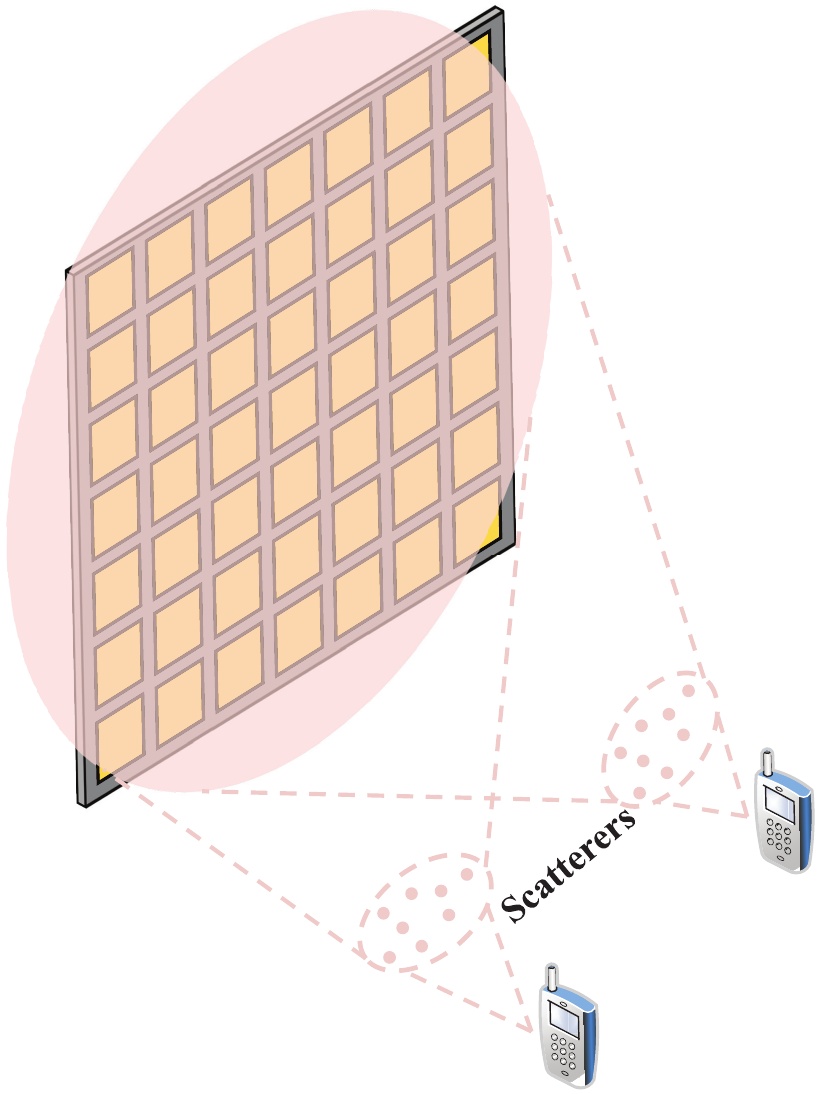}
	   \label{Figure: ELAA_VR_Model1}	
    }
   \subfigure[Non-stationary MIMO.]
    {
        \includegraphics[height=0.3\textwidth]{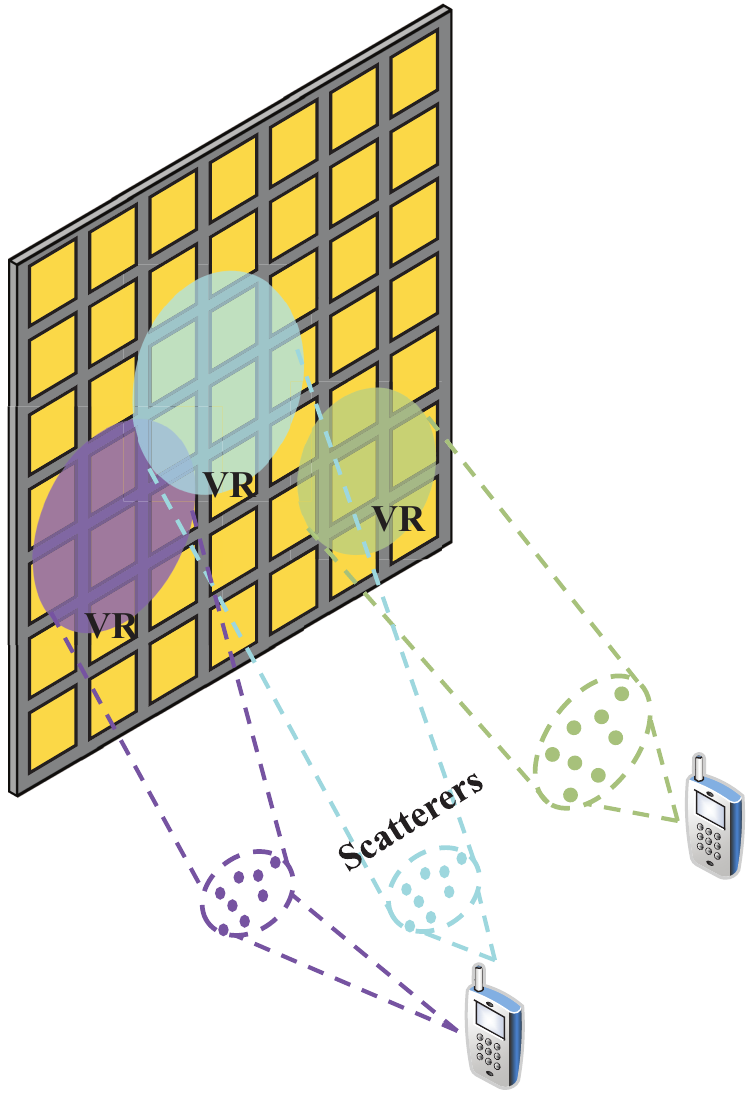}
	   \label{Figure: ELAA_VR_Model2}	
    }
\caption{VRs of stationary MIMO and non-stationary MIMO.}
    \label{Figure: ELAA_VR_Model}
\end{figure}

In conventional far-field MIMO systems, the array aperture is typically much smaller than the propagation distance. Consequently, the entire array can be treated as a single point, and signals transmitted from different locations on the array experience equal path loss and have common angles of departure and arrival (AoDs/AoAs). Furthermore, the entire array is visible to users or scatterer clusters, and the radiated power is evenly distributed across the array, as shown in {\figurename} {\ref{Figure: ELAA_VR_Model1}}. This phenomenon causes the channel to exhibit \textbf{\emph{spatial stationarity}} in the far-field region.

However, recent channel measurements indicate that \textbf{\emph{non-stationary}} channel characteristics become increasingly prominent as the array aperture expands significantly \cite{nurmela2015deliverable,bourdoux2015d1,flordelis2019massive}. One aspect of spatial non-stationarity is that different antenna elements can detect distinct sets of scatterers or users and experience variations in power and delay. This is especially pronounced in near-field channels \cite{wang2018survey}, where two main issues related to non-stationarity emerge.

\subsubsection{Spherical Wavefront} \label{Channel modeling of NFC: Spherical Wavefront}
In NFC systems, the distance between the transceiver and scatterer may be less than the Rayleigh distance. Consequently, it is crucial to consider spherical, rather than planar, wavefronts.

\subsubsection{Visibility Region (VR)} \label{Channel modeling of NFC: Visibility Region (VR)}
\begin{figure}[!t]
    \centering
    \subfigure[UPL.]
    {
        \includegraphics[height=0.35\textwidth]{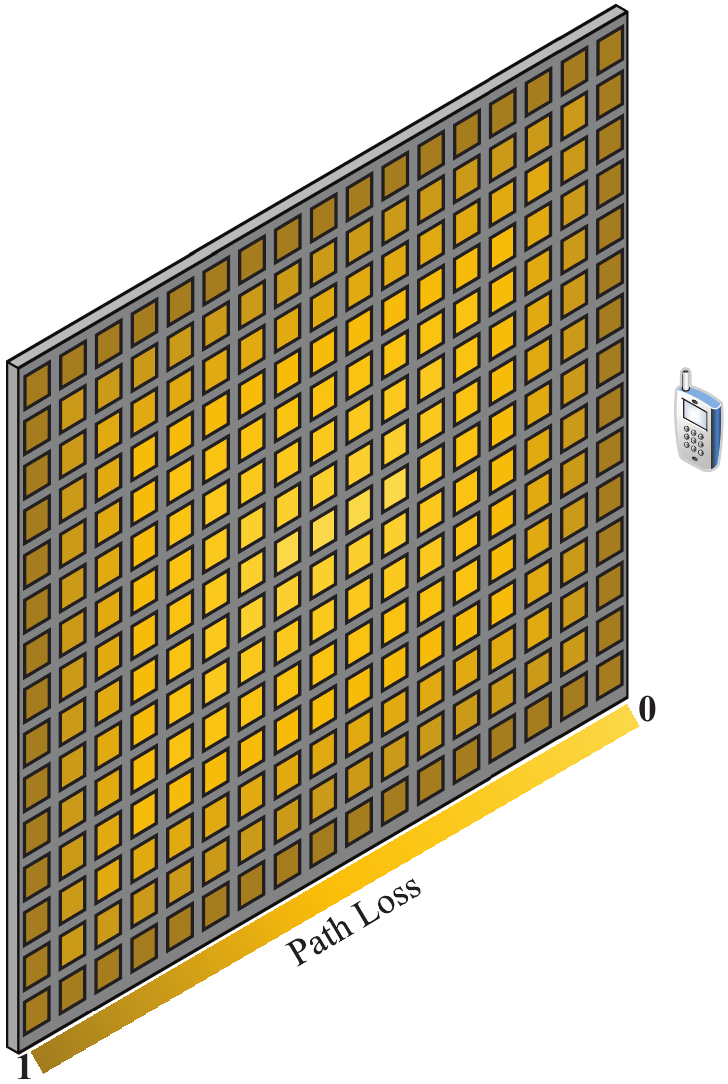}
	   \label{Figure: ELAA_VR_Property1}	
    }
   \subfigure[Blockage due to obstacles.]
    {
        \includegraphics[height=0.35\textwidth]{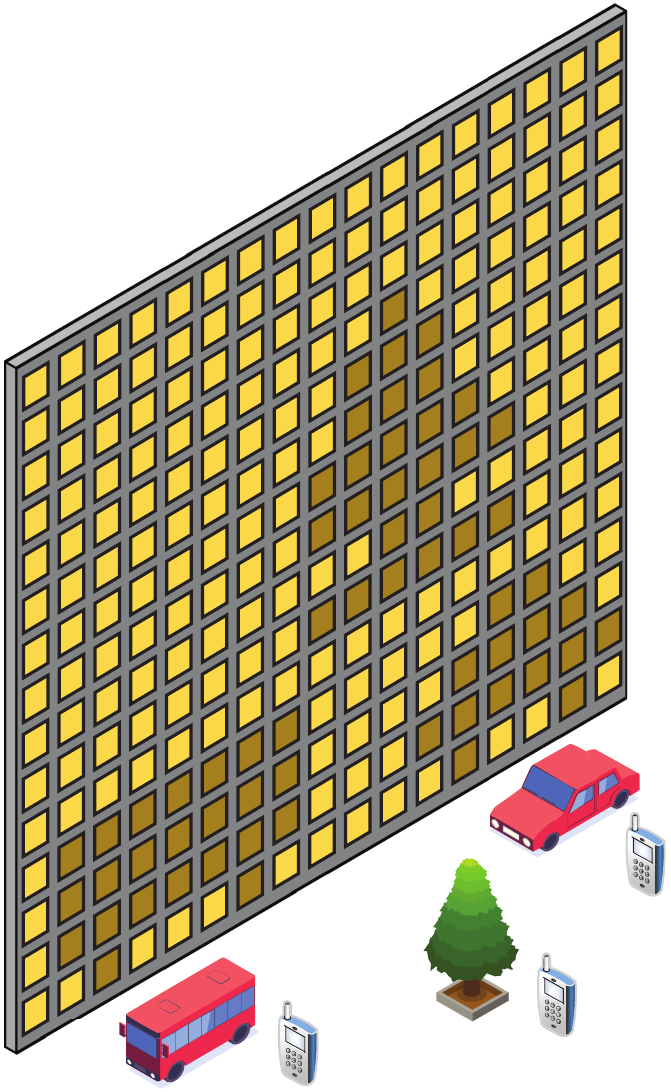}
	   \label{Figure: ELAA_VR_Property2}	
    }
\caption{Manifestations of the VR in NFC. In (a), light and dark squares represent antennas whose path losses are strong and weak, respectively. In (b), light and dark squares represent antennas whose channels are visible and blocked, respectively.}
    \label{Figure: ELAA_VR_Property}
\end{figure}

NFC systems feature large-aperture arrays, and different segments of the array will have distinct perspectives of the propagation environment. As a result, these segments can observe the same channel paths but with varying power levels or even completely different paths, as confirmed by recent empirical findings \cite{martinez2014towards,gao2015massive,gao2015massivemimo}. Under these conditions, users or scatterers can only perceive a portion of the array due to rapid signal attenuation and the substantial array size. This portion is referred to as the VR. The VR highlights the uneven distribution of channel power across the array and arises from two main factors
\begin{enumerate}[i)]
\item \emph{Unequal Path Loss (UPL):} When the propagation distance is comparable to or less than the array aperture, significant variations in path loss occur across different pairs of transmit and receive antennas due to varying distances between them, as shown in {\figurename} {\ref{Figure: ELAA_VR_Property1}}. Channel power is mainly captured by antennas near the source, while those farther away experience considerably weaker signals
\item \emph{Blockage due to Obstacles:} Obstacles such as trees, vehicles, and infrastructure can obstruct the channel between the array and a user. Unlike far-field scenarios, where the entire channel may be blocked, near-field scenarios often involve the blockage of only a portion of the array. The obstructed subarrays tend to reflect the contours of the blocking objects. As shown in {\figurename} {\ref{Figure: ELAA_VR_Property2}}, the blocked subarrays (depicted in dark colors) exhibit patterns resembling the obstacles that cause them. This uneven distribution of channel power due to blockage is independent of UPL variations.
\end{enumerate}
Each user or scatterer has a unique VR, and the locations of VRs for different users or scatterers may be separate, partially overlapping, or fully overlapping. As a result, different areas of the array may fall within the VRs of various scatterers or users, leading to variations in signal power, angular power spectra, and power delay profiles across the antennas.

These points collectively highlight the \emph{spatial non-stationarity} inherent in near-field channels, which distinguishes them from far-field channels. This property has been mathematically validated in \cite{dong2022near} and empirically confirmed through measurements in \cite{wang2022characteristics} and \cite{feng2022mutual}.

\subsection{Line-of-Sight Channel Models} \label{Channel modeling of NFC: LoS Channel Models}
The expansion of the near-field region mainly occurs in the millimeter-wave (mmWave) and sub-Terahertz (THz) frequency bands, resulting in channels dominated by LoS propagation. The following section reviews LoS models tailored for near-field channels supported by SPD and CAP arrays, respectively. For brevity, this discussion focuses on narrowband channels.
\subsubsection{Spatially-Discrete Arrays}\label{Channel modeling of NFC: SPD-NFC}
The narrowband LoS spatial response for SPD arrays is characterized by a complex-valued channel coefficient linking the transmit point $\mathbf{t}$ to the receive point $\mathbf{r}$, which comprises amplitude and phase. In NFC, the phase component is generally expressed as ${\rm{e}}^{-{\rm{j}}\frac{2\pi}{\lambda}\lVert{\mathbf{r}}-{\mathbf{t}}\rVert}$, where $\lambda$ denotes the wavelength. Several methods exist for modeling the amplitude component. Broadly, when modeling amplitude in SPD-NFC, three essential properties are considered:
\begin{enumerate}[i)]
  \item \emph{Free-Space Path Loss (FSPL)}: This represents the power loss due to free-space radiation.
  \item \emph{Effective Aperture Loss (EAL)}: This accounts for power loss resulting from the mismatch between the direction of incident signals and the normal direction of the receiving antenna's maximum effective aperture. It reflects deviations from optimal reception conditions \cite{balanis2012advanced,balanis2016antenna}.
  \item \emph{Polarization Loss (PL)}: This represents power loss due to polarization effects, influenced by the inner product between the polarization vectors of the receiving and transmitting antennas \cite{balanis2012advanced,balanis2016antenna}. PLs indicate the degree of signal polarization alignment or misalignment.
\end{enumerate}
The deployment of ELAAs in NFC creates a high-dimensional channel environment. Understanding this complex channel requires analyzing the relationship between the array aperture and the distance between transceivers. The characterization of amplitude components for SPD arrays can be clarified using two distinct models.

$\bullet$ \emph{Uniform Spherical-Wave (USW) Model:} This model applies when the array aperture is relatively small compared to the propagation distance, with the receiver situated in the USW region \cite{yaghjian1986overview,liu2023near}. In this scenario, the channel response exhibits \emph{uniform spherical wave} characteristics, and amplitude variations across the receive aperture can generally be disregarded. As a result, FSPL, EAL, and PL remain consistent for any pair of transmit and receive points at the transceivers.

In contrast to the amplitude component, the received signal phases depend precisely on the distances between the transmit and receive antennas. Specifically, the phase component between $\mathbf{t}$ and $\mathbf{r}$ is proportional to the propagation distance $\lVert{\mathbf{r}}-{\mathbf{t}}\rVert$, manifesting as a square root function concerning the antenna index. In the USW region, where the propagation distance significantly exceeds the dimensions of each SPD antenna, the ``\emph{Fresnel approximation}'' effectively approximates this square root term as a quadratic term concerning the antenna index using a Maclaurin series expansion \cite{bohagen2007design,liu2023near,selvan2017fraunhofer}. Alongside the quadratic phase approximation, a piecewise-far-field channel with piecewise-linear phase properties was introduced in \cite{cui2021near} to accurately approximate the near-field channel. This approximation can be considered a piecewise linearization of the classical near-field channel model. These findings suggest that spatial non-stationarity in the USW model primarily manifests in the phase component of each complex-valued channel response.

The USW model was initially proposed to characterize LoS propagation in MIMO systems with widely spaced antennas \cite{driessen1999capacity}. Later, \cite{bohagen2006modeling} used it to analyze the singular values of a $2\times2$ near-field MIMO channel. This work was further expanded to cover arbitrary near-field LoS MIMO channels employing uniform linear arrays (ULAs) \cite{bohagen2009spherical}. Due to its analytical tractability, the USW model has been widely adopted in various research on NFC, including channel estimation \cite{le2019massive,wei2021channel,cui2022channel,lu2023near} and beam training \cite{wei2022codebook,cui2022rainbow,zhang2022fast} algorithms. Additionally, \cite{wu2023multiple} used this model to demonstrate the asymptotic orthogonality of near-field beamforming vectors in the range domain when the number of antennas approaches infinity. However, this conclusion is not entirely accurate, as the USW model cannot be applied to an infinitely large array \cite{ding2023resolution}. In these studies, the USW models considered the effects of the spherical wavefront but ignored the influence of the VR. A significant step in this direction was taken in \cite{han2023towards}, which introduced the concept of VR into the USW model by incorporating a binary VR mask that multiplies the array response vector.

$\bullet$ \emph{Non-Uniform Spherical-Wave (NUSW) Model}: When the array aperture is comparable to the propagation distance, i.e., the receiver is within the \emph{NUSW region}, the channel response exhibits a distinctive non-uniform spherical wave characteristic. In this context, both amplitude and phase vary significantly across the receive aperture \cite{liu2023near}. Therefore, accurately modeling the phase component and accounting for FSPL requires considering the exact propagation distance between each transmit and receive point \cite{liu2023near}. Furthermore, due to the closely matched array size and propagation distance, the receiver perceives signals from different array elements at varying angles \cite{liu2023near}. This leads to dynamic fluctuations in EAL and PL across the array. Compared to the USW model, the NUSW model offers a more detailed yet accurate representation when the array aperture is similar to the propagation distance. Spatial non-stationarity in the NUSW model manifests in both the phase and amplitude components of each complex-valued channel response.

The first NUSW channel model, introduced in \cite{driessen1999capacity}, was based on ray tracing. This model was later applied to short-range MIMO channel modeling \cite{jiang2003distributed}, where it helped define a threshold distance below which the spherical-wave model was essential for accurate performance estimation \cite{jiang2005spherical}. Since then, the NUSW model has been extensively applied in short-range MIMO communications, including femtocells and IEEE 802.11 (WiFi) \cite{nishimori2011transmission}, among others.

The NUSW model made its debut in near-field LoS conditions in \cite{zhou2015spherical}, where the authors demonstrated that spherical wavefronts were more effective than planar wavefronts in decorrelating spatial channels. However, they only considered FSPL by treating array elements as dimensionless points. To advance this, \cite{lu2021communicating} incorporated EAL variations into the signal amplitude model, focusing on hypothetical isotropic antennas with a constant directional gain pattern that remains independent of the direction of signal incidence. Building on this, \cite{feng2023near} introduced a more comprehensive directional gain pattern for each antenna element, determined by the elevation and azimuth angles of the incident signal. Using this approach, a more general NUSW model was proposed for LoS propagation in extremely large-scale RIS communications. The authors of \cite{zhi2023performance} extended previous research by considering the influence of polarization mismatch. This new NUSW model applies when both the receiving-mode polarization vector and the normalized electric current vector are aligned along the same axis \cite{zhi2023performance}. In \cite{liu2023near}, an improved NUSW model was proposed that accommodates arbitrary polarization modes and current directions. To incorporate the influence of the VR into the NUSW model, a binary VR mask vector can be added, resulting in a reduced-dimensionality channel model \cite{han2023towards}. This simplification can help streamline transceiver design, as shown in \cite{han2020channel,zhu2021bayesian}.

\begin{table*}[!t]
\caption{Contributions on LoS Channel Modeling for SPD-NFC}
\label{tab:Section_Performance_Analysis_SPD_Channel_LoS_Table}
\centering
\resizebox{0.95\textwidth}{!}{
\begin{tabular}{|l|l|l|lll|lc|l|}
\hline
\multirow{2}{*}{\textbf{Category}}    & \multirow{2}{*}{\textbf{Ref.}} & \multirow{2}{*}{\textbf{Array}} & \multicolumn{3}{c|}{\textbf{Amplitude}}                            & \multicolumn{2}{c|}{\textbf{VR}}       & \multirow{2}{*}{\textbf{Characteristics}} \\ \cline{4-8}
                             &                       &                        & \multicolumn{1}{c|}{\textbf{FSPL}} & \multicolumn{1}{c|}{\textbf{EAL}} & \textbf{PL} & \multicolumn{1}{c|}{\textbf{UPL}} & \textbf{Blockage} &                                  \\ \hline
\multirow{4}{*}{\textbf{Uniform}}     & {\cite{bohagen2009spherical}}               & ULA                    & \multicolumn{1}{c|}{\ding{52}}    & \multicolumn{1}{l|}{\ding{56}}   & \ding{56}  & \multicolumn{1}{c|}{\ding{56}}   & \ding{56}  & The first application of the USW model to near-field MIMO                                 \\ \cline{2-9} 
                             & {\cite{bohagen2007design}}               & ULA                    & \multicolumn{1}{c|}{\ding{52}}    & \multicolumn{1}{l|}{\ding{56}}   & \ding{56}  & \multicolumn{1}{c|}{\ding{56}}   & \ding{56}  & Fresnel approximation-based phase shifts                                \\ \cline{2-9} 
                             & {\cite{wu2023multiple}}               & UPA                    & \multicolumn{1}{c|}{\ding{52}}    & \multicolumn{1}{l|}{\ding{56}}   & \ding{56}  & \multicolumn{1}{c|}{\ding{56}}   & \ding{56}  & Fresnel approximation-based phase shifts                                \\ \cline{2-9} 
                             & {\cite{han2023towards}}               & ULA\&UPA                    & \multicolumn{1}{c|}{\ding{52}}    & \multicolumn{1}{l|}{\ding{56}}   & \ding{56}  & \multicolumn{1}{c|}{\ding{56}}   & \ding{52}  & The generation of the VR comes from the blockages                                \\ \hline
\multirow{6}{*}{\textbf{Non-Uniform}} & {\cite{zhou2015spherical}}               & ULA                    & \multicolumn{1}{c|}{\ding{52}}    & \multicolumn{1}{l|}{\ding{56}}   & \ding{56}  & \multicolumn{1}{c|}{\ding{52}}   & \ding{56}  & The first application of the NUSW model to near-field MIMO                                \\ \cline{2-9} 
                             & {\cite{han2020channel}}               & ULA                    & \multicolumn{1}{c|}{\ding{52}}    & \multicolumn{1}{l|}{\ding{56}}   & \ding{56}  & \multicolumn{1}{c|}{\ding{52}}   & \ding{52}  & The generation of the VR comes from both UPL and blockages                                \\ \cline{2-9}
                             & {\cite{lu2021communicating}}               & UPA                    & \multicolumn{1}{c|}{\ding{52}}    & \multicolumn{1}{l|}{\ding{52} (Isotropic)}   & \ding{56}  & \multicolumn{1}{c|}{\ding{52}}   & \ding{56}  & A clear model for effective aperture area of isotropic antennas                                \\ \cline{2-9} 
                             & {\cite{feng2023near}}               & UPA                    & \multicolumn{1}{c|}{\ding{52}}    & \multicolumn{1}{l|}{\ding{52} (Arbitrary)}   & \ding{56}  & \multicolumn{1}{c|}{\ding{52}}   & \ding{56}  & An LoS model for extremely large-scale IRS communications                                \\ \cline{2-9} 
                             & {\cite{zhi2023performance}}               & UPA                    & \multicolumn{1}{c|}{\ding{52}}    & \multicolumn{1}{l|}{\ding{52} (Isotropic)}   & \ding{52} (Uni-polarized)  & \multicolumn{1}{c|}{\ding{52}}   & \ding{56}  & A clear model for polarization mismatch of uni-polarized antennas                                \\ \cline{2-9} 
                             & {\cite{liu2023near}}               & UPA                    & \multicolumn{1}{c|}{\ding{52}}    & \multicolumn{1}{l|}{\ding{52} (Isotropic)}   & \ding{52} (Tri-polarized)  & \multicolumn{1}{c|}{\ding{52}}   & \ding{56}  & A general model that considers FSPL, EAL, and arbitrary PL                                \\ \hline
\end{tabular}}
\end{table*}


For reference, the primary contributions to near-field LoS channel modeling for SPD arrays are summarized in Table \ref{tab:Section_Performance_Analysis_SPD_Channel_LoS_Table}.
\subsubsection{Continuous-Aperture Arrays} \label{Channel modeling of NFC: CAP-NFC}
CAP-NFC employs a continuous array model consisting of an infinite number of antennas separated by infinitesimal distances. Unlike the SPD array, which delivers finite-dimensional signal vectors, the CAP array supports a continuous distribution of source currents ${\mathbf{j}}({\mathbf{t}})\in{\mathbbmss{C}}^{3\times1}$ within the transmit aperture ${\mathcal{V}}_{T}$. This results in the generation of an electric radiation field ${\mathbf{e}}({\mathbf{r}})\in{\mathbbmss{C}}^{3\times1}$ at the receive aperture ${\mathcal{V}}_{R}$, as shown in {\figurename} \ref{fig:na}. The spatial LoS channel impulse response between any two points $({\mathbf{r}},{\mathbf{t}})$ on the CAP transceiver arrays is described by the tensor Green's function ${\mathbf{G}}({\mathbf{r}},{\mathbf{t}})\in{\mathbbmss{C}}^{3\times3}$ \cite{harrington2001time}. This function connects the transmitter's current distribution to the receiver's electric field via a spatial integral \cite{jackson1999classical}, accurately modeling EM characteristics in free space and representing the channel response between transceivers, similar to the channel matrix in SPD-NFC. The tensor Green's function includes triple polarization, where each element of ${\mathbf{G}}({\mathbf{r}},{\mathbf{t}})$ is the scalar Green's function between one polarization at the receive point and one polarization at the source point. Additionally, Green's function can account for effective aperture, polarization mismatch, and VR. If the receiver is situated in the USW region, the uniform channel power approximation simplifies Green's function.

As a pioneering attempt, Miller \cite{miller1998spatial} used the scalar Green's function to describe LoS propagation between two volumetric CAP arrays in both the radiating near-field and far-field regions. Subsequent authors \cite{hu2017potential,hu2017cramer,hu2018beyond} expanded this model by incorporating the effects of the projected aperture, focusing on LoS propagation between a CAP array and a point in its radiating near-field. Using the Fresnel approximation, \cite{miller2000communicating} proposed a simplified scalar Green's function model for the USW region, later utilized in near-field STAR-RIS communications \cite{xu2022modeling}. The authors of \cite{piestun1999degrees,piestun2000electromagnetic,wan2023mutual} then expanded the scalar model into a tensor-based model to explore the DoFs in CAP-NFC.

Using Fourier transforms, the authors of \cite{poon2005degrees} expanded the tensor Green's function into a summation of three terms. The first term corresponds to the radiating near and far fields, while the remaining two terms pertain to the reactive near field. They demonstrated that the power of the latter two terms decays rapidly and does not significantly contribute to EM radiation. By excluding these two terms, \cite{dardari2020communicating} derived a simplified tensor Green's function LoS model for communication between pairs of active intelligent surfaces. This model specifically addresses the radiating near-field region and considers the effects of effective aperture and uni-polarization mismatch. This model has been adopted in \cite{bjornson2020power} to characterize LoS propagation for large-aperture passive reflecting surfaces. The authors of \cite{liu2023near} further extended this model to account for polarization mismatch in tri-polarized antennas. Each antenna in SPD-NFC can be considered a small CAP array, allowing its EM properties to be effectively characterized using Green's functions. Building on this insight, \cite{wei2023tri} introduced a tensor Green's function LoS model tailored to SPD multiuser MIMO channels, particularly for applications within the USW region.

A summary of the main contributions regarding near-field LoS channel modeling for CAP arrays is given in Table \ref{tab:Section_Performance_Analysis_CAP_Channel_LoS_Table}.

\begin{table*}[!t]
\caption{Contributions on Green's Function-Based LoS Channel Modeling for CAP-NFC}
\label{tab:Section_Performance_Analysis_CAP_Channel_LoS_Table}
\centering
\resizebox{0.95\textwidth}{!}
{\begin{tabular}{|l|l|ccc|cc|l|l|}
\hline
\multirow{2}{*}{\textbf{Category}} & \multirow{2}{*}{\textbf{Ref.}} & \multicolumn{3}{c|}{\textbf{Amplitude}}                            & \multicolumn{2}{c|}{\textbf{Expansion Terms}}                      & \multirow{2}{*}{\textbf{Region of Application}} & \multirow{2}{*}{\textbf{Characteristics}} \\ \cline{3-7}
                          &                       & \multicolumn{1}{c|}{\textbf{FSPL}} & \multicolumn{1}{c|}{\textbf{EAL}} & \textbf{PL} & \multicolumn{1}{c|}{\textbf{radiating}} & \textbf{Reactive} &                                        &                                  \\ \hline
\multirow{4}{*}{\textbf{Scalar}}   & {\cite{miller1998spatial}}               & \multicolumn{1}{c|}{\ding{52}}    & \multicolumn{1}{c|}{\ding{56}}   & \ding{56}  & \multicolumn{1}{c|}{\ding{52}}             & \ding{56}              &  The radiating near-field region                                      & The first application of scalar Green's function to near field
                                \\ \cline{2-9} 
                          & {\cite{hu2017potential,hu2017cramer,hu2018beyond}}               & \multicolumn{1}{c|}{\ding{52}}    & \multicolumn{1}{c|}{\ding{52}}   & \ding{56}  & \multicolumn{1}{c|}{\ding{52}}             & \ding{56}              &  The radiating near-field region                                      & The influence of EAL is considered                                \\ \cline{2-9} 
                          & {\cite{miller2000communicating}}               & \multicolumn{1}{c|}{\ding{52}}    & \multicolumn{1}{c|}{\ding{56}}   & \ding{56}  & \multicolumn{1}{c|}{\ding{52}}             & \ding{56}              &  USW region                                       & Fresnel approximation-based propagation distance                                \\ \cline{2-9} 
                          & {\cite{xu2022modeling}}               & \multicolumn{1}{c|}{\ding{52}}    & \multicolumn{1}{c|}{\ding{56}}   & \ding{56}  & \multicolumn{1}{c|}{\ding{52}}             & \ding{56}              &  USW region                                       & A near-field LoS model for STAR-RIS communications                                \\ \hline
\multirow{5}{*}{\textbf{Tensor}}   & {\cite{piestun1999degrees,piestun2000electromagnetic,wan2023mutual}}               & \multicolumn{1}{c|}{\ding{52}}    & \multicolumn{1}{c|}{\ding{56}}   & \ding{56}  & \multicolumn{1}{c|}{\ding{52}}             & \ding{52}              &   The whole near-field region                                     & A tensor Green's function-based model for all EM regions                                \\ \cline{2-9} 
                          & {\cite{dardari2020communicating}}               & \multicolumn{1}{c|}{\ding{52}}    & \multicolumn{1}{c|}{\ding{52}}   & \ding{52}  & \multicolumn{1}{c|}{\ding{52}}             & \ding{56}              & The radiating near-field region                                      & The influence of EAL and uni-polarization mismatch is considered                                \\ \cline{2-9} 
                          & {\cite{bjornson2020power}}               & \multicolumn{1}{c|}{\ding{52}}    & \multicolumn{1}{c|}{\ding{52}}   & \ding{52}  & \multicolumn{1}{c|}{\ding{52}}             & \ding{56}              & The radiating near-field region                                      & An LoS model for extremely large-scale IRS communications                                \\ \cline{2-9} 
                          & {\cite{liu2023near}}               & \multicolumn{1}{c|}{\ding{52}}    & \multicolumn{1}{c|}{\ding{52}}   & \ding{52}  & \multicolumn{1}{c|}{\ding{52}}             & \ding{56}              & The radiating near-field region                                      & A general model that considers FSPL, EAL, and arbitrary PL                                \\ \cline{2-9} 
                          & {\cite{wei2023tri}}               & \multicolumn{1}{c|}{\ding{52}}    & \multicolumn{1}{c|}{\ding{56}}   & \ding{56}  & \multicolumn{1}{c|}{\ding{52}}             & \ding{52}              & USW region                                       & A tensor Green's function-based LoS model for MU-MIMO                                \\ \hline
\end{tabular}}
\end{table*}

\subsection{Non-Line-of-Sight Channel Models}\label{Channel modeling of NFC: NLoS Channel Models}
Practical wireless propagation is often influenced by scattering effects, which results in NLoS conditions \cite{imoize2021standard}. Next, we explore near-field NLoS channel models.
\subsubsection{Spatially-Discrete Arrays} \label{Channel modeling of NFC NLoS Channel Models: SPD-NFC}
In SPD-NFC, two primary NLoS propagation channel modeling approaches exist:
\begin{itemize}
  \item[$\diamond$] \textbf{\emph{Physical propagation-based stochastic models (PPBSMs)}}: These models characterize EM propagation by incorporating key physical propagation parameters, including the AoA/AoD, multipath delay, scatterer distribution, radar cross-section (RCS) of scatterers, system bandwidth, and antenna configurations (such as antenna types, mutual coupling, antenna pattern, polarization, and array geometry). Together, these parameters describe the near-field NLoS channel \cite{imoize2021standard}.
  \item[$\diamond$] \textbf{\emph{Correlation-based stochastic models (CBSMs)}}: In contrast, CBSMs characterize the channel impulse response as a function of unstructured channel statistics that do not rely on the physical parameters of individual multipath rays. While CBSMs are relatively straightforward to simulate, they may offer limited insights.
\end{itemize}
PPBSMs are particularly useful in the deployment or optimization of site-specific radio systems. They also play a crucial role in channel evaluation during system design for specific reference cases \cite{imoize2021standard}. Meanwhile, CBSMs are widely used in simulation environments, such as link- and system-level simulations, due to their ease of generation and ability to represent various channel responses \cite{imoize2021standard}.
\begin{figure}[!t]
    \centering
    \subfigure[CCM.]
    {
        \includegraphics[width=0.45\textwidth]{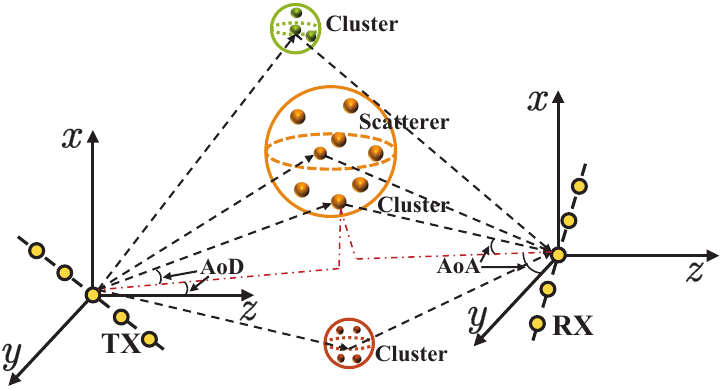}
	   \label{Figure: ELAA_NLOS_CCM}	
    }
   \subfigure[FDC model.]
    {
        \includegraphics[width=0.45\textwidth]{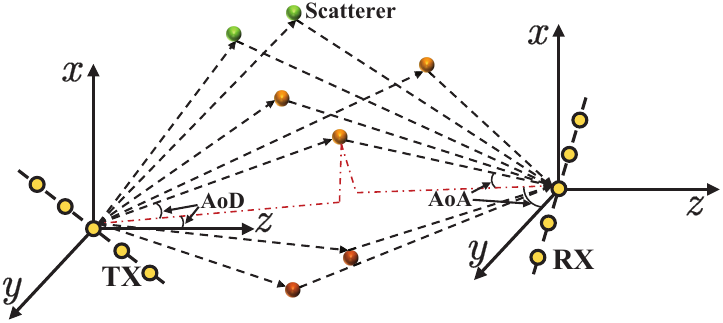}
	   \label{Figure: ELAA_NLOS_FDC}	
    }
\caption{Illustration of the extended S-V model.}
    \label{Figure: ELAA_VR_Property}
\end{figure}

$\bullet$ \emph{Physical Propagation-based Stochastic Models:} The increased FSPL inherent to extremely high-frequency propagation in NFC results in limited spatial selectivity and scattering. The extended Saleh-Valenzuela (S-V) model \cite{spencer2000modeling}, also known as the clustered channel model (CCM) \cite{gustafson2013mm}, is one of the most widely used models to describe such a wireless propagation environment. In this model, scatterers are grouped into clusters based on their respective delays and AoAs/AoDs. Each cluster is treated as a sub-channel, characterized by specific delay profiles and spatial signatures. The aggregation of these sub-channels creates the overall NLoS channel response, as depicted in {\figurename} \ref{Figure: ELAA_NLOS_CCM}. Mathematically, the model describes the channel response in terms of three components: the \emph{transmit array response}, the \emph{receive array response}, and the \emph{complex gain associated with each ray} in every scattering cluster. The complex gain is often modeled as a complex Gaussian variate, reflecting the RCS of the corresponding scatterer \cite{moustakas2000communication}. In adapting the CCM to SPD-NFC, the array response vectors are replaced with the LoS channel vectors detailed in Table \ref{tab:Section_Performance_Analysis_SPD_Channel_LoS_Table}. For example, \cite{wu2017general} proposed a CCM-based NLoS channel model for near-field MIMO channels by incorporating the NUSW model and varying FSPL. In \cite{wu2017general}, another CCM-based NLoS channel model was proposed by incorporating the effects of EAL, PL, and VR.

While the CCM captures most of the physical information of NLoS channels, it can be computationally intensive. To address this, a streamlined variant known as the finite-dimensional channel (FDC) model was developed. In the FDC model, each cluster is simplified to contain only a single scatterer \cite{sayeed2002deconstructing}. The model assumes that the channel's characteristics are primarily defined by a limited number of dominant paths, each with its own delay and AoA/AoD, as shown in {\figurename} \ref{Figure: ELAA_NLOS_FDC}. The FDC model is known for its simplicity and computational efficiency, making it a popular choice in current SPD-NFC research \cite{liu2023near}. For instance, \cite{cui2022channel} and \cite{zhou20154} utilized the FDC model to represent NLoS conditions in SPD-NFC multiple-input single-output (MISO) and MIMO systems, respectively. In these studies, the USW model was used to characterize the channel response between the SPD array and scatterers. As an improvement, \cite{han2020channel} introduced a more versatile NUSW-based NLoS model specifically tailored for near-field MISO channels.

The formulation of a PPBSM-based channel response requires various pieces of information, such as array geometry and RCS patterns. Notably, the RCS pattern is influenced not only by the AoA/AoD but also by the distances of scatterer clusters from the antenna array. This characterization is often referred to as the \emph{power location spectrum (PLOS)} \cite{dong2022near}. The CCM and FDC models generally include a finite number of scatterers, corresponding to a discrete \emph{PLOS} function. However, some research endeavors adopt a continuous \emph{PLOS} function to construct a PPBSM-based channel response, which requires a continuous geometrical distribution of scatterers. For instance, \cite{dong2022near} proposed a near-field PPBSM by employing a one-ring scatterer distribution \cite{abdi2002space}.

$\bullet$ \emph{Correlation-based Stochastic Models:} CBSM is another approach for modeling near-field NLoS propagation. Unlike PPBSMs, CBSMs focus on characterizing the second-order statistics of channels between different antenna pairs, such as correlation or covariance \cite{imoize2021standard}. This characterization plays a crucial role in aspects like developing optimal transmission strategies based on statistical channel knowledge \cite{tulino2006capacity}. Moreover, CBSMs aid in analyzing key performance metrics such as the signal-to-interference-plus-noise ratio (SINR) \cite{ali2019linear} and ergodic capacity \cite{li2015capacity}.

\begin{figure}[!t]
    \centering
    \subfigure[Single-scattering model.]
    {
        \includegraphics[width=0.4\textwidth]{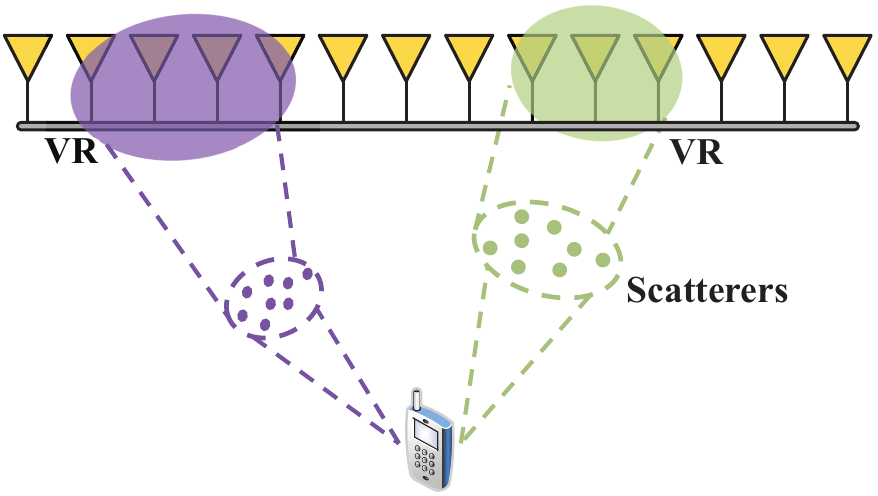}
	   \label{Figure: ELAA_NLOS_VR_Correlation_Case1}	
    }
   \subfigure[Double-scattering model.]
    {
        \includegraphics[width=0.4\textwidth]{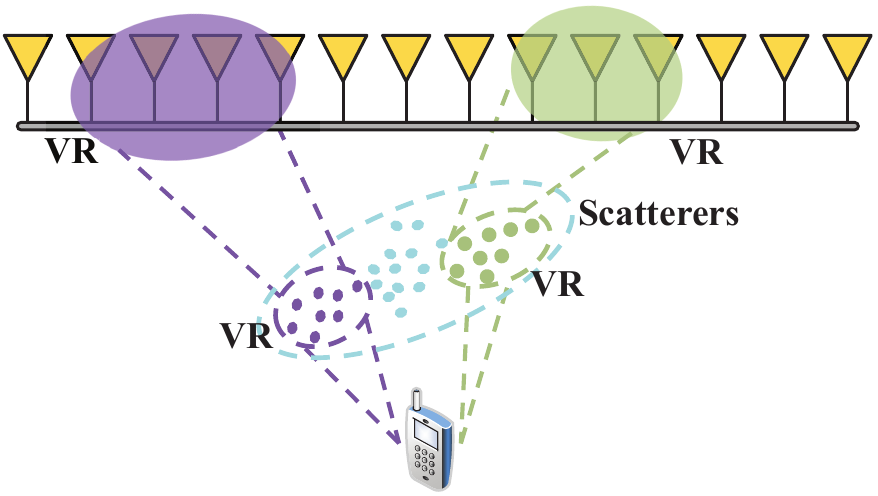}
	   \label{Figure: ELAA_NLOS_VR_Correlation_Case2}	
    }
\caption{Illustration of spatial non-stationary correlation.}
    \label{Figure: ELAA_NLOS_VR_Correlation}
\end{figure}

In far-field channels, spatial correlation is based on the plane-wave assumption and is determined by the \emph{power angular spectrum (PAS)} \cite{abdi2002space}. Here, the correlation coefficient between each antenna pair depends on their relative position within the array \cite{abdi2002space}. In NFC systems, however, spatial correlation between array elements is influenced by their absolute positions, reflecting the spatial non-stationarity \cite{dong2022near}. Accurately computing a near-field correlation matrix requires the \emph{PLOS}, which is computationally intensive. To address this, some researchers have proposed constructing the near-field correlation matrix from its far-field counterpart while accounting for spatial non-stationarity across the array aperture, such as the impact of the VR \cite{de2020non}. Two distinct approaches exist for constructing near-field spatial correlation, each tailored to a specific scenario, as illustrated in {\figurename} {\ref{Figure: ELAA_NLOS_VR_Correlation}}.

The first method for constructing spatial correlation matrices is effective when the influence of scatterer clusters is not emphasized. As depicted in {\figurename} {\ref{Figure: ELAA_NLOS_VR_Correlation_Case1}}, this approach, known as the \emph{single-scattering model (SSM)}, assumes that the user can observe all scatterers. In this scenario, the near-field spatial correlation matrix is derived by applying a binary VR mask to its far-field counterpart; for more details, see \cite{ali2019linear}. Building on this method, several spatially correlated channel models have been developed to incorporate VR.

For instance, the authors of \cite{rodrigues2020low} utilized a VR mask to transform a stationary independent and identically distributed (i.i.d.) correlation model into a non-stationary near-field counterpart. This adaptation was specifically developed for designing low-complexity message-passing NFC detection algorithms, assuming an isotropic scattering environment. Advancing further, \cite{croisfelt2021accelerated} introduced an independent but not identically distributed (i.n.i.d.) correlation model for near-field MISO channels. This model offers greater precision by accounting for the UPL from different antenna elements. Previous studies typically relied on a diagonal correlation matrix to construct their near-field counterparts, which inadequately captures comprehensive correlation effects. To improve on this, \cite{yang2020uplink} introduced a correlated model for near-field uplink MISO channels using the \emph{PAS}-based correlation matrix established in \cite{mckay2006random}. This model, which incorporates Gaussian angular spread, was further employed to design VR-aware user scheduling algorithms. Additionally, \cite{amiri2021uncoordinated} applied a one-ring model to calculate the correlation matrix by adopting a binary VR mask to describe near-field spatial correlation. As a more flexible alternative, \cite{marinello2020antenna} proposed an i.n.i.d. correlation model for near-field MISO channels, where the diagonal elements of the VR mask are not restricted to binary values. 

The second method relies on the \emph{double-scattering model (DSM)} that works when the influence of scatterer clusters is highlighted, as depicted in {\figurename} {\ref{Figure: ELAA_NLOS_VR_Correlation_Case2}}, where users can only observe a portion of the scatterers. In this scenario, the clusters of scatterers act as a virtual antenna array \cite{guerra2022clustered}. Constructing a near-field correlation matrix under these conditions requires two VR masks: one to select the scatterers visible to the user and another to select the antennas visible to the clusters. This correlation-based DSM was first introduced in \cite{li2015capacity} to examine the impact of non-wide sense stationarity on the capacity of uplink massive MIMO channels. Recently, it has been employed in NFC systems for tasks such as analyzing ergodic channel capacity \cite{guerra2022clustered} and designing low-complexity distributed receivers \cite{amiri2021distributed}. In these applications, the spatial correlation matrix of the channel is influenced by factors such as angular spread and antenna spacing.

\begin{table*}[!t]
\caption{Contributions on NLoS Channel Modeling for SPD-NFC}
\label{tab:Section_Performance_Analysis_SDP_Channel_NLoS_Table}
\centering
\resizebox{0.95\textwidth}{!}
{\begin{tabular}{|l|l|l|c|c|l|l|}
\hline
\multirow{2}{*}{\textbf{Category}} & \multirow{2}{*}{\textbf{Ref.}} & \multirow{2}{*}{\textbf{Approach}} & \multirow{2}{*}{\textbf{Channel}} & \multirow{2}{*}{\textbf{VR}} & \multirow{2}{*}{\textbf{Correlation}} & \multirow{2}{*}{\textbf{Characteristics}} \\
                          &                       &                           &                          &                     &                              &                                  \\ \hline
\multirow{6}{*}{\textbf{PPBSM}}    & {\cite{wu2017general}}               & CCM                         & MIMO                        & \ding{56}                   & ---                            & A 3-D non-stationary model accounting for the influence of FSPL                                \\ \cline{2-7} 
                          & {\cite{han2023towards}}               & CCM                         & MISO                        & \ding{52}                   & Kronecker model                            & A general model accounting for the influence of FSPL, EAL, and PL                                \\ \cline{2-7} 
                          & {\cite{cui2022channel}}               & FDC                         & MISO                        & \ding{56}                   & Kronecker model                            & The USW model is employed to characterize the channel response                                \\ \cline{2-7} 
                          & {\cite{zhou20154}}               & FDC                         & MIMO                        & \ding{56}                   & ---                            & The USW model is employed to characterize the channel response                                \\ \cline{2-7} 
                          & {\cite{han2020channel}}               & FDC                         & MISO                        & \ding{52}                   & Kronecker model                            & The NUSW model is employed to characterize the channel response                                \\ \cline{2-7}
                          & {\cite{dong2022near}}               & PLOS                         & MISO                        & \ding{56}                   & One-ring model                            & The scatterers are located on a ring subject to von-Mises distribution                                \\ \hline
\multirow{7}{*}{\textbf{CBSM}}     & {\cite{ali2019linear}}               & SSM                         & MISO                        & \ding{52}                   & Arbitrary                            & A framework to generate near-field correlation from the far-field one                                \\ \cline{2-7} 
                          & {\cite{rodrigues2020low}}               & SSM                         & MISO                        & \ding{52}                   & i.i.d. model                            & An isotropic environment with uniformly distributed independent scatterers                                \\ \cline{2-7} 
                          & {\cite{croisfelt2021accelerated}}               & SSM                         & MISO                        & \ding{52}                   & i.n.i.d. model                            & The influence of UPL from different antenna elements is considered                                \\ \cline{2-7} 
                          & {\cite{yang2020uplink,nishimura2020grant}}               & SSM                         & MISO                        & \ding{52}                   & Kronecker model                            & The PAS-based correlation matrix with Gaussian angular spread distribution                                 \\ \cline{2-7} 
                          & {\cite{amiri2019message,amiri2021uncoordinated}}               & SSM                         & MISO                        & \ding{52}                   & Kronecker model                            & A one-ring model is used to characterize the scattering environment                                \\ \cline{2-7} 
                          & {\cite{marinello2020antenna,de2021quasi}}               & SSM                         & MISO                        & \ding{52}                   & i.n.i.d. model                            & The diagonal elements of the VR mask matrix are not restricted to 0 and 1                               \\ \cline{2-7} 
                          & {\cite{li2015capacity,guerra2022clustered,amiri2021distributed}}               & DSM                         & MISO                        & \ding{52}                   & Kronecker model                            & The spatial correlation is influenced by angular spread and antenna spacing                                \\ \hline
\end{tabular}}
\end{table*}

Existing CBSMs for near-field NLoS channels primarily address MISO channels but are adaptable to MIMO scenarios. When constructing correlation matrices, prior research typically employs models such as i.i.d., i.n.i.d., and the Kronecker model \cite{ozcelik2003deficiencies}. Additional models, such as the Weichselberger model \cite{weichselberger2006stochastic}, virtual channel representation (VCR) \cite{sayeed2002deconstructing}, and rough surface (RoS) \cite{xu2009novel}, are also applicable. A summary of the key contributions related to near-field NLoS channel modeling for SDP arrays is provided in Table \ref{tab:Section_Performance_Analysis_SDP_Channel_NLoS_Table}.

\begin{figure}[!t]
 \centering
\includegraphics[height=0.25\textwidth]{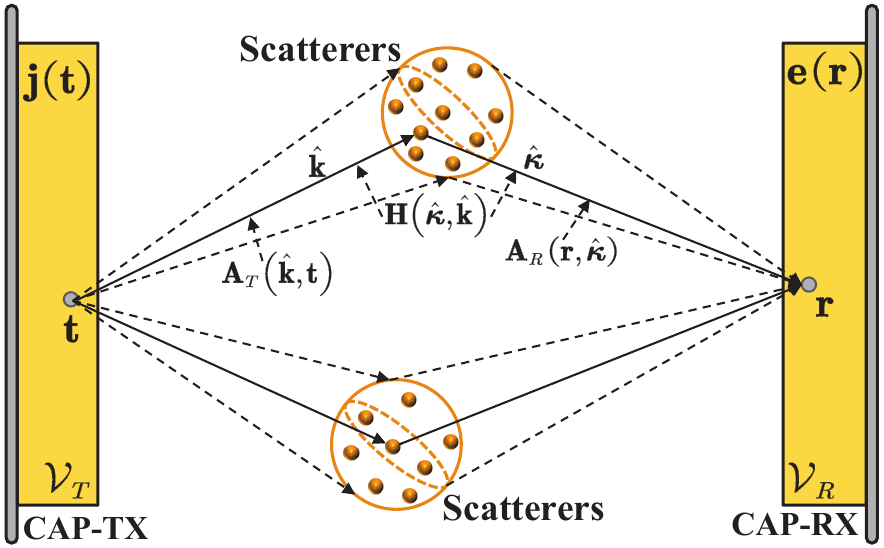}
\caption{Illustration of PPBSMs for CAP-NFC in NLoS propagations.}
\label{Figure: CAP_NFC_NLoS System Model}
\end{figure}

\subsubsection{Continuous-Aperture Arrays}\label{Channel modeling of NFC NLoS Channel Models: CAP-NFC}
Next, we focus on NLoS channel models applicable to CAP-NFC by investigating PPBSMs and CBSMs, which are prominent in current research. 

$\bullet$ \emph{Physical Propagation-based Stochastic Models:} As previously discussed, PPBSMs characterize the propagation environment by considering critical physical parameters such as AoAs/AoDs and the scattering RCS pattern. For CAP-NFC, the PPBSM-based channel response comprises three fundamental components: \romannumeral1) the \emph{transmit array response} ${\mathbf{A}}_T(\hat{\mathbf{k}},{\mathbf{t}})$ that maps the excitation current distribution to the radiated field pattern, \romannumeral2) the \emph{receive array response} ${\mathbf{A}}_R({\mathbf{r}},\hat{\bm{\kappa}})$ that translates the incident field pattern to the induced current distribution, and \romannumeral3) the \emph{scattering response} ${\mathbf{H}}(\hat{\bm{\kappa}},\hat{\mathbf{k}})$ that provides channel gain and polarization information between the transmit direction $\hat{\mathbf{k}}$ and receive direction $\hat{\bm{\kappa}}$, as illustrated in {\figurename} {\ref{Figure: CAP_NFC_NLoS System Model}}. Unlike in SPD-NFC, the array responses for CAP-NFC are characterized using Green's functions, and the scattering response, sandwiched between these array responses, represents the attenuation and polarization of each ray, maintaining the desirable property of being array-independent.

To accurately construct the scattering response, a comprehensive understanding of the \emph{PLOS} of the scattering environment is required. For instance, \cite{poon2005degrees} proposed a tensor Green's function-based NLoS model for communications between two CAP arrays that considers non-uniform power distribution. This model calculates the scattering response by employing a ray-tracing approach \cite{zwick2002stochastic} and grouping the paths into clusters, similar to \cite{heddergott2000statistical}. This NLoS model represents a specific instance of the CCM. Note that the PPBSM involves using two Green's functions (viz. ${\mathbf{A}}_R({\mathbf{r}},\hat{\bm{\kappa}})$ and ${\mathbf{A}}_T(\hat{\mathbf{k}},{\mathbf{t}})$), which increases computational complexity. Moreover, the presence of an arbitrary number of paths within each cluster reduces the model's analytical tractability. Consequently, these factors have limited the adoption and simulation of PPBSM-based CAP-NFC models, despite their accuracy.

$\bullet$ \emph{Correlation-based Stochastic Models:} While PPBSMs can offer precise insights into signal propagation within specific environments \cite{sarkar2003survey}, their dependence on numerical EM solvers renders them highly site-specific. This limitation has spurred the development of CBSMs.

The authors of \cite{wan2024near} explored the spatial correlation in a near-field CAP-MISO channel by considering inhomogeneous media and employing EM scattering theory. Unlike the correlation matrices used in SPD arrays, the channel correlation function for CAP arrays is constructed using Gaussian random fields to model scattering statistics. The findings in \cite{wan2024near}, directly derived from the vector wave equation, are precise yet complex. In 2020, Pizzo \emph{et al.} introduced a more accessible wavenumber-domain (angle-domain) statistical model for CAP-NFC \cite{pizzo2020spatially}. This model builds upon the fact that every spherical wave can be decomposed exactly into an infinite number of plane waves. It represents the transmit and receive fields as a two-dimensional (2D) Fourier plane-wave spectrum. The input spectrum is then transformed into an output spectrum of plane waves via a linear scattering kernel integral operator \cite{marzetta2018spatially,pizzo2022spatial}. This operator encapsulates the scattering mechanism by linking all incident plane waves to every received plane wave, allowing the entire channel response to be expressed as a four-dimensional (4D) Fourier plane-wave spectrum, hence termed as the Fourier plane-wave based stochastic model (FPBSM). The spatial correlation of the entire channel is then characterized by the correlation function of the scattering kernel \cite{pizzo2022spatial}.

By partitioning the wavenumber domain into equally spaced discrete angular sets, the 4D Fourier plane-wave channel model can be discretized into a Fourier plane-wave series expansion dominated by finite terms. Consequently, the authors of \cite{pizzo2022fourier} proposed leveraging these dominant terms to accurately approximate the entire continuous spatial response as the array size becomes sufficiently large. Therefore, the correlation function simplifies into a correlation matrix, and the statistical channel model aligns with the Weichselberger model. In \cite{pizzo2022fourier}, numerical results are provided for various scattering environments, including isotropic, CCM-based, and 3D von Mises-Fisher distributions. This model was further refined by \cite{liu2022effect,wang2022electromagnetic} to address more practical scenarios, incorporating mutual coupling at the transceivers and accounting for phenomena like antenna pattern distortion and antenna efficiency. Subsequently, \cite{zhang2023capacity} developed a discrete angular response model for the single-user MIMO channel by assuming a wrapped Gaussian azimuth angle distribution and a truncated Laplacian elevation angle distribution.

The single-user FPBSM was expanded to a multiuser context in \cite{wei2022multi}, proving useful in optimizing energy efficiency for near-field downlink multi-user systems \cite{bahanshal2023holographic} and in cell-free configurations with multiple ELAA-based access points \cite{lei2023uplink}. This model is also instrumental in formulating antenna selection algorithms \cite{liu2023antenna} and power control strategies \cite{liu2023uplink} for uplink cell-free systems, all based on the assumption of scattering separability. Advancing beyond this assumption, \cite{zhang2023fundamental} introduced a model with a non-separable correlation structure to analyze channel capacity. These studies focus on scalar EM fields corresponding physically to acoustic propagation. An extension to tensor EM channels, without considering environmental randomness, is shown in \cite{zhang2023pattern}.

\begin{table*}[!t]
\caption{Contributions on NLoS Channel Modeling for CAP-NFC}
\label{tab:Section_Performance_Analysis_CAP_Channel_NLoS_Table}
\centering
\resizebox{0.95\textwidth}{!}
{\begin{tabular}{|l|l|l|l|l|c|l|}
\hline
\multirow{2}{*}{\textbf{Category}} & \multirow{2}{*}{\textbf{Ref.}} & \multirow{2}{*}{\textbf{Channel}} & \multirow{2}{*}{\begin{tabular}[l]{@{}l@{}}\textbf{Scattering}\\ \textbf{Separability}\end{tabular}} & \multirow{2}{*}{\textbf{Angular Response}} & \multirow{2}{*}{\begin{tabular}[l]{@{}l@{}}\textbf{Evanescent}\\ \textbf{Waves}\end{tabular}} & \multirow{2}{*}{\textbf{Characteristics}} \\
                          &                       &                          &                                                                                    &                                   &         &                         \\ \hline
\textbf{PPBSM}                     & {\cite{poon2005degrees}}               & SU                        & ---                                                                                  & CCM      & \ding{56}                            & A tensor Green’s function-based model for CAP-NFC within the USW region                                \\ \hline
\textbf{CBSM}                     & {\cite{wan2024near}}               & SU-MISO                        & Separable                                                                                  & CCM      & \ding{56}                            & A statistical model based on the assumption of Gaussian random fields                                \\ \hline
\multirow{11}{*}{\textbf{FPBSM}}     & {\cite{marzetta2018spatially,pizzo2022spatial}}               & SU                        & Both                                                                                  & Arbitrary    & \ding{52}                              & The first application of 4D Fourier plane-wave representation to CAP-NFC                                \\ \cline{2-7} 
                          & {\cite{pizzo2020spatially}}               & SU                        & Separable                                                                                  & Isotropic      & \ding{52}                            & The first application of 4D Fourier plane-wave series expansion to CAP-NFC                                \\ \cline{2-7} 
                          & {\cite{pizzo2022fourier}}               & SU-MIMO                        & Both             & Isotropic/CCM/3D von Mises-Fisher \cite{pizzo2022spatial}                                                                      & \ding{56}                                 & The resulting model is the Weichselberger model                                \\ \cline{2-7} 
                          & {\cite{liu2022effect}}               & SU-MIMO                        & Separable             & Isotropic                                                                      & \ding{56}                                  & The influence of scan impedance mismatch caused by coupling effect is introduced                                \\ \cline{2-7} 
                          & {\cite{wang2022electromagnetic}}               & SU-MIMO                        & Separable             & 3D von Mises-Fisher                                                                      & \ding{56}                                  & The influence of antenna pattern distortion and efficiency decrease is introduced                                \\ \cline{2-7} 
                          & {\cite{zhang2023capacity}}               & SU-MIMO                        & Separable            & Wrapped Gaussian/truncated Laplacian                                                                    & \ding{56}                                  & Standard 3D MIMO channel models \cite{zhang2014three} are used to model the angle distribution                                \\ \cline{2-7} 
                          & {\cite{wei2022multi}}               & MU-MIMO                        & Separable            & Isotropic                                                                       & \ding{56}                                 & The first application of 4D Fourier plane-wave series expansion to MU-MIMO                                \\ \cline{2-7} 
                          & {\cite{bahanshal2023holographic}}               & MU-MIMO                        & Separable            & 3D von Mises-Fisher                                                                       & \ding{56}                                 & The achievable ergodic rate is analyzed                                \\ \cline{2-7} 
                          & {\cite{lei2023uplink,liu2023antenna,liu2023uplink}}               & MU-MIMO                        & Separable             & Isotropic                                                                      & \ding{56}                                   & The application of Fourier plane-wave series expansion to uplink cell-free systems    \\ \cline{2-7} 
                          & {\cite{zhang2023fundamental}}               & MU-MIMO                        & Non-separable             & Arbitrary                                                                      & \ding{56}                                   & The capacity is analyzed by considering the non-separable correlation structure     \\ \cline{2-7} 
                          & {\cite{zhang2023pattern}}               & MU-MIMO                        & ---             & ---                                                                      & \ding{56}                                   & Tensor EM channels without randomness are considered                      \\ \hline
\end{tabular}}
\end{table*}

A summary of the main contributions in near-field NLoS channel modeling for CAP arrays is outlined in Table \ref{tab:Section_Performance_Analysis_CAP_Channel_NLoS_Table}, where ``SU'' denotes single-user and ``MU'' denotes multi-user.
\subsection{Hybrid Line-of-Sight/Non-Line-of-Sight Models}\label{Channel modeling of NFC: Hybrid LoS/NLoS Models}
Channel modeling for both LoS and NLoS propagation has been reviewed. However, in practical scenarios, hybrid propagation channels that encompass both LoS and NLoS links may be encountered.
\subsubsection{Spatially-Discrete Arrays}\label{Channel modeling of NFC: Hybrid LoS/NLoS Models: SPD-NFC}
In the context of SPD-NFC, generating hybrid LoS/NLoS channels is straightforward. This hybrid model combines the LoS component, typically modeled using the USW or NUSW approaches, with the NLoS component, often represented by PPBSMs or CBSMs. An example of this hybrid approach is detailed in \cite{lu2023near}. 
\subsubsection{Continuous-Aperture Arrays}\label{Channel modeling of NFC: Hybrid LoS/NLoS Models: CAP-NFC}
For CAP-NFC, there are two approaches to modeling hybrid LoS/NLoS propagation channels, depending on the use of Green's functions to describe the EM characteristics. When Green's functions are used, hybrid LoS/NLoS channels can be established by combining the LoS model from Section \ref{Channel modeling of NFC: CAP-NFC} with the PPBSM model from Section \ref{Channel modeling of NFC NLoS Channel Models: CAP-NFC}. An alternative method is based on the Fourier plane-wave representation, which accommodates both LoS and NLoS propagation channels \cite{pizzo2022spatial}. In this approach, both the LoS and NLoS channels are modeled as Fourier plane-wave expansions. The hybrid propagation channel is then constructed by merging these two components. A practical implementation of the FPBSM-based hybrid model is outlined in \cite{zhang2023fundamental}.

\begin{table*}[!t]
\caption{Summary of Channel Modeling Approaches for NFC}
\label{tab:Section_Performance_Analysis_Channel_modeling_Table}
\centering
\resizebox{0.95\textwidth}{!}
{\begin{tabular}{|l|l|l|l|l|l|}
\hline
\multirow{2}{*}{\textbf{Antenna}} & \multirow{2}{*}{\textbf{Propagation}} & \multirow{2}{*}{\textbf{Section}} & \multirow{2}{*}{\textbf{Approach}} & \multirow{2}{*}{\textbf{Ref.}} & \multirow{2}{*}{\textbf{Characteristics}} \\
                         &                              &                          &                           &                       &                                  \\ \hline
\multirow{5}{*}{\textbf{SPD}}     & \multirow{2}{*}{LoS}         & \multirow{2}{*}{\ref{Channel modeling of NFC: SPD-NFC}}       & USW                       & \cite{bohagen2009spherical}                     & Amplitude variation across the receive
aperture can be safely disregarded                                \\ \cline{4-6} 
                         &                              &                          & NUSW                      & \cite{zhou2015spherical}                     & Both the amplitude and phase exhibit pronounced variations across the receive aperture                                \\ \cline{2-6} 
                         & \multirow{2}{*}{NLoS}        & \multirow{2}{*}{\ref{Channel modeling of NFC NLoS Channel Models: SPD-NFC}}       & PPBSM                     & \cite{cui2022channel}                     & This model is based on the full knowledge of power location spectrum                                \\ \cline{4-6} 
                         &                              &                          & CBSM                      & \cite{ali2019linear}                     & The influence of the VR is taken into consideration                                \\ \cline{2-6} 
                         & Hybrid                       & \ref{Channel modeling of NFC: Hybrid LoS/NLoS Models: SPD-NFC}                        & LoS+NLoS                  & \cite{lu2023near}                     & This model can be constructed by simply adding the LoS component to the NLoS component                                \\ \hline
\multirow{5}{*}{\textbf{CAP}}     & LoS                          & \ref{Channel modeling of NFC: CAP-NFC}                        & Green's function          & \cite{miller1998spatial}                     & Scalar or Tensor Green's functions are used to characterize the EM characteristics                              \\ \cline{2-6} 
                         & \multirow{3}{*}{NLoS}        & \multirow{3}{*}{\ref{Channel modeling of NFC NLoS Channel Models: CAP-NFC}}       & PPBSM                     & \cite{poon2005degrees}                     & Three components: the transmit/receive array response and the complex gain associated with each ray                                \\ \cline{4-6} 
                         &                              &                          & CBSM                     & \cite{wan2024near}                     & Rooted in the EM scattering theory and the assumption of Gaussian random field                                \\ \cline{4-6}
                         &                              &                          & FPBSM                     & \cite{pizzo2022spatial}                     & Rooted in the fact that wave propagation can be expressed in terms of plane waves                                \\ \cline{2-6} 
                         & \multirow{2}{*}{Hybrid}      & \multirow{2}{*}{\ref{Channel modeling of NFC: Hybrid LoS/NLoS Models: CAP-NFC}}       & Green's function          & \cite{poon2005degrees}                     & The EM characteristics of LoS and NLoS components are both denoted by Green's functions                                \\ \cline{4-6} 
                         &                              &                          & FPBSM                     & \cite{zhang2023fundamental}                     & The LoS and NLoS components are both represented by Fourier-plane wave series expansion                                \\ \hline
\end{tabular}}
\end{table*}

\subsection{Discussions and Outlook}
We have studied near-field channel modeling by highlighting the issue of spatial non-stationarity. For easy reference, we have summarized the reviewed channel models in Table \ref{tab:Section_Performance_Analysis_Channel_modeling_Table}. Despite extensive research, numerous open problems remain in this area. Below, we elaborate on two major points.
\begin{itemize}
  \item \textit{\textbf{Non-Stationary FPBSMs for NFC:}} Current research on FPBSMs typically models the radiating near field as a zero-mean, spatially-stationary, and correlated Gaussian scalar random field \cite{pizzo2020spatially}. This approach effectively captures the EM characteristics of spherical wavefronts. However, it neglects the impact of spatial non-stationarity arising from the VR. Considering the influence of VR is crucial, especially for accurately modeling angular power distribution. As of now, a comprehensive analysis of non-stationarity effects and methods to integrate these effects into FPBSMs remains an open research challenge.      
  \item \textit{\textbf{Hybrid-Field Channel Modeling:}} In practice, some scatterers are located within the near field, while others are in the far field. It is therefore essential to consider the unique channel characteristics of these hybrid-field propagation scenarios. Channel components contributed by near- and far-field scatterers should be modeled separately. Research in hybrid-field channel modeling is still nascent, with only a few recent studies emerging \cite{wei2021channel, cui2022channel, hu2022hybrid}. These studies mainly focus on FDC-based modeling schemes and further efforts are required to develop comprehensive hybrid-field models.
\end{itemize}

\section{Performance Analysis of Near-Field Communications}\label{Section: Performance Analysis of NFC}
In this section, we review the literature concerning near-field performance analysis using the channel models discussed in Section \ref{Sec:CMoN}. We consider three key metrics, including: \romannumeral1) DoFs and EDoFs, \romannumeral2) power scaling law, and \romannumeral3) transmission rate.
\subsection{Degrees-of-Freedom and Effective Degrees-of-Freedom}\label{Section: Performance Analysis of NFC: DoF and EDoF}
The number of DoFs in a system offers valuable insights into the independent signal dimensions available for information transmission over a given wireless channel. Understanding the near-field DoF characteristics reveals the superior data capacity of NFC compared to FFC \cite{ouyang2023near}. Motivated by this, we will review recent progress in analyzing the DoFs for NFC, with a particular focus on single-user near-field MIMO channels to provide clarity.
\subsubsection{Line-of-Sight Channels}\label{Section: Performance Analysis of NFC: DoF and EDoF: LoS Channels}
This section explores the DoFs for near-field LoS channels supported by SPD and CAP arrays. 

$\bullet$ \emph{Spatially-Discrete Arrays:}
For a channel matrix ${\mathbf{H}}$, the number of spatial DoFs is determined by the number of positive singular values of ${\mathbf{H}}$ \cite{tse2005fundamentals}. In the near field, spherical waves exhibit non-linear variations in phase shifts and power levels across each link, resulting in a full-rank channel matrix. Consequently, near-field channels can have more spatial DoFs than far-field channels \cite{ouyang2023near}. A straightforward approach to leverage this is to deploy more antennas, thereby increasing the DoFs. However, with a fixed array aperture, increasing the number of antennas decreases the spacing between them, potentially causing antennas to become indistinguishable and resulting in significant mutual coupling. This could reduce channel capacity, which means that enhancing DoFs does not always lead to improved system performance. Thus, adapting the concept of DoFs to near-field channels is essential.

\begin{figure}[!t]
 \centering
\includegraphics[width=0.45\textwidth]{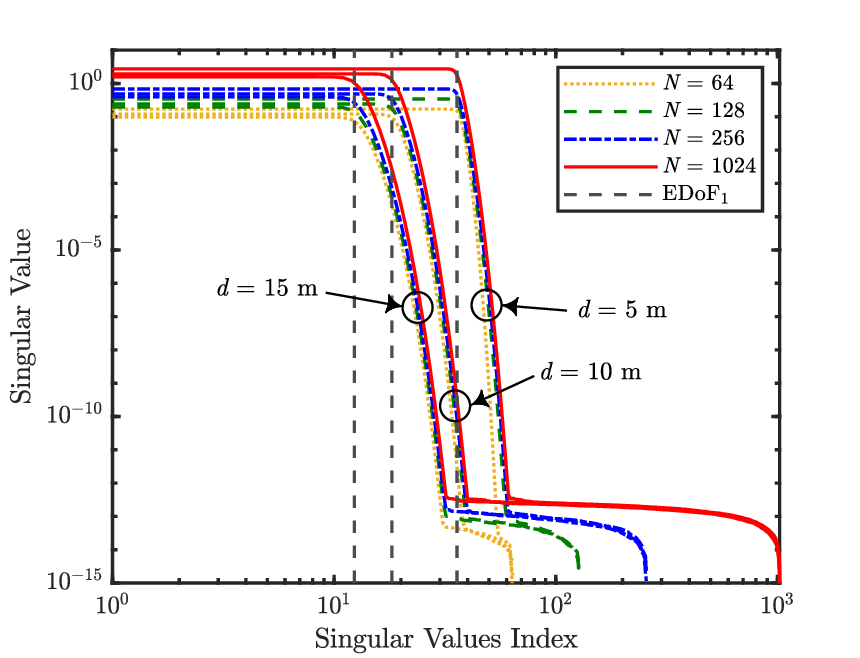}
\caption{Illustration of the singular values and ${\mathsf{EDoF}}_1$ for near-field SPD-MIMO LoS channels. The simulation parameters can be found in \cite{ouyang2023near}.}
\label{Figure: EIT_EDoF1}
\end{figure}

As confirmed in \cite{bucci1989degrees,xu2006electromagnetic,migliore2006role}, the singular values of ${\mathbf{H}}$ exhibit a \emph{step-like} behavior. Specifically, the ordered singular values follow a two-stage pattern, where they remain roughly constant at first before rapidly decreasing to zero, as shown in {\figurename} {\ref{Figure: EIT_EDoF1}}. The singular value at this transition point is often referred to as \emph{the number of EDoFs}, denoted as ${\mathsf{EDoF}}_1$. This pattern is more pronounced for arrays with large apertures. Leveraging this behavior, the EDoFs of near-field SPD-MIMO systems can be approximated by ${\mathsf{EDoF}}_2=({\mathsf{tr}}({\mathbf{HH}}^{\mathsf{H}})/\lVert{\mathbf{HH}}^{\mathsf{H}}\rVert_{\rm{F}})^2$ \cite{ouyang2023near}.

Building on this foundation, the authors of \cite{yuan2021electromagnetic} simulated the EDoFs of a tensor Green's function-based free-space LoS SPD-MIMO channel. Considering EM waves propagating only above ground, a Sommerfeld identity-based approach was used to compute the Green's function within the resulting half-space and assess the EDoFs \cite{jiang2022electromagnetic}. This analysis revealed that the effect of half-space propagation on EDoFs is significant. In these works, EDoFs were calculated numerically without closed-form expressions. The authors of \cite{xie2023performance} went a step further by deriving a closed-form expression for the number of EDoFs in a Fresnel approximation-based NUSW SPD-MIMO channel. The above results highlight two common insights: first, \emph{EDoFs can be increased by reducing the propagation distance or enlarging the aperture size}. Second, \emph{reducing antenna spacing within a fixed aperture ultimately causes the number of EDoFs to reach a saturation point, which serves as an upper limit for SPD-NFC}. This limiting value coincidentally matches the number of EDoFs for CAP-NFC.

$\bullet$ \emph{Continuous-Aperture Arrays:}
The DoFs for CAP-MIMO depend on the behavior of the singular values of Green's function, which also show a \emph{step-like} pattern. The number of EDoFs corresponds to the distinct inflection point in the curve of ordered singular values \cite{ouyang2023near}.

The simplest approach to evaluating the EDoFs for CAP-MIMO is using the ${\mathsf{EDoF}}_2$-based approximation. The authors of \cite{jiang2022electromagnetic} extended this approximation from SPD-MIMO to CAP-MIMO. Using the new formula, they simulated the EDoFs between two continuous-aperture linear arrays in a half-space. A closed-form expression for ${\mathsf{EDoF}}_2$ was later derived for linear arrays using the Fresnel approximation \cite{xie2023performance} and further extended to planar arrays \cite{wang2024analytical}.

Besides this method, various other techniques exist for approximating the EDoFs of CAP-MIMO channels. In \cite{piestun2000electromagnetic}, a framework was proposed for determining the singular values of the tensor Green's function and associated singular functions. These functions form two complete sets of orthogonal bases---one set associated with the transmit aperture and the other with the receive aperture---similar to the singular vectors of an SPD-MIMO channel matrix.

Miller \cite{miller2000communicating} explored a scalar version of this framework in a paraxial setup\footnote{In our paper, the term ``paraxial setup'' denotes a scenario in which the midpoints of the transmitter and receiver are perfectly aligned along the same line, although they are not necessarily parallel to each other \cite{bartoli2023spatial}.} and considered parallel CAP arrays within the USW region. Assuming these conditions, \cite{miller2000communicating} used the Fresnel approximation to compute the singular value decomposition (SVD) of Green's operator. The singular functions were linked to prolate spheroidal wave functions (PSWFs) \cite{slepian1965some}, which allows the number of EDoFs to be calculated based on the energy concentration property of the PSWFs. A similar method was adopted in \cite{pu2014effects} to derive a closed-form expression for the EDoFs and singular functions of a linear array-based CAP-MIMO channel in a paraxial setup but with different orientations. However, in the more complex non-paraxial setting, calculating the SVD of Green's function becomes challenging. Recent advancements have addressed specific scenarios \cite{ruiz2023degrees}, where the authors examined NFC between two non-paraxial 2D CAP arrays within the USW region. Using a quartic approximation, they derived the PSWF-related singular functions for particular non-paraxial cases.

The framework developed in \cite{piestun2000electromagnetic} requires the exact SVD of the Hermitian kernel of Green's function, which is computationally demanding. To address this, Miller \cite{miller2000communicating} proposed an approximation method inspired by Gabor's diffraction theory, which divides a continuous 3D volumetric array into smaller, non-overlapping volumes. The singular functions are approximated as a constant within a small volume and zero outside it, with the number of EDoFs estimated by counting these volumes. This simplified approach works well when the volumetric array has a uniform thickness. Inspired by this technique, the method was also used to analyze the EDoFs of STAR-RIS-based NFC in a paraxial setup \cite{xu2022modeling}. Additionally, the authors of \cite{decarli2021communication} leveraged diffraction theory and the energy-focusing property of near-field beamforming to iteratively construct approximate singular functions for NFC between two non-paraxial continuous linear arrays. They derived an approximate expression for the number of EDoFs when one of the array apertures is significantly smaller than the propagation distance.

Landau's eigenvalue theorem \cite{landau1975szego} is another commonly used method to approximate the EDoFs. The theorem suggests that the singular values of band-limited functions experience a \emph{sharp transition} from near-one to near-zero, marking the asymptotic dimension of the band-limited function space \cite{franceschetti2015landau}. This behavior resembles the \emph{step-like} pattern seen in Green's function's singular values, as Green's function is inherently spatially bandlimited. Using the Fresnel approximation in a paraxial setup, \cite{pizzo2022landau} demonstrated that the scalar Green's function in a near-field CAP-MIMO system adheres to Landau's operator. This allowed them to apply Landau's eigenvalue theorem to obtain a closed-form expression for the EDoFs, akin to the expression derived in \cite{miller2000communicating}. This approach has also been used to compute the EDoFs for NFCs involving continuous linear arrays \cite{do2023parabolic} and RIS-aided continuous planar arrays \cite{do2022line}, both of which rely on the Fresnel approximation but not requiring paraxial arrangements. Furthermore, \cite{ruiz2023degrees} utilized Landau's eigenvalue theorem to determine the EDoFs of a near-field channel between two non-paraxial continuous planar arrays. They provided closed-form expressions for the singular functions and established a general upper bound for the EDoFs when the array aperture is comparable to the propagation distance.

The rationale for applying Landau's eigenvalue theorem lies in the fact that Green's function is band-limited in the spatial domain. Similar to time-domain waveforms with finite bandwidth, EM channels are spatially band-limited due to the low-pass filtering effect that occurs during propagation \cite{bucci1989degrees}. In this analogy, time is replaced by space, and frequency is replaced by spatial frequency (or wavenumber) \cite{pizzo2022nyquist}. According to the Nyquist sampling theorem and Fourier theory, a band-limited signal can be expressed using a sampling series. The series coefficients represent the field's values at the Nyquist intervals, and the number of required samples matches the field's EDoFs \cite{migliore2006role}. On this basis, \cite{dardari2020communicating} derived a closed-form expression for the EDoFs between two CAP surfaces when one surface's aperture is much smaller than the propagation distance. Other studies have examined the EDoFs of near-field channels between continuous linear arrays using the ``cut-set integral'' approach \cite{franceschetti2011degrees}, which approximates Green's function with a carefully designed band-limited signal. The EDoFs of Green's function are then derived from the Nyquist rate of this approximated signal \cite{ding2022degrees}. However, this method applies exclusively to one-dimensional (1D) CAP volumes.

\begin{table*}[!t]
\caption{Contributions on EDoF Analysis for LoS Channels}
\label{tab:Section_Performance_Analysis_DoF_LoS_Table}
\centering
\resizebox{0.95\textwidth}{!}
{\begin{tabular}{|l|l|l|l|l|l|l|ll|}
\hline
\textbf{Antenna}              & \textbf{Ref.}    & \textbf{Tx}      & \textbf{Rx}      & \begin{tabular}[c]{@{}l@{}}\textbf{Paraxial} \\ \textbf{Deployment}\end{tabular} & \begin{tabular}[c]{@{}l@{}}\textbf{Fresnel}\\ \textbf{Approximation}\end{tabular} & \textbf{Eigenfunctions} & \multicolumn{2}{l|}{\textbf{Characteristics}} \\ \hline
\multirow{4}{*}{\textbf{SPD}} & {\cite{yuan2021electromagnetic}} & {Arbitrary} & {Arbitrary} & {Arbitrary}                                                        & {\ding{56}}                                                         & {SVD}        & \multicolumn{2}{l|}{An {${\mathsf{EDoF}}_2$-based approximation} used in the free space}         \\ \cline{2-9} 
                     & {\cite{jiang2022electromagnetic}} & {Linear} & {Linear} & {Arbitrary}                                                        & {\ding{56}}                                                         & {SVD}        & \multicolumn{2}{l|}{An {${\mathsf{EDoF}}_2$-based approximation} used in the half-space}         \\ \cline{2-9} 
                     & {\cite{shen2023electromagnetic}} & {Arbitrary} & {Arbitrary} & {Arbitrary}                                                        & {\ding{56}}                                                         & {SVD}        & \multicolumn{2}{l|}{An {{${\mathsf{EDoF}}_2$-based approximation} used in a PE-based channel}}         \\ \cline{2-9} 
                     & {\cite{xie2023performance}} & {Linear} & {Linear} & {Paraxial}                                                        & {\ding{52}}                                                         & {SVD}        & \multicolumn{2}{l|}{An {{${\mathsf{EDoF}}_2$-based} closed-form expressions for SPD-MIMO}}         \\ \hline
\multirow{13}{*}{\textbf{CAP}} & {\cite{jiang2022electromagnetic}} & {Linear} & {Linear} & {Arbitrary}                                                        & {\ding{56}}                                                         & {\ding{56}}        & \multicolumn{2}{l|}{A calculation formula of {{${\mathsf{EDoF}}_2$}} for CAP-MIMO}         \\ \cline{2-9} 
                     & {\cite{xie2023performance}} & {Linear} & {Linear} & {Paraxial}                                                        & {\ding{52}}                                                         & {\ding{56}}        & \multicolumn{2}{l|}{{{${\mathsf{EDoF}}_2$-based} closed-form expressions for CAP-MIMO}}         \\ \cline{2-9} 
                     & {\cite{piestun2000electromagnetic} } & {Arbitrary} & {Arbitrary} & {Arbitrary}                                                        & {\ding{56}}                                                        & {\ding{56}}        & \multicolumn{2}{l|}{A general yet computationally intensive framework}         \\ \cline{2-9} 
                     & {\cite{miller2000communicating}} & {Planar} & {Planar} & {Paraxial}                                                        & {\ding{52}}                                                         & {PSWF-related}        & \multicolumn{2}{l|}{The closed-form {{SVD of scalar Green's operator}}}         \\ \cline{2-9} 
                     & {\cite{pu2014effects}} & {Linear} & {Linear} & {Paraxial}                                                        & {\ding{52}}                                                         & {PSWF-related}        & \multicolumn{2}{l|}{SVD of Green's operator with different orientations} \\ \cline{2-9}     
                     & {\cite{ruiz2023degrees}} & {Planar} & {Planar} & {Arbitrary}                                                        & {\ding{56}}                                                         & {PSWF-related}        & \multicolumn{2}{l|}{Landau's eigenvalue theorem used for planar arrays}\\ 
                     \cline{2-9}     
                     & {\cite{xu2022modeling}} & {Planar} & {Planar} & {Paraxial}                                                        & {\ding{52}}                                                         & {Small volume-based}        & \multicolumn{2}{l|}{{A small volume-based method used for STAR-RIS}}\\
                     \cline{2-9}     
                     & {\cite{decarli2021communication}} & {Linear} & {Linear} & {Arbitrary}                                                        & {\ding{52}}                                                         & {Beamfocusing-based}        & \multicolumn{2}{l|}{A diffraction theory and beamfocusing-based method}  \\\cline{2-9}     
                     & {\cite{pizzo2022landau}} & {Planar} & {Planar} & {Paraxial}                                                        & {\ding{52}}                                                         & {\ding{56}}        & \multicolumn{2}{l|}{The first application of Landau's theorem to CAP-NFC}         
                     \\ \cline{2-9}     
                     & {\cite{do2023parabolic}} & {Linear} & {Linear} & {Arbitrary}                                                        & {\ding{52}}                                                         & {\ding{56}}        & \multicolumn{2}{l|}{Landau's eigenvalue theorem used for continuous linear arrays}\\ \cline{2-9}     
                     & {\cite{do2022line}} & {Planar} & {Planar} & {Arbitrary}                                                        & {\ding{52}}                                                         & {\ding{56}}        & \multicolumn{2}{l|}{Landau's eigenvalue theorem used for RIS-assisted CAP-MIMO}
                     \\ \cline{2-9}     
                     & {\cite{dardari2020communicating}} & {Planar} & {Planar} & {Paraxial}                                                        & {\ding{52}}                                                         & {\ding{56}}        & \multicolumn{2}{l|}{A 2D sampling theory-based method}
                     \\ \cline{2-9}     
                     & {\cite{ding2022degrees}} & {Linear} & {Linear} & {Arbitrary}                                                        & {\ding{56}}                                                         & {\ding{56}}        & \multicolumn{2}{l|}{A sampling theory-based ``cut-set integral'' method}
                     \\\hline
\end{tabular}}
\end{table*}

Overall, the ${\mathsf{EDoF}}_2$-based approximation is the simplest method to estimate the number of EDoFs, though it provides little insight into the optimal singular functions. Like SPD-MIMO, \emph{the number of EDoFs for CAP-MIMO is proportional to the product of the CAP arrays' apertures and inversely proportional to the transmission distance}. A summary of the primary contributions to the analysis of EDoFs for near-field LoS channels can be found in Table \ref{tab:Section_Performance_Analysis_DoF_LoS_Table}.
\subsubsection{Non-Line-of-Sight Channels}
Let us now consider the DoFs of NLoS channels by incorporating the influence of scattering. Similar to LoS channels, the singular values of an NLoS channel matrix or spatial response function in a richly scattered environment exhibit a \emph{step-like} property \cite{xu2006electromagnetic,xu2009capacity}. The index at the knee singular value, referred to as the number of EDoFs, is denoted as ${\mathsf{EDoF}}_1$. This step-like pattern is even more pronounced in near-field channels supported by large-aperture arrays.

$\bullet$ \emph{Spatially-Discrete Arrays:}
For SPD-MIMO NLoS channels, the step-like behavior suggests approximating ${\mathsf{EDoF}}_1$ with ${\mathsf{EDoF}}_2=({\mathsf{tr}}({\mathbf{HH}}^{\mathsf{H}})/\lVert{\mathbf{HH}}^{\mathsf{H}}\rVert_{\rm{F}})^2$. This method was used in \cite{zhang2022electromagnetic} to estimate ${\mathsf{EDoF}}_1$ of a tensor Green's function-based SPD-NFC channel that included both direct and reflected paths. It was also employed to assess ${\mathsf{EDoF}}_1$ of SPD-NFC in three representative 2D inhomogeneous environments: keyholes, cylindrical scatterers, and cavities \cite{yuan2022electromagnetic,yuan2022electromagneticd}. These studies treated the entire channel deterministically, disregarding its inherent randomness. When randomness is considered, the number of EDoFs can be determined from the correlation matrix of the stochastic channel. In this context, a new formula was proposed for ${\mathsf{EDoF}}_2$, where ${\mathbf{HH}}^{\mathsf{H}}$ is replaced by its expectation ${\mathbbmss{E}}\{{\mathbf{HH}}^{\mathsf{H}}\}$ that represents spatial correlation at the receiver \cite{wang2021improvement}. However, this formula only applies to channels without transmit correlation and requires more transmit than receive antennas. For further improvement, \cite{yuan2023effects} used the spatial correlation matrix of the entire MIMO channel to calculate ${\mathsf{EDoF}}_2$ by replacing ${\mathbf{HH}}^{\mathsf{H}}$ with ${\bm\Phi}={\mathbbmss{E}}\{{\mathsf{vec}}\{{\mathbf{H}}\}{\mathsf{vec}}\{{\mathbf{H}}\}^{\mathsf{H}}\}$. This formula is limited because it does not approximate ${\mathsf{EDoF}}_1$. For instance, consider an $N_{\mathsf{r}}\times N_{\mathsf{t}}$ standard complex Gaussian random matrix $\mathbf{H}$. In this case, ${\mathsf{EDoF}}_1$ equals $\min\{N_{\mathsf{r}},N_{\mathsf{t}}\}$. However, the correlation matrix ${\bm\Phi}$ is an $N_{\mathsf{r}}N_{\mathsf{t}}\times N_{\mathsf{r}}N_{\mathsf{t}}$ identity matrix, leading to an overestimated ${\mathsf{EDoF}}_1$ value of $N_{\mathsf{r}}N_{\mathsf{t}}$, which is much larger than the true value $\min\{N_{\mathsf{r}},N_{\mathsf{t}}\}$.

An alternative approach, presented in \cite{pizzo2022fourier}, analyzed the EDoFs of near-field SPD-MIMO fading channels based on a Fourier plane-wave model. As discussed in Section \ref{Channel modeling of NFC NLoS Channel Models: CAP-NFC}, the Fourier plane-wave model describes the spatial response in the angular domain, which leverages the dominant coupling coefficients between each transmit and receive angular set to capture most of the channel information. Consequently, the number of EDoFs in the SPD-MIMO channel can be approximated by the EDoFs of the \emph{angular-domain channel matrix} or \emph{coupling matrix} \cite{pizzo2022fourier}. This model implies that the EDoFs are influenced by the angular power distribution. Building on this approach, \cite{ji2023extra} analyzed the EDoFs of an isotropic fading SPD-NFC channel while considering the impact of evanescent waves. A heuristic formula was proposed to approximate the number of EDoFs as a function of the propagation distance. The study shows that evanescent waves can enhance the EDoFs and subsequently improve the NFC channel capacity.

In addition to defining the rank of the channel matrix as the number of DoFs, other researchers have considered the rank of the transmit/receive correlation matrix as a measure of the DoFs. This consideration arises from the symmetrical treatment of the transmitter and receiver, with the overall number of spatial DoFs being the minimum of the two \cite{heath2018foundations}. The authors of \cite{pizzo2020degrees} analyzed the EDoFs of the receive correlation matrix in a Fourier plane-wave model-based channel. By recognizing that the receiver-side angular domain is primarily defined by dominant sampling points, the authors evaluated the EDoFs for radiating near-field MIMO channels with linear, planar, and volumetric arrays. This approach is limited to isotropic scattering environments but could be extended to non-isotropic cases. The Fourier plane-wave model in \cite{pizzo2020degrees} differs slightly from that in \cite{pizzo2022fourier}; the former associates randomness with the receiver's plane-wave spectrum, while the latter attributes it to coupling coefficients. The results from \cite{pizzo2020degrees} were also applied to evaluate the EDoFs of the correlation matrix in an RIS-based channel \cite{bjornson2020rayleigh}. 

While using dominant samples to represent the entire angular domain is asymptotically lossless as the aperture size approaches infinity, the results in \cite{pizzo2020degrees,bjornson2020rayleigh} might be compromised for finite array apertures. To address this limitation, \cite{sun2021small} proposed a heuristic formula to approximate the number of EDoFs in an RIS-based communication system with a finite aperture. The formula's mathematical structure relies on heuristic assumptions, with parameters determined through numerical fitting. The study in \cite{sun2021small} demonstrated that, with a fixed array aperture, increasing the number of antennas can sometimes lead to a decrease in the EDoFs. Subsequent research in \cite{sun2022characteristics} further explained this observation by revealing that the correlation matrix examined in \cite{sun2021small} shows a symmetric Toeplitz block structure under isotropic scattering conditions. As a result, the singular values of the correlation matrix behave similarly to the spectral sampling points of the matrix. Numerical results also showed that larger antenna spacing leads to more pronounced spectrum aliasing, primarily due to the presence of evanescent waves generated by the array. Additionally, \cite{sun2022characteristics} discussed the impact of mutual coupling between antennas on the EDoFs.

$\bullet$ \emph{Continuous-Aperture Arrays:}
We next shift our focus to the EDoFs of CAP-NFC under NLoS conditions.

Analyzing the EDoFs for a CAP-MIMO NLoS channel is more challenging than for its LoS counterpart because the NLoS channel involves at least two Green's operators. To address this challenge, the authors of \cite{hanlen2006wireless} proposed using two pre-defined sets of orthogonal basis functions to transform the spatial response function into an infinitely large matrix. Each matrix element corresponds to a coupling coefficient between a transmission mode within the transmit aperture and a reception mode within the receive aperture. To estimate the number of EDoFs of the NLoS channel, an SVD is performed on a suitably truncated version of the matrix. The authors made the simplifying assumption that each scatterer acts as a purely reflective plane, making the channel model easier to manage.

The work in \cite{hanlen2006wireless} provides a general framework for approximating the EDoFs for CAP-NFC NLoS channels. This framework relies on a high-dimensional coupling matrix to ensure accuracy, making it computationally intensive. An alternative approach utilizes the Nyquist sampling theorem. Since EM fields are band-limited spatial signals, they can be represented through finite, non-redundant sampling points in the wavenumber or angular domain. This allows the Nyquist number to approximate the EDoFs, as suggested in \cite{bucci1987spatial}. Leveraging this space-wavenumber relationship, the authors in \cite{poon2005degrees} analyzed the EDoFs of cluster fading channels supported by linear, circular, and spherical CAP arrays. They demonstrated that the number of EDoFs is the product of the wavenumber and array aperture. In \cite{pizzo2022landau}, a similar result was confirmed using Landau's eigenvalue theorem, which provides a more generalized formula for measuring the EDoFs across various scattering environments. Extension to non-monochromatic (wide-band) environments and apertures with arbitrary geometries is presented in \cite{franceschetti2015landau}.

For convenience, the primary contributions to the EDoF analysis in near-field NLoS channels are summarized in Table \ref{tab:Section_Performance_Analysis_DoF_NLoS_Table}, where ``FPWD'' refers to Fourier plane-wave decomposition. The collective findings indicate that the number of EDoFs in NLoS CAP-NFC channels is proportional to the product of the CAP array apertures. 

\begin{table*}[!t]
\caption{Contributions on EDoF Analysis for NLoS Channels}
\label{tab:Section_Performance_Analysis_DoF_NLoS_Table}
\centering
\resizebox{0.95\textwidth}{!}
{\begin{tabular}{|l|l|l|l|l|l|l|ll|}
\hline
\textbf{Antenna}              & \textbf{Ref.}    & \textbf{Tx}      & \textbf{Rx}    & \textbf{Scattering} & \begin{tabular}[c]{@{}l@{}}\textbf{Randomness due} \\ \textbf{to Scatterers}\end{tabular} & \textbf{Eigenfunctions} & \multicolumn{2}{l|}{\textbf{Characteristics}} \\ \hline
\multirow{10}{*}{\textbf{SPD}} & {\cite{zhang2022electromagnetic}} & {Planar} & {Planar} & {FDC}                                                        & {\ding{56}}                                                         & {SVD}        & \multicolumn{2}{l|}{${\mathsf{EDoF}}_2$-based approximation used for NFC with direct and reflected paths}         \\ \cline{2-9} 
                     & {\cite{yuan2022electromagnetic,yuan2022electromagneticd}} & {Arbitrary} & {Arbitrary} & {Isotropic}                                                        & {\ding{56}}                                                         & {SVD}        & \multicolumn{2}{l|}{${\mathsf{EDoF}}_2$-based approximation used for inhomogeneous permittivity profiles}         \\ \cline{2-9} 
                     & {\cite{wang2021improvement}} & {Volumetric} & {Volumetric} & {Arbitrary}                                                        & {\ding{52}}                                                         & {SVD}        & \multicolumn{2}{l|}{A variant for ${\mathsf{EDoF}}_2$ for SPD-MIMO with single-side correlation}         \\ \cline{2-9} 
                     & {\cite{yuan2023effects}} & {Linear} & {Linear} & {Arbitrary}                                                        & {\ding{52}}                                                         & {SVD}        & \multicolumn{2}{l|}{A loose variant for ${\mathsf{EDoF}}_2$ by repalcing ${\mathbf{HH}}^{\mathsf{H}}$ with ${\mathbbmss{E}}\{{\mathsf{vec}}\{{\mathbf{H}}\}{\mathsf{vec}}\{{\mathbf{H}}\}^{\mathsf{H}}\}$}         \\ \cline{2-9} 
                     & {\cite{pizzo2022fourier}} & {Planar} & {Planar} & {FPBSM}                                                        & {\ding{52}}                                                         & SVD or FPWD        & \multicolumn{2}{l|}{Fourier theory-based space-wavenumber relation for near-field MIMO}\\ \cline{2-9} 
                     & {\cite{ji2023extra}} & {Planar} & {Planar} & {Isotropic}                                                        & {\ding{52}}                                                         & SVD or FPWD        & \multicolumn{2}{l|}{A space-wavenumber relation-based method used for NFC with evanescent waves}\\ \cline{2-9} 
                     & {\cite{pizzo2020degrees}} & {\ding{56}} & {Arbitrary} & {Isotropic}                                                        & {\ding{52}}                                                         & SVD or FPWD        & \multicolumn{2}{l|}{A space-wavenumber relation-based method used for correlation matrices}\\ \cline{2-9} 
                     & {\cite{bjornson2020rayleigh}} & {\ding{56}} & {Planar} & {Isotropic}                                                        & {\ding{52}}                                                         & SVD or FPWD        & \multicolumn{2}{l|}{\begin{tabular}[c]{@{}l@{}}{A space-wavenumber relation-based method used for RIS-side correlation}\end{tabular}}\\ \cline{2-9} 
                     & {\cite{sun2021small}} & {\ding{56}} & {Planar} & {Isotropic}                                                        & {\ding{52}}                                                         & SVD or FPWD        & \multicolumn{2}{l|}{A space-wavenumber relation-based method used for finite-aperture RIS}\\ \cline{2-9} 
                     & {\cite{sun2022characteristics}} & {\ding{56}} & {Planar} & {Isotropic}                                                        & {\ding{52}}                                                         & SVD or FPWD        & \multicolumn{2}{l|}{A block-Toeplitz structure-based power spectrum method for RIS}\\ \hline
\multirow{4}{*}{\textbf{CAP}} & {\cite{hanlen2006wireless}} & {Arbitrary} & {Arbitrary} & {FDC}                                                        & {\ding{56}}                                                         & SVD        & \multicolumn{2}{l|}{A truncated coupling matrix-based framework}        \\ \cline{2-9} 
                     & {\cite{poon2005degrees}} & {Uni-polarized} & {Uni-polarized} & {CCM}                                                        & {\ding{56}}                                                         & {PSWF-related}        & \multicolumn{2}{l|}{A space-wavenumber relation-based method used for cluster fading channel}         \\ \cline{2-9} 
                     & {\cite{pizzo2022landau} } & {Planar} & {Planar} & {Arbitrary}                                                        & {\ding{56}}                                                        & {\ding{56}}        & \multicolumn{2}{l|}{Landau's eigenvalue theorem used for CAP planar arrays}         \\ \cline{2-9} 
                     & {\cite{franceschetti2015landau}} & {Arbitrary} & {Arbitrary} & {Arbitrary}                                                        & {\ding{56}}                                                         & {\ding{56}}        &   \multicolumn{2}{l|}{Landau's eigenvalue theorem used for wideband CAP-NFC}       
                     \\\hline
\end{tabular}}
\end{table*}

\subsection{Power Scaling Law}
One advantage of using antenna arrays is that the signal-to-noise ratio (SNR) increases with the number of antennas ($N$). Research on massive MIMO has shown that the SNR scales with $N$ when optimal beamforming techniques are applied, implying that the received SNR increases at a rate of ${\mathcal{O}}(N)$. Conversely, the transmit power required to achieve a target SNR decreases as ${\mathcal{O}}(N^{-1})$. This phenomenon is known as the ``\emph{power scaling law}'' \cite{ngo2013energy}. Recent findings suggest that when optimally configured, the SNR can grow at a rate of ${\mathcal{O}}(N^2)$ when using an RIS with $N$ elements \cite{wu2019intelligent,basar2019wireless}. However, these results are not entirely rigorous, as they were derived under far-field assumptions and do not account for spatial non-stationarity, which becomes significant for large apertures. If the SNR were to increase at rates of ${\mathcal{O}}(N)$ or ${\mathcal{O}}(N^2)$, the received power could exceed the transmit power, violating the law of energy conservation \cite{liu2023near}. These considerations have motivated researchers to explore the power scaling law further in NFC systems.
\subsubsection{Continuous-Aperture Arrays}
One of the earliest contributions to the power scaling law for CAP-NFC was presented in \cite{hu2017potential}. This study derived a closed-form expression for the received SNR in an uplink NFC channel between a single-antenna user and a paraxially deployed CAP array, considering both FSPL and EAL. As the array aperture approaches infinity, the received SNR converges to $\frac{P}{2\sigma^2}$, where $P$ and $\sigma^2$ represent transmission and noise power, respectively. This result aligns intuitively with the principle of energy conservation since half of the isotropically transmitted power reaches the CAP array, and the other half propagates away \cite{hu2018beyond}.

The work in \cite{hu2017potential} was further extended to uplink NFC channels between a single-antenna user and an SPD array in \cite{bjornson2019demystifying}. Assuming edge-to-edge deployment of antenna elements to maximize the receive aperture area, the asymptotic SNR also converges to $\frac{P}{2\sigma^2}$. In this setup, the edge-to-edge deployment makes the SPD array functionally equivalent to a CAP array. Simultaneously, Dardari \cite{dardari2020communicating} introduced the NUSW model to analyze the received SNR in NFC between a small CAP array and a larger CAP array in a paraxial scenario. This work considered the effects of FSPL, EAL, and PL. As the aperture size of the larger CAP array approached infinity, the received SNR converged to a constant $\frac{P}{3\sigma^2}$ rather than $\frac{P}{2\sigma^2}$.

Subsequently, \cite{bjornson2020power} analyzed the received SNR in near-field uplink MISO channels using SPD arrays, half-duplex relay-aided single-input single-output (SISO) channels, and the SNR upper bound for near-field RIS-aided SISO channels. In all these cases, the received SNR converged to a constant value as the array aperture approached infinity. These results were based on zero antenna spacing in the SPD array, which is essentially a CAP array.

\subsubsection{Spatially-Discrete Arrays}
Regarding the power scaling law for SPD-NFC, \cite{lu2021does} analyzed the received SNR of in uplink SPD-MISO near-field channel using a linear array with non-zero element spacing. This study considered the influence of FSPL in a non-paraxial scenario and established that the asymptotic SNR is related to the AoA, antenna spacing, and the propagation distance between the user and the array center. 

The analysis was then extended to a planar SPD array-based near-field MISO channel in \cite{lu2021communicating}, where the influence of the varying EAL was also considered. This work suggested that the asymptotic received SNR is given by $\frac{P}{2\sigma^2}\epsilon$, where $\epsilon$ represents the array occupation ratio. When $\epsilon = 1$, the SPD array becomes a CAP array, aligning the result with \cite{hu2017potential}. The numerical results obtained in \cite{lu2021communicating} indicate that the power scaling law derived under the USW model contradicts the law of energy conservation. In \cite{zheng2022simultaneous}, the SNR of a near-field MISO channel was also analyzed considering FSPL and EAL, which showed that the NUSW model---solely considers FSPL---also results in a power scaling law that contravenes the law of energy conservation. This corroborates earlier observations presented in \cite{bjornson2020power}. Furthermore, the methodologies applied in \cite{lu2021does} and \cite{lu2021communicating} were extended to derive the SNR for a near-field MISO channel relying on a modular linear ELAA. Unlike the typically assumed collocated configuration, modular arrays feature regularly spaced modules separated by distances much larger than the signal wavelength \cite{li2022near}. In this context, a variable FSPL is considered, and the asymptotic SNR is found to be inversely proportional to the module separation \cite{li2022near}.

As previously mentioned, \cite{bjornson2020power} derived an upper bound for the received SNR in a near-field RIS-aided SISO channel using the Cauchy inequality, which is known to be loose. To improve upon this, the authors of \cite{feng2021wireless} proposed a tighter bound for the SNR by considering the impact of variable FSPL and EAL. Their research revealed that the asymptotic SNR converges to a constant related to the propagation distances between the RIS and the transceivers. While prior works on projected effective aperture focused on isotropic antennas with a constant directional gain pattern, the researchers in \cite{zeng2022reconfigurable} introduced a more versatile directional gain pattern for each antenna element, determined by the elevation and azimuth angles of the incident signal. Building on this enhancement, they analyzed the SNR in SPD-NFC supported by reconfigurable refractive surfaces (RRSs) \cite{zeng2022reconfigurable}. However, while \cite{zeng2022reconfigurable} observed the convergence of the received SNR to a constant, it did not provide a closed-form expression. Advancing this research, the authors of \cite{feng2023near} derived closed-form lower and upper bounds for the received SNR in RIS-aided near-field SISO channels. This work accounted for the effects of FSPL and a generic directional gain pattern, with numerical simulations confirming the bounds' accuracy. However, these studies did not account for PL.

To address this gap, the authors of \cite{zhi2023performance} extended previous research by incorporating EM polarization effects and variations in signal amplitude in the NUSW model. With this model, they derived a closed-form expression for the SNR in NFC between a single-antenna device and a planar SPD array. As the array aperture approaches infinity, the SNR converges to $\frac{P}{3\sigma^2}\epsilon$, similar to the result in \cite{bjornson2020power}. This NUSW model is applicable when the polarization vector and the normalized electric current vector are aligned along the same axis. To provide a broader scope, the authors of \cite{liu2023near} introduced an improved NUSW model for arbitrary polarization modes and current directions. Based on this model, they derived a power scaling law for both SPD-NFC and CAP-NFC. They also demonstrated theoretically that the USW model could yield a logarithmic power scaling law of ${\mathcal{O}}(\log{N})$, which violates the law of energy conservation.

The results above indicate that to ensure the derived power scaling law adheres to the law of energy conservation, it is crucial to account for the influence of FSPL in linear arrays, and both FSPL and EAL in planar arrays. However, there is an exception for planar arrays, as demonstrated in \cite{deng2022hdma}. In this study, the authors analyzed the SNR of a near-field downlink MISO channel supported by RRSs. Their derivation considered only the influence of FSPL to model spatial non-stationarity. Despite this, the numerical results indicated that the received SNR also converges to a constant as the array aperture approaches infinity. This phenomenon is attributed to the RRS's ability to produce distinct directional gain patterns for each antenna element. Although the EAL impact was not directly included in the channel model, it was indirectly accounted for through the variable patterns. Consequently, the resulting SNR increased with larger array apertures until it eventually reached a constant value. 

In summary, the primary contributions to the analysis of the near-field power scaling law are outlined in Table \ref{tab:Section_Performance_Analysis_PSL_LoS_Table}. 

\begin{table*}[!t]
\caption{Contributions on Power Scaling Law Analysis for NFC}
\label{tab:Section_Performance_Analysis_PSL_LoS_Table}
\centering
\resizebox{0.95\textwidth}{!}
{\begin{tabular}{|l|l|lll|l|l|ll|}
\hline
\multirow{2}{*}{\textbf{Antenna}} & \multirow{2}{*}{\textbf{Ref.}} & \multicolumn{3}{l|}{\textbf{Spatial Non-Stationarity}}                                        & \multirow{2}{*}{\textbf{Array}} & \multirow{2}{*}{\begin{tabular}[c]{@{}l@{}}\textbf{Paraxial} \\ \textbf{Deployment}\end{tabular}} & \multicolumn{2}{l|}{\multirow{2}{*}{\textbf{Characteristics}}} \\ \cline{3-5}
                         &                       & \multicolumn{1}{l|}{\textbf{FSPL}}    & \multicolumn{1}{l|}{\textbf{EAL}}     & \textbf{PL}      &                        &                                                                                 & \multicolumn{2}{l|}{}                                 \\ \hline
\multirow{4}{*}{\textbf{CAP}}     & {\cite{hu2017potential}}               & \multicolumn{1}{l|}{\ding{52}} & \multicolumn{1}{l|}{\ding{52}} & {\ding{56}} &  Planar                      & {Paraxial}                                                                         & \multicolumn{2}{l|}{One of the earliest contributions to the near-field power scaling law}                          \\ \cline{2-9} 
                         & {\cite{bjornson2019demystifying}}               & \multicolumn{1}{l|}{\ding{52}} & \multicolumn{1}{l|}{\ding{52}} & {\ding{56}} &  Planar                      & {Paraxial}                                                                         & \multicolumn{2}{l|}{Power scaling law for an SPD array with zero antenna spacing}                          \\ \cline{2-9} 
                         & {\cite{dardari2020communicating}}               & \multicolumn{1}{l|}{\ding{52}} & \multicolumn{1}{l|}{\ding{52}} & {\ding{52}} &  Planar                      & {Paraxial}                                                                         & \multicolumn{2}{l|}{NFC between a small CAP array and an infinitely large one}                          \\ \cline{2-9} 
                         & {\cite{bjornson2020power}}               & \multicolumn{1}{l|}{\ding{52}} & \multicolumn{1}{l|}{\ding{52}} & {\ding{52}} &  Planar                      & {Paraxial}                                                                         & \multicolumn{2}{l|}{Power scaling law for MISO, half-duplex relay-aided SISO, and RIS-aided SISO channels}                          \\ \hline
\multirow{10}{*}{\textbf{SPD}}     & {\cite{lu2021does}}               & \multicolumn{1}{l|}{\ding{52}} & \multicolumn{1}{l|}{\ding{56}} & {\ding{56}} &  Linear                      & {Arbitrary}                                                                         & \multicolumn{2}{l|}{Relation between the power scaling law and AoA/antenna spacing/propagation distance}                          \\ \cline{2-9}  
                         & {\cite{lu2021communicating}}               & \multicolumn{1}{l|}{\ding{52}} & \multicolumn{1}{l|}{\ding{52}} & {\ding{56}} &  Planar                      & {Arbitrary}                                                                        & \multicolumn{2}{l|}{Relation between the power scaling law and array occupation ratio}                         \\ \cline{2-9} 
                         & {\cite{zheng2022simultaneous}}               & \multicolumn{1}{l|}{\ding{52}} & \multicolumn{1}{l|}{\ding{52}} & {\ding{56}} &  Planar                      & {Arbitrary}                                                                        & \multicolumn{2}{l|}{Power scaling law based on the NUSW model with varying FSPL}                          \\ \cline{2-9} 
                         & {\cite{li2022near}}               & \multicolumn{1}{l|}{\ding{52}} & \multicolumn{1}{l|}{\ding{56}} & {\ding{56}} & Linear                      & {Arbitrary}                                                                        & \multicolumn{2}{l|}{Power scaling law for modular linear arrays}                          \\ \cline{2-9} 
                         & {\cite{feng2021wireless}}               & \multicolumn{1}{l|}{\ding{52}} & \multicolumn{1}{l|}{\ding{52}} & {\ding{56}} &  Planar                      & {Arbitrary}                                                                        & \multicolumn{2}{l|}{Tight power scaling bounds for near-field RIS-aided SISO channels}                          \\ \cline{2-9} 
                         & {\cite{zeng2022reconfigurable}}               & \multicolumn{1}{l|}{\ding{52}} & \multicolumn{1}{l|}{\ding{52}} & {\ding{56}} &  Planar                      & {Arbitrary}                                                                        & \multicolumn{2}{l|}{Power scaling law with generic directional antenna gain patterns for RRS-aided NFC}                          \\ \cline{2-9} 
                         & {\cite{feng2023near}}               & \multicolumn{1}{l|}{\ding{52}} & \multicolumn{1}{l|}{\ding{52}} & {\ding{56}} &  Planar                      & {Arbitrary}                                                                        & \multicolumn{2}{l|}{Power scaling law with generic directional antenna gain patterns for RIS-aided NFC}                          \\ \cline{2-9} 
                         & {\cite{zhi2023performance}}               & \multicolumn{1}{l|}{\ding{52}} & \multicolumn{1}{l|}{\ding{52}} & {\ding{52}} &  Planar                      & {Arbitrary}                                                                        & \multicolumn{2}{l|}{Power scaling law with impact of antenna polarization}                          \\ \cline{2-9} 
                         & {\cite{liu2023near}}               & \multicolumn{1}{l|}{\ding{52}} & \multicolumn{1}{l|}{\ding{52}} & {\ding{52}} &  Planar                      & {Arbitrary}                                                                        & \multicolumn{2}{l|}{General expressions that apply to arbitrary polarization modes and current directions}                          \\ \cline{2-9} 
                         & {\cite{deng2022hdma}}               & \multicolumn{1}{l|}{\ding{52}} & \multicolumn{1}{l|}{\ding{56}} & {\ding{56}} &  Planar                      & {Arbitrary}                                                                        & \multicolumn{2}{l|}{The impact of EAL reflected through the variant antenna patterns}                          \\ \hline
\end{tabular}}
\end{table*}

\subsection{Transmission Rate}
Our focus now turns to another crucial performance metric: the transmission rate. This is a key determinant of a system's throughput and spectral efficiency. The supremum of error-free transmission rate, known as \emph{channel capacity}, is achieved through specific channel coding/decoding and precoding/equalizing designs. However, identifying optimal channel codes and precoders/equalizers is challenging. As a result, research efforts often concentrate on determining achievable transmission rates. Analyzing transmission rates depends on the channel model. Our next objective is to review existing literature investigating transmission rates in both LoS and NLoS models.
\subsubsection{Line-of-Sight Channels}
Let us first analyze transmission rates in near-field LoS channels supported by SPD and CAP arrays.

$\bullet$ \emph{Spatially-Discrete Arrays:}
The work in \cite{wu2023multiple} analyzed the sum rate of a near-field downlink multiuser MISO channel using zero-forcing (ZF) precoding and the USW model. A closed-form approximation for the sum rate was derived using Bessel integral functions, suggesting that channel correlation between different users decreases as the number of array elements increases. However, this result is imprecise due to the limitations of the USW model as the number of antennas approaches infinity, as discussed in \cite{ding2023resolution}. As explained in Section \ref{Section: Performance Analysis of NFC: DoF and EDoF: LoS Channels}, the singular values of near-field SPD-MIMO channels exhibit a \emph{step-like} pattern. Leveraging this property, the authors of \cite{xie2023performance} introduced ${\mathsf{EDoF}}_2$ as an approximation for the EDoFs. This approach provided closed-form solutions based on dominant singular values, allowing them to derive a closed-form expression for the capacity of a near-field single-user MIMO channel with isotropic inputs and uniform power allocation. Both works assumed a fixed orientation and antenna spacing for the SPD array. In contrast, \cite{do2020reconfigurable} analyzed the channel capacity of a near-field single-user MIMO channel with reconfigurable uniform linear arrays. They derived a general upper bound for the information-theoretic capacity, considering arbitrary array orientation and antenna spacing, as well as the associated conditions for achieving this capacity bound. Their findings highlighted the significant influence of array orientation and antenna spacing on capacity. Building on this, they proposed a method for identifying the optimal array configuration with a low-complexity implementation. Their subsequent work \cite{do2022line} extended this analysis to the capacity of an RIS-assisted near-field single-user MIMO channel, providing a general upper bound expressed as a function of the singular values of the transmitter-to-RIS and RIS-to-receiver channel matrices.

The above works were based on the USW model. Extensions to the NUSW model were explored in \cite{zhou2015spherical,de2020near,lu2021near}. \cite{zhou2015spherical} analyzed the sum rate of a two-user near-field uplink MISO system using a minimum mean squared error (MMSE) equalizer with successive interference cancellation (SIC) for optimal reception. This work was expanded in \cite{de2020near}, where parallel maximum ratio combining (MRC) and MMSE equalization schemes were considered. Numerical results in both studies indicated that the sum rate reaches a saturation point as the number of receive antennas increases. While the MMSE scheme quickly converges to an interference-free scenario, MRC presents a notable performance gap. An extension to a more general multiuser scenario incorporating ZF decoding was presented in \cite{lu2021near}. In these three works, the sum rate was linked to channel correlation between different users. However, deriving a closed-form expression or approximation for this correlation remains challenging under the NUSW model. As a practical alternative, these studies provided numerical results showing that the correlation between distinct near-field users diminishes as the number of array elements increases.

The above works modeled the spatial response between SPD antennas as a complex coefficient, derived from the scalar Green's function. The authors of \cite{wei2023tri} investigated the sum rate of a near-field downlink multiuser MIMO channel using the tensor Green's function to describe LoS propagation between SPD antennas. Unlike the scalar model, the tensor model introduces two major sources of interference: cross-polarization and inter-user interference. To mitigate these sources, the authors designed a ZF-based dual-layer precoding approach. The first layer addresses cross-polarization, while the second layer manages interference from other users in co-polarized channels. The work in \cite{wei2023tri} assumes parallel arrays, while \cite{gong2023holographic} analyzed the capacity of a near-field single-user tri-polarized MIMO channel with isotropic inputs and equal transmit power allocation (ETPA) in non-paraxial and non-parallel deployments. They used the Fresnel approximation to simplify the channel matrix and derived an upper bound on capacity, which serves as a performance limit for the system.

$\bullet$ \emph{Continuous-Aperture Arrays:}
Next, we will shift our attention to CAP-NFC.

In \cite{hu2018beyond}, the authors analyzed the channel capacity of a near-field uplink multiuser MISO channel where uniformly distributed users communicate with a CAP array under a paraxial setup. The users are arranged in a uniform grid parallel to the CAP array. A spatio-temporal MRC-based equalizer approximated the channel as a sinc-function-like inter-symbol interference (ISI) channel. Using this model, the authors studied the channel capacity averaged across the number of terminals and the grid area while also identifying optimal user spacing. Although the derivations in \cite{hu2018beyond} were mathematically rigorous, they might not be easily accessible to all readers. To address this, the authors of \cite{xie2023performance} introduced a simplified ${\mathsf{EDoF}}_2$-based approximation to calculate dominant singular values and channel capacity between two linear CAP arrays. This approach resulted in a straightforward closed-form expression for the channel capacity, accounting for the geometrical randomness of the CAP transmit array. A similar method was employed in \cite{do2022line}, where Landau's theorem was used to estimate the number of EDoFs in an RIS-aided near-field CAP-MIMO channel. Building on these findings and leveraging the step-like characteristics of the channel's singular values, the authors derived a closed-form expression for the capacity upper bound.

The results discussed above involve various approximations and simplifications. A significant advancement was made in \cite{wan2023mutual}, which derived an exact expression for the channel capacity of a tensor Green's function-based MIMO channel between two continuous linear arrays. By modeling the communication process as random fields, the authors employed the Mercer expansion \cite{mercer1909xvi} and Fredholm determinant \cite{simon2005trace} to derive comprehensive expressions for the channel capacity under both white-noise and colored-noise scenarios. Building on this framework, the research in \cite{wan2023can} incorporated the effects of mutual coupling. Theoretical analysis revealed a key insight: as the antenna density increases without mutual coupling, the difference in channel capacity between CAP-MIMO and SPD-MIMO systems diminishes progressively, eventually converging to zero.

The channel capacity expressions in \cite{wan2023mutual} and \cite{wan2023can} are based on the assumption of uniform transmit power allocation. It is well-known that achieving channel capacity requires acquiring the SVD of the channel response and applying an appropriate power allocation algorithm. However, as previously discussed in Section \ref{Section: Performance Analysis of NFC: DoF and EDoF: LoS Channels}, decomposing the Green's operator is computationally intensive. 

To address this challenge, \cite{sanguinetti2022wavenumber} introduced a wavenumber-division multiplexing (WDM) scheme in which the spatially continuous transmitted currents and received electric fields are represented using Fourier basis functions. This representation enables deriving the transmission rate based on the resulting wavenumber-domain channel response. As the array aperture approaches infinity, these spatial Fourier basis functions become asymptotically orthogonal in the wavenumber domain. However, due to the infinite spatial extent of the EM channel, WDM cannot provide non-interfering communication modes or a diagonal wavenumber-domain channel response. Consequently, linear equalizers based on SVD, MMSE, and MRC are essential for reducing interference \cite{sanguinetti2022wavenumber}. While \cite{sanguinetti2022wavenumber} assumed a paraxial and parallel deployment scenario, the results were extended to non-parallel deployment configurations in \cite{d2022performance}. 

\begin{table*}[!t]
\caption{Contributions on Transmission Rate Analysis for LoS Channels} 
\label{tab:Section_Performance_Analysis_TR_LoS_Table}
\centering
\resizebox{0.95\textwidth}{!}
{\begin{tabular}{|l|l|l|l|l|l|l|l|}
\hline
\textbf{Antenna}              & \textbf{Ref.}    & \textbf{Scenarios} & \textbf{Array}   & \begin{tabular}[c]{@{}l@{}}\textbf{Paraxial} \\ \textbf{Deployment}\end{tabular} & \textbf{Precoding} & \textbf{Equalization} & \textbf{Characteristics} \\ \hline
\multirow{9}{*}{\textbf{SPD}} & {\cite{wu2023multiple}} & {MU-DL-MISO}   & Planar & {Arbitrary}                                                        & {ZF}   & {---}    & A closed-form sum rate under the USW model         \\ \cline{2-8} 
                     & {\cite{xie2023performance}} & {SU-MIMO}   & {Linear} & {Paraxial}                                                        & Isotropic   & {MMSE-SIC}    & An ${\mathsf{EDoF}}_2$-based sum rate approximation for USW models \\ \cline{2-8} 
                     & {\cite{do2020reconfigurable}} & {SU-MIMO}   & {Linear} & Paraxial                                                        & {SVD}   & {SVD}    & A capacity upper bound between two reconfigurable ULAs        \\ \cline{2-8} 
                     & {\cite{do2022line}} & {SU-MIMO}   & {Planar} & {Paraxial}                                                        & {SVD}   & {SVD}    & A capacity upper bound for RIS-aided MIMO-NFC         \\ \cline{2-8} 
                     & {\cite{zhou2015spherical}} & {TU-UL-MISO}   & {Linear} & {Arbitrary}                                                        & {ETPA}   & {MMSE-SIC}    & A simulated sum rate under the NUSW model          \\ \cline{2-8} 
                     & {\cite{de2020near}} & {TU-UL-MISO}   & {Planar} & {Arbitrary}                                                        & {ETPA}   & MRC/MMSE    & Relation between sum rate and channel correlation among users          \\ \cline{2-8} 
                     & {\cite{lu2021near}} & {MU-UL-MISO}   & {Planar} & {Arbitrary}                                                        & {ETPA}   & ZF/MRC/MMSE    & A closed-form sum rate under the NUSW model           \\ \cline{2-8} 
                     & {\cite{wei2023tri}} & {MU-DL-MIMO}   & {Planar} & Arbitrary                                                        & ZF   & {---}    & The sum rate for tensor Green's function-based multiuser NFC            \\ \cline{2-8} 
                     & {\cite{gong2023holographic}} & {SU-MIMO}   & {Planar} & {Arbitrary}                                                        & SVD/ETPA   & {SVD}    & A capacity upper bound for near-field tri-polarized CAP-MIMO\\ \hline
\multirow{7}{*}{\textbf{CAP}} & {\cite{hu2018beyond}} & {MU-DL-MISO}   & {Planar} & {Paraxial}                                                        & {ETPA}   & {MRC}    & The channel approximated by a sinc-function-like ISI channel         \\ \cline{2-8} 
                     & {\cite{xie2023performance}} & {SU-MIMO}   & {Linear} & {Paraxial}                                                        & {ETPA}   & {MMSE-SIC}    & ${\mathsf{EDoF}}_2$-based channel capacity approximation         \\ \cline{2-8} 
                     & {\cite{do2022line}} & {SU-MIMO}   & {Planar} & {Arbitrary}                                                        & {SVD}   & {SVD}    & Landau’s theorem-based capacity bound for RIS-aided MIMO         \\ \cline{2-8} 
                     & {\cite{wan2023mutual}} & {SU-MIMO}   & {Linear} & Paraxial/Parallel                                                       & SVD/ETPA  & {SVD}    & A CAP-MIMO capacity expression based on random fields        \\ \cline{2-8} 
                     & {\cite{wan2023can}} & {SU-MIMO}   & {Linear} & Paraxial/Parallel                                                       & SVD/ETPA   & {SVD}    & A CAP-MIMO capacity expression with mutual coupling \\ \cline{2-8} 
                     & {\cite{sanguinetti2022wavenumber}} & {SU-MIMO}   & {Linear} & Paraxial/Parallel                                                        & {WDM}   & SVD/MRC/MMSE    & Transmission rates of WDM for parallel arrays            \\ \cline{2-8} 
                     & {\cite{d2022performance}} & {SU-MIMO}   & {Linear} & {Arbitrary}                                                        & {WDM}   & SVD/MRC/MMSE    & Transmission rates of WDM for arbitrary arrays            \\ \hline
\end{tabular}}
\end{table*}

For easy reference, the key contributions related to transmission rate analysis in near-field LoS channels are summarized in Table \ref{tab:Section_Performance_Analysis_TR_LoS_Table}. ``SU'', ``TU'', and ``MU'' represent single-user, two-user, and multiuser, respectively. ``DL'' and ``UL'' represent downlink and uplink, respectively.

\subsubsection{Non-Line-of-Sight Channels}
Next, we focus on the transmission rate in NLoS channels, where our primary objective is to analyze the mean transmission rate, considering the stochastic nature of small-scale fading. This measure is commonly referred to as the \emph{average transmission rate (ATR)} or \emph{ergodic transmission rate (ETR)}. In the following section, we will review existing analyses of ATR and ETR, with a particular emphasis on both \emph{PPBSMs (physical propagation-based stochastic models)} and \emph{CBSMs (correlation-based stochastic models)}.

$\bullet$ \emph{Physical Propagation-based Stochastic Models:} The statistics of PPBSMs arise from the randomness of scatterers' reflection coefficients and spatial distribution. In \cite{liu2023near}, closed-form expressions were derived for the ATR in a near-field downlink single-user MISO channel for both finite-dimensional channel (FDC) models and clustered channel models (CCMs). The NUSW model was used to describe the array response between the array and the scatterers. Extending this analysis to the MIMO scenario remains challenging, and to date, a closed-form expression for ATR in MIMO is still unavailable, with only bounds established using Jensen's inequality \cite{yang2019ergodic}. The inherently complex nature of PPBSMs makes ATR analysis challenging, which is why most research focuses on ATR analysis under CBSMs.

$\bullet$ \emph{Correlation-based Stochastic Models:} Compared to PPBSMs, CBSMs are more analytically tractable, as they directly model the channel mean and covariance.

Considering the blockages that affect VRs, \cite{ali2019linear} analyzed the ATR for a downlink multiuser MISO channel by accounting for near-field channel correlation based on the single-scattering model (SSM). Using high-dimensional random matrix theory, \cite{ali2019linear} derived deterministic expressions for the SINR achieved with ZF and maximum ratio transmission (MRT) precoders, as well as closed-form expressions for the average sum rates (ASRs). The results showed that, under optimal conditions where users' VRs do not overlap, the SINR achieved with MRT and ZF can surpass that of a stationary channel. These findings were expanded in \cite{marinello2020antenna}, which proposed an energy-efficient antenna selection strategy. Using the same channel model, \cite{xu2023low} analyzed the ASR achieved by regularized ZF (RZF) precoding and provided a low-complexity regularization design method.

The above studies focused on downlink scenarios. The uplink sum rate of a near-field multiuser MISO channel was analyzed in \cite{yang2020uplink} using the SSM-based correlation model. For MRC equalization, an approximate ASR expression was derived using Mullen's and Jensen's inequalities \cite{mullen1967teacher}. For linear MMSE (LMMSE) equalization, closed-form ASR expressions were obtained for scenarios where users' VRs either completely or partially overlap. These results assume perfect knowledge of each user's VR. In contrast, \cite{zhang2022average} introduced a novel VR estimation algorithm and analyzed a deterministic equivalent of the ASR for a near-field downlink multiuser MISO channel with MRT precoding. In \cite{li2015capacity}, the authors analyzed the ASR in a near-field uplink multiuser MISO channel using the double-scattering model (DSM) and MMSE-SIC equalization. They applied the principal minor determinant expansion theorem \cite{grant2002rayleigh} and the Binet-Cauchy formula \cite{browne2018introduction} to derive a closed-form upper bound for the sum rate. Using the same scattering model, \cite{guerra2022clustered} examined the ASR of a near-field uplink multiuser MISO channel with MRC, ZF, and MMSE equalizers.

Apart from the two VR-based models discussed above, another approach for modeling near-field multipath fading is the Fourier plane-wave-based stochastic model (FPBSM). Initially developed for CAP arrays, this model also applies to SPD arrays to describe small-scale fading between antenna pairs. The SPD-MIMO channel essentially resembles a Weichselberger fading channel, a more general model that includes the Kronecker model as a special case \cite{ozcelik2003deficiencies}. Building on this, researchers have used random matrix theory to analyze the channel capacity of a near-field single-user MIMO system with isotropic inputs and MMSE-SIC equalization, resulting in an analytically tractable expression for the ATR \cite{pizzo2022fourier}. This framework can be extended to analyze near-field multiple-access channels, as discussed in \cite{couillet2011deterministic}.

In \cite{zhang2023capacity}, the analysis from \cite{pizzo2022fourier} was extended to include non-uniform transmit power allocation, and the channel capacity of a near-field single-user MIMO system was numerically analyzed using a wrapped Gaussian azimuth angle distribution and a truncated Laplacian elevation angle distribution for scatterers. These contributions were then applied to the downlink multiuser MIMO scenario in \cite{wei2022multi}, where closed-form approximations were derived for the ASR achieved by ZF and MRT precoding. In this work, users are assumed to have perfect channel information, which is difficult to achieve in practice. Acknowledging this limitation, \cite{bahanshal2023holographic} analyzed the achievable ASR of a downlink multiuser MIMO system where each user only knows the channel mean. These studies were further extended to the cell-free scenario in \cite{lei2023uplink}, where the achievable uplink ATR was numerically evaluated. Most of the aforementioned works focus on the mean of the transmission rate. In \cite{zhang2023fundamental}, the authors also examined the second-order statistics of the transmission rate in a near-field single-user MIMO channel under FPBSM. Using the central limit theorem, they showed that the transmission rate follows a Gaussian distribution as the number of antennas tends to infinity.

\begin{table*}[!t]
\caption{Contributions on Transmission Rate Analysis for NLoS Channels}
\label{tab:Section_Performance_Analysis_TR_NLoS_Table}
\centering
\resizebox{0.95\textwidth}{!}
{\begin{tabular}{|l|l|l|l|l|l|l|l|}
\hline
\textbf{Category}               & \textbf{Ref.}    & \textbf{Scenarios} & \textbf{Array}   & \textbf{Correlation} & \textbf{Precoding} & \textbf{Equalization} & \textbf{Characteristics} \\ \hline
\multirow{2}{*}{\textbf{PPBSM}} & {\cite{liu2023near}} & {SU-DL-MISO }   & {Planar} & {Kronecker}     & {MRT}   & {---}    & A closed-form ATR         \\ \cline{2-8} 
                       & {\cite{yang2019ergodic}} & {SU-MIMO}   & {Linear} & {FDC}     & Isotropic   & {MMSE-SIC}    & Closed-form upper bounds of the ATR         \\ \hline
\multirow{13}{*}{\textbf{CBSM}}  & {\cite{ali2019linear}} & {MU-DL-MISO}   & {Arbitrary} & Single-Scattering     & {ZF/MRT}   & {---}    & Deterministic equivalents of SINRs and ASRs         \\ \cline{2-8} 
                       & {\cite{marinello2020antenna}} & {MU-DL-MISO}   & {Linear} & Single-Scattering     & {ZF/MRT}   & {---}    & Deterministic ASRs-based antenna selection design         \\ \cline{2-8} 
                       & {\cite{xu2023low}} & {MU-DL-MISO}   & {Arbitrary} & Single-Scattering     & {RZF}   & {---}    & A low-complexity regularizer design to improve the ASR         \\ \cline{2-8} 
                       & {\cite{yang2020uplink}} & {MU-UL-MISO}   & {Linear} & Single-Scattering     & {ETPA}   & MRC/LMMSE    & Closed-form approximations for ASRs         \\ \cline{2-8} 
                       & {\cite{zhang2022average}} & {MU-DL-MISO}   & {Arbitrary} & Single-Scattering     & {MRT}   & {---}    & A VR estimator along with the achieved sum rate         \\ \cline{2-8} 
                       & {\cite{li2015capacity}} & {MU-UL-MISO}   & {Linear} & {Double-Scattering}     & {ETPA}   & {MMSE-SIC}    & A closed-form bound of the average capacity\\ \cline{2-8} 
                       & {\cite{guerra2022clustered}} & {MU-UL-MISO}   & Planar & {Double-Scattering}  & {ETPA}   & MRC/ZF/MMSE   & Numerically analyzed ASRs\\ \cline{2-8} 
                       & {\cite{pizzo2022fourier}} & {SU-MIMO}   & {Planar} & FPBSM  & {ETPA}   & {MMSE-SIC}   & A random matrix theory-based ATR expression\\ \cline{2-8} 
                       & {\cite{zhang2023capacity}} & {SU-MIMO}   & {Planar} & FPBSM  & {SVD}   & {SVD}   & Channel capacity under two non-isotropic scattering channels\\ \cline{2-8} 
                       & {\cite{wei2022multi}} & {MU-DL-MIMO}   & {Planar} & FPBSM  & {ZF\&MRT}   & {---}   & Closed-form approximations for ASRs\\ \cline{2-8} 
                       & {\cite{bahanshal2023holographic}} & {MU-DL-MIMO}   & {Planar} & FPBSM  & {MRT}   & {---}   & Only the channel expectation available for the users\\ \cline{2-8} 
                       & {\cite{lei2023uplink}} & {MU-UL-MIMO}   & {Planar} & FPBSM  & {ETPA}   & {Arbitrary}   & ASRs for near-field cell-free channels\\ \cline{2-8} 
                       & {\cite{zhang2023fundamental}} & {SU-MIMO}   & {Planar} & FPBSM  & {ETPA}   & {MMSE-SIC}   & Second-order statistics of the transmission rate\\ \hline
\end{tabular}}
\end{table*}

For reference, the key contributions to rate analysis in near-field NLoS channels are summarized in Table \ref{tab:Section_Performance_Analysis_TR_NLoS_Table}.

\subsection{Discussions and Outlook}
In this section, we have reviewed mainstream contributions to the performance analysis of NFC. Our primary focus has been on three widely used performance metrics: DoFs, power scaling law, and transmission rate. It is concluded that the spatial non-stationarity brought by near-field effects has a significant impact on the performance of NFC systems, which must be taken into account to derive accurate results. We believe that the insights gained from our review can offer a foundational understanding of the fundamental performance limits of NFC. Additionally, we hope this synthesis will serve as an initial guide for researchers intending to contribute to this evolving field. However, numerous open research problems remain within this domain, some of which are outlined below.
\begin{itemize}
  \item \textit{\textbf{EDoFs Analysis for Fading Channels}} Current research on EDoFs in fading channels does not establish the explicit relationship between EDoFs and the propagation distances between transceivers or scatterers. There are two key reasons for this gap. First, many studies focus on the NLoS component introduced by scatterers while overlooking the LoS component, which is closely linked to propagation distance. Second, when modeling scatterers, most research assumes spherical wavefronts but neglects other effects of spatial non-stationarity in the near field, including FSPL and EAL, e.g., \cite{poon2005degrees,pizzo2022fourier}. Accounting for more spatial non-stationary effects should clarify how EDoFs relate to propagation distances.
  \item \textit{\textbf{Power Scaling Law for General Setups:}} Current research on power scaling laws has mainly focused on single-user LoS channels. For a comprehensive study, insights should be extended to encompass multiuser channels and NLoS propagation. Although power scaling laws for traditional multiuser massive MIMO fading channels are well-documented, they are based on far-field assumptions that overlook spatial non-stationarity, leading to conclusions that violate the principle of energy conservation. While there is no apparent reason to doubt that the asymptotic trends in Table \ref{tab:Section_Performance_Analysis_PSL_LoS_Table} could extend to NLoS propagation---where the SNR or SINR stabilizes at a constant value as the array aperture increases---research in this area remains open and challenging.  
  \item \textit{\textbf{Information-Theoretic Aspects for CAP-NFC:}} A thorough understanding of the information-theoretic aspects of NFC is crucial for its practical implementation. Given the differences in channel and signal models between NFC and FFC, further exploration of NFC's information-theoretic limits is needed. Current research in this area is mainly focused on SPD-NFC, e.g., \cite{zhao2024modeling,zhao2024channel}. While some efforts have been made to study the information-theoretic limits for CAP-NFC, most of these studies are limited to single-user or two-user scenarios; see \cite{wan2023mutual,zhao2024continuous} for more details. Future research should explore near-field broadcast channels, multiple-access channels, wiretap channels, and other scenarios under a more general setup.
\end{itemize}





\section{Signal Processing for Near-Field Communications}\label{Section: Signal Processing for NFC}
As detailed in the preceding section, when contrasting NFC with FFC, fundamental disparities emerge in their EM characteristics and channel models. As a result, the signal processing techniques employed in NFC diverge significantly from those applied in FFC. In this section, we will provide an overview of pertinent research endeavors from three distinct vantage points: channel estimation, beamforming design, and low-complexity beam training.

\subsection{Near-Field Channel Estimation}
Numerous wireless applications crucially rely on channel state information (CSI) to mitigate the impact of factors such as fading, noise, and interference during data transmission. Given the stochastic nature of wireless channels, channel estimation is typically carried out prior to transmission to acquire the CSI. Traditional channel estimation techniques, including the least-squares (LS) and the LMMSE methods, have been employed for unstructured channels \cite{van1995channel}. However, as the number of antennas has substantially increased, the LS and LMMSE methods often incur significant computational overhead. Moreover, these methods fail to identify the positions of scatterers within the channel, limiting the potential for advanced transceiver design. These challenges motivate the consideration of channel estimation for structured channels. Structured channels are usually encountered in high-frequency bands like mmWave and THz, where sparse scattering leads to a reduced number of channel parameters. Leveraging the structure of far-field channels, many low-overhead channel estimation methods have been proposed for FFC, making use of approaches such as compressive sensing \cite{bajwa2010compressed, alkhateeb2014channel, lee2016channel} and parametric estimation \cite{guo2017millimeter, liao20172d, shafin2017angle}. However, for NFC, the spherical-wave propagation results in channel structures distinct from those encountered in FFC. Consequently, existing structured far-field channel estimation methods are no longer suitable for NFC. 

\begin{figure}[!t]
    \centering
    \includegraphics[width=0.45\textwidth]{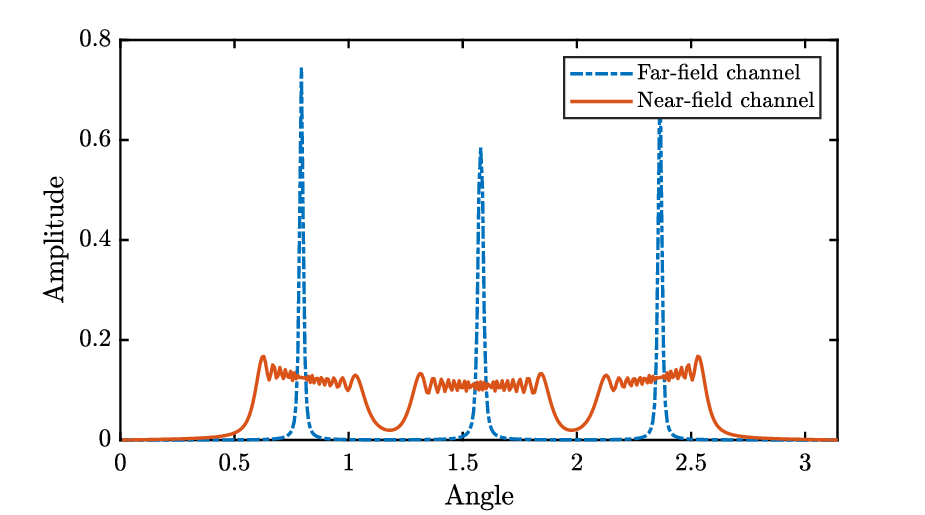}
    \caption{Angular-domain representations of far-field and near-field channels, where the BS is equipped with $N=400$ antenna ULA and operates at $28$ GHz. There are three scatterers for the far-field (100 m) and near-field (5 m) channels.}
    \label{fig:channel_sparsity}
\end{figure}

\subsubsection{Compressive Sensing}
Compressive sensing leverages the sparsity of the channel in a particular domain to enhance channel estimation. In FFC, angular-domain sparsity is typically exploited, based on which the classical orthogonal matching pursuit (OMP) algorithm can be used to recover the angular-domain far-field channel \cite{lee2016channel}. In NFC, the assumption of angular-domain sparsity assumption no longer applies due to spherical-wave propagation, as illustrated in Fig. \ref{fig:channel_sparsity}. \textcolor{black}{Specifically, the far-field channel exhibits discernible peaks in the angular domain, whereas the near-field channel is spread across a wider angular region due to the variation in angle-of-arrival across the array.} To address this challenge, the authors of \cite{le2019massive} refined the far-field OMP method by directly replacing the far-field array response vector with one for the near-field. To reduce the complexity, a sequential method was further proposed in which the conventional far-field OMP is adopted for angle estimation, followed by a refined OMP for range estimation. The authors of \cite{han2020channel} further refined the far-field OMP method by taking the non-stationary property of the near-field channel into account, developing subarray-wise and scatterer-wise near-field channel estimation methods. A more detailed theoretical analysis of near-field compressive sensing in \cite{cui2022channel} unveiled an alternative form of sparsity in the polar domain, which encompasses joint range and angle information. Based on polar-domain sparsity, \cite{cui2022channel} introduced a novel polar-domain OMP to recover the near-field channel. In \cite{guo2023compressed}, the authors explored near-field channel estimation for a planar antenna array based on 3D compressive sensing, considering the elevation-azimuth angles and range. To tackle this complex problem, they proposed a triple parametric decomposition approach that effectively partitions the 3D compressive sensing task into multiple 1D subproblems. Another range-parameterized angular-domain representation of the near-field channel was conceived in \cite{zhang2022near} that addresses the storage burden and high column coherence of the polar-domain representation. By exploiting this new representation, a dictionary-learning OMP channel estimation method was developed. As a further advance, \cite{cao2023efficient} proposed a sparse Bayesian learning (SBL) method for near-field channel estimation based on compressive sensing that requires a lower pilot overhead and computational burden than conventional compressive sensing. Inspired by classical OMP, \cite{cui2023near} proposed a bilinear pattern detection approach for the near-field wideband channel estimation, where the spatial-wideband effect was effectively solved. Another near-field wideband OMP approach was devised in \cite{elbir2023near}, which is based on a beam-squint-aware dictionary.

Considering the practical scenario where both users and scatterers within a communication channel can reside in either the near- or far-field regions, some research efforts have been dedicated to addressing the challenges of hybrid-field channel estimation. In particular, a hybrid-field OMP channel estimation method was proposed in \cite{wei2021channel} that demonstrated reduced estimation errors compared to both far-field and near-field OMP methods. In an effort to mitigate estimation complexity, the authors of \cite{hu2022hybrid} introduced an approach that combines support detection and OMP, and enables the successive estimation of far-field and near-field paths. However, these methods typically require prior knowledge of the proportion of far-field and near-field paths. To address this challenge, an enhanced hybrid-field OMP method was introduced in \cite{yang2023practical} that starts with a coarse estimation based on far-field OMP and subsequently refines the estimate by adjusting the proportion of far-field and near-field paths. Hybrid-field channel estimation for the MIMO scenario was studied in \cite{tarboush2023cross}, and a reduced dictionary method was developed to first determine which model is appropriate for each user.


\begin{figure*}[!t]
    \centering
    \includegraphics[width=0.95\textwidth]{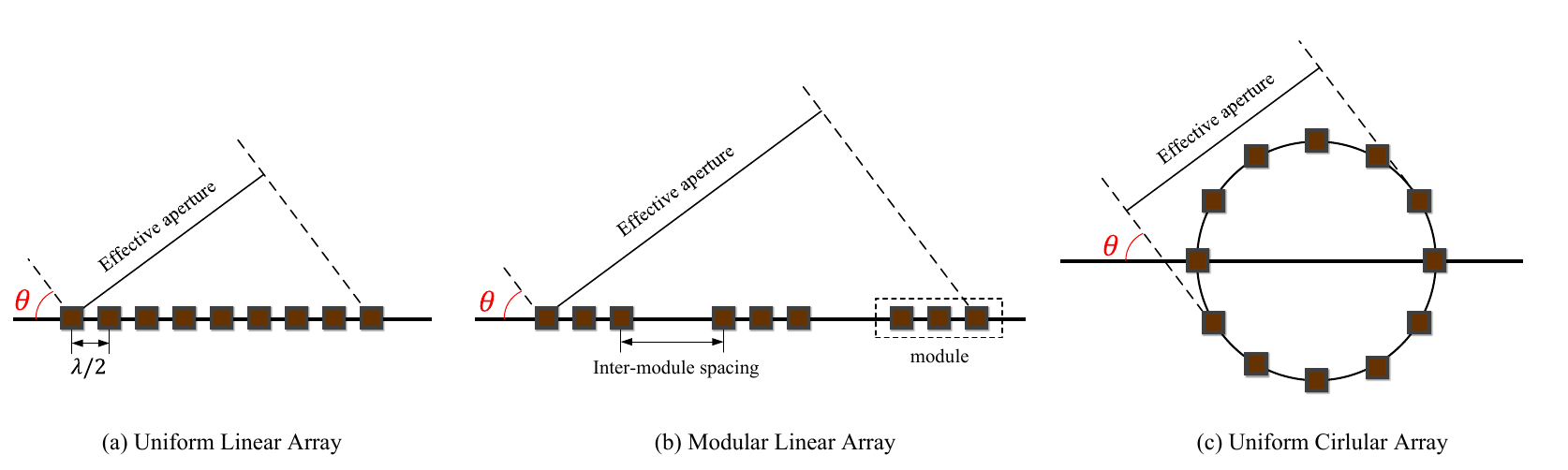}
    \caption{Illustration of array geometries for NFC.}
    \label{fig:array_geo}
\end{figure*}

\subsubsection{Parametric Estimation}
Parametric estimation is based on the assumption that the communication channel can be characterized by a relatively small set of parameters, such as angles of departure and arrival \cite{swindlehurst2022channel}. By estimating these parameters, the communication channels can be reconstructed. Parametric estimation is a mainstream channel estimation paradigm that runs parallel to compressive sensing. The pros and cons of the parametric estimation method compared to compressive sensing are summarized below. For compressive sensing, channel estimation involves building a channel codebook in a selected domain and applying a sparse recovery algorithm such as OMP. The codebook contains the samples of channels in this selected domain, which are also referred to as on-grid points. The estimation accuracy is limited by the resolution of these on-grid points. In contrast to compressive sensing, parametric estimation methods typically estimate the channel parameters using a search over the continuous parameter space. Thus, parametric estimation approaches can be regarded as “off-grid” methods exhibiting higher accuracy, but typically at the cost of higher complexity. Compared to FFC, the key challenge of parametric estimation in NFC is related to the estimation of the range information. The authors of \cite{cui2022channel} proposed an iterative joint range and angle estimation algorithm for all paths in a near-field MISO channel based on the maximum likelihood principle, which demonstrated a better performance than compressive sensing. In \cite{seo2021cumulant}, an approach based on cumulant matrices was introduced for the sequential estimation of angle and range for NLoS scatterers in the near-field MIMO channel. For LoS-dominated near-field MIMO channels, \cite{ghermezcheshmeh2023parametric} divided a large array into several tiles such that the far-field assumption holds at each tile, and then applied parametric far-field channel estimation methods to each tile. Considering both LoS and NLoS paths, \cite{lu2023near} devised a two-stage channel estimation method. In the first stage, the angle and range information for the LoS path was estimated and iteratively refined. Subsequently, in the second stage, the NLoS path parameters were estimated using near-field compressive sensing. Finally, a sub-array-based channel estimation algorithm was developed in \cite{zhu2023sub} for near-field MIMO communications. In particular, the antenna arrays at both the transmitter and receiver were partitioned into multiple sub-arrays, with the channel between each sub-array approximated as a far-field channel. Subsequently, the estimates from each pair of subarrays are combined to determine the overall channel.


\subsubsection{Machine Learning}
Machine learning provides a promising set of tools to facilitate low-complexity near-field channel estimation, and has attracted significant attention. For example, the authors of \cite{lei2023channel} proposed a multiple residual dense network (MRDN) to estimate the near-field channels by exploiting polar-domain sparsity. Then, they adopted atrous spatial pyramid pooling to enhance the performance of the MRDN by capturing the multi-scale information of the input. In \cite{zhang2023model}, a model-based deep learning approach was introduced in which the near-field channel estimation was first formulated as a compressive sensing problem and then solved by the learning iterative shrinkage and threshold algorithm. The authors of \cite{chen2021hybrid} adopted a hybrid spherical- and plane-wave model, where the antenna array is divided into multiple subarrays whose channel parameters are first estimated using a deep convolutional neural network and used to recover the overall near-field channel. Federated learning was employed in \cite{elbir2023near} to facilitate OMP-based near-field channel estimation in wideband systems. To address the hybrid-field channel estimation problem, \cite{yu2023adaptive} conceived a general deep-learning-based framework exploiting iterative channel estimators. A fixed point network was proposed for implementing each iteration, consisting of a closed-form linear estimator and a deep-learning-based non-linear estimator. 


\subsection{Near-Field Beamforming}
Transmit beamforming employs an antenna array to direct a signal towards a specific receiver, rather than broadcasting it in all directions. It can be achieved by adjusting the weights and phases of the antenna elements to create constructive and destructive interference patterns in the desired directions or locations. Beamforming can improve the quality, capacity, and reliability of wireless communication by enhancing the desired signal and mitigating interference. As discussed in previous sections, in contrast to FFC, an antenna array in NFC can resolve not only the direction of the signal but also the distance over which it has propagated. As a result, near-field beamforming can be designed in both the angle and range domains and has greater flexibility than far-field beamforming. In the following, we first review array geometries and control techniques for near-field beamforming, followed by a review of existing near-field beamforming designs.

\begin{figure*}[!t]
    \centering
    \includegraphics[width=0.85\textwidth]{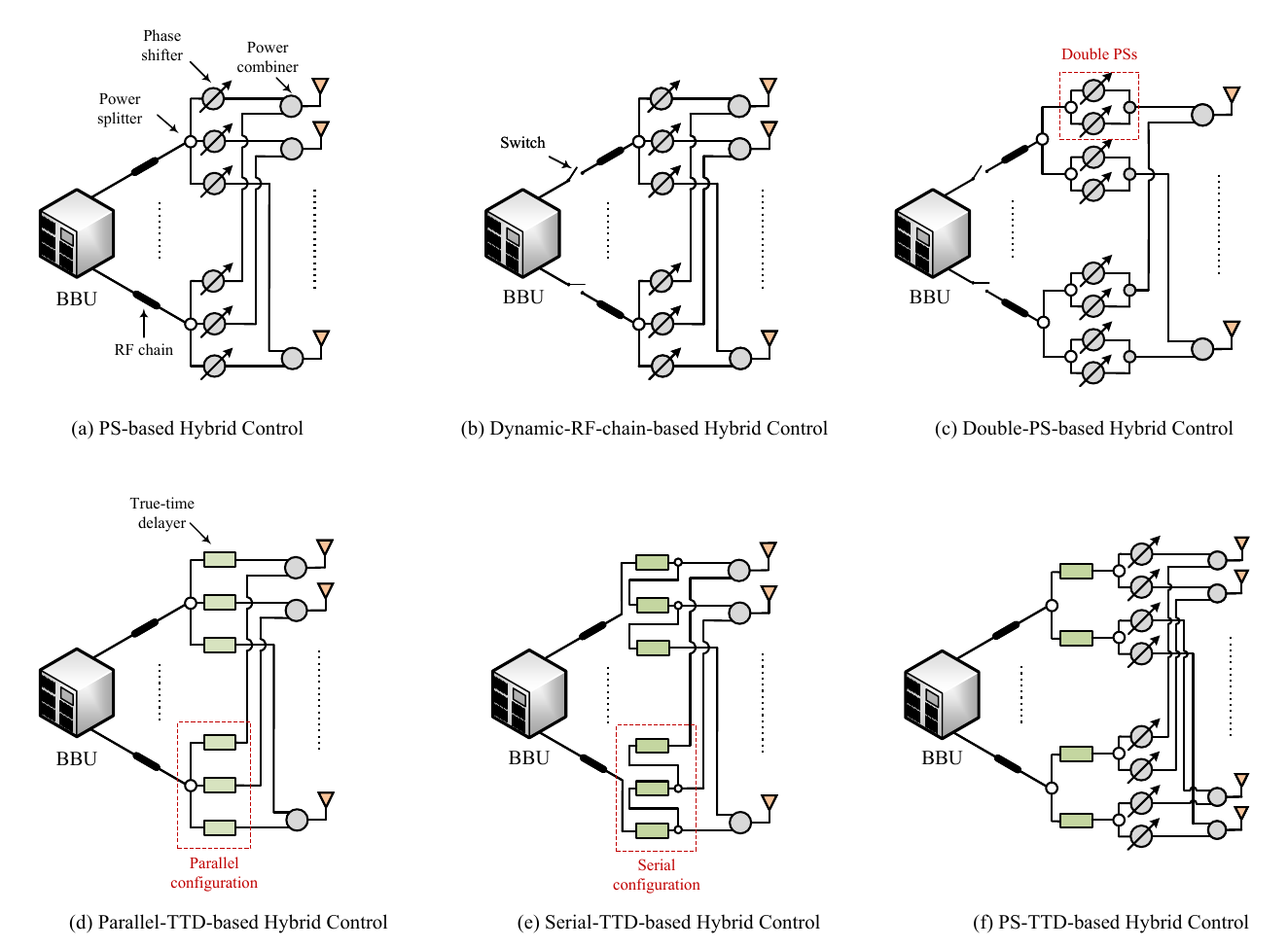}
    \caption{Illustration of hybrid digital and analog control techniques for NFC.}
    \label{fig:control}
\end{figure*}

\subsubsection{Array Geometries and Control Techniques}\label{Section: Array Geometries and Control Techniques}
Array geometries play a pivotal role in shaping the spatial distribution of the beamformer radiation pattern pattern in terms of beamwidth and grating lobes. Furthermore, different control techniques in the analog and digital domains can be employed to realize beamforming, contingent upon the specific demands of the system and the physical limitations. 

$\bullet$ \emph{Array Geometries:} Recall that compared to the far-field case, near-field beamforming introduces an additional dimension related to distance. This additional dimension plays a crucial role in enhancing communication performance, as we will detail in upcoming sections. As a result, a key goal in designing the array geometry in NFC is to expand the near-field region with respect to (w.r.t.) the antenna array, which enables a larger number of communication users to reap the advantages of near-field beamforming. According to the expression for Rayleigh distance, $2A^2/\lambda$, the array aperture is the key factor in determining the size of the near-field region. To avoid grating lobes, ULAs and uniform planar arrays (UPAs) are typically designed to satisfy the Nyquist sampling criterion in the spatial domain \cite{van2002optimum, mailloux2017phased}, i.e., the antenna spacing should be no larger than half of the wavelength, $\lambda /2$, as depicted in Fig. \ref{fig:array_geo}(a). However, for NFC, such a constraint may limit the aperture of the antenna array, thus confining the near-field region. As a remedy, a modular array \cite{li2022near}, which is also referred to as a widely-spaced multi-subarray (WSMS) \cite{song2018two, yan2021joint}, has been proposed. For the example illustrated in Fig. \ref{fig:array_geo}(b), the linear array is divided into multiple modules, and is referred to as modular linear array (MLA). Within each module, the antenna spacing satisfies the half-wavelength criteria, while the inter-module spacing is typically much larger than the signal wavelength. Therefore, compared to a conventional collocated array, a modular array exhibits a larger near-field region, but may incur grating lobes due to the wide spacing between subarrays. The concept of the modular array was first introduced in \cite{song2018two} and \cite{yan2021joint} for mmWave and THz communications, respectively, where the authors posited plane wave propagation across each individual module and spherical wave propagation across different modules. Furthermore, the authors of \cite{li2022near} studied the complete near-field modeling and performance analysis of the modular array design and showed that it exhibits a more significant near-field effect that is more suitable for NFC. Additionally, to address the issue of grating lobes for the modular array, the authors of \cite{li2023multi} conceived a user grouping and scheduling method, where different groups of users are not scheduled in the same resource blocks if they are located in each other's grating lobe. Another geometry used to enlarge the near-field region is the uniform circular array (UCA) \cite{wu2023enabling}, as shown in Fig. \ref{fig:array_geo}(c). For linear and planar arrays, the near-field region is non-uniform in all directions. For instance, the ULA exhibits an ellipsoidal near-field region that is largest for $\theta = \pi/2$, and gradually diminishes to zero as $\theta \rightarrow 0$ or $\pi$ \cite{lu2021communicating}. This phenomenon is due to the fact that the linear antenna arrangement leads to different \emph{effective apertures} in different directions. However, for a UCA, its rotationally symmetric geometry results in the same effective aperture in all directions \cite{wu2023enabling}. Therefore, a user has the potential to benefit from the near-field beamforming even if $\theta \rightarrow 0$ or $\pi$, and the near-field region is enlarged.


$\bullet$ \emph{Control Techniques:} In NFC systems, achieving a large array aperture necessitates the use of a substantial number of antenna elements. However, this presents a challenge for fully digital control due to hardware limitations and power constraints \cite{heath2016overview}. A prevalent solution to this issue is to reduce the number of RF chains through hybrid digital and analog control, as illustrated in Fig. \ref{fig:control}. To clarify, the digital control takes place at a baseband unit (BBU), which then transmits signals to the antenna elements through a limited number of RF chains and an analog control network. Fig. \ref{fig:control}(a) depicts the conventional hybrid control architecture, where a fixed number of active RF chains is employed, and phase shifters (PSs) are used for analog control \cite{heath2016overview, bjornson2023twenty}. However, this conventional hybrid control architecture may not be suitable for NFC for two primary reasons. 

\begin{figure*}[!t]
    \centering
    \subfigure[ULA, $90^\circ$]{
        \includegraphics[width=0.3\textwidth]{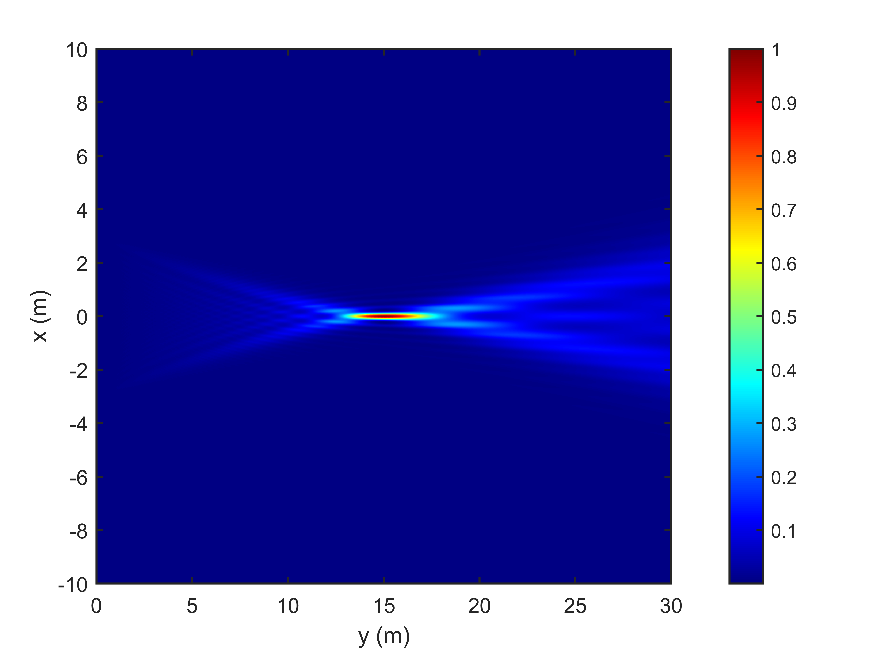}
    }
    \subfigure[MLA, $90^\circ$]{
        \includegraphics[width=0.3\textwidth]{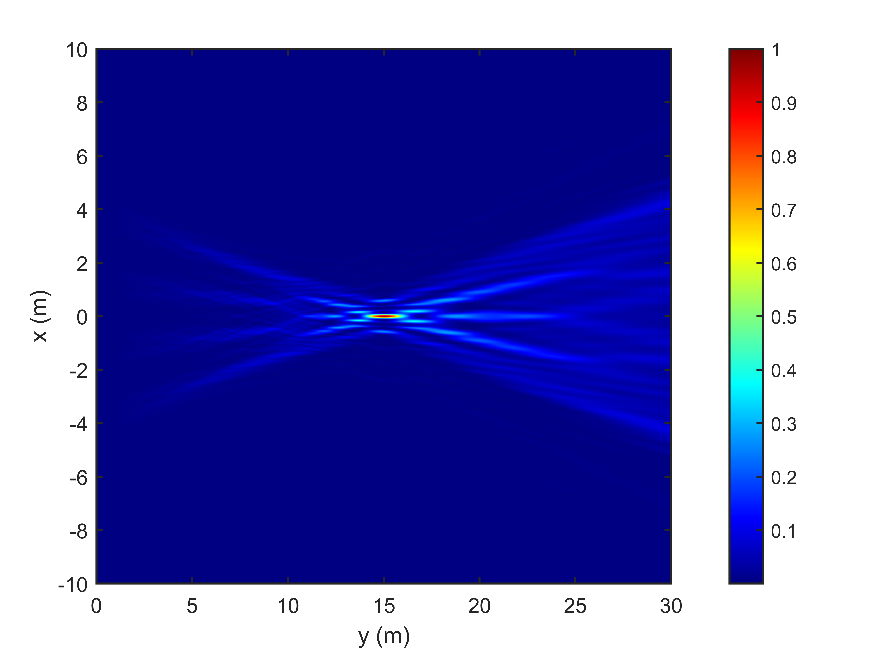}
    }
    \subfigure[UCA, $90^\circ$]{
        \includegraphics[width=0.3\textwidth]{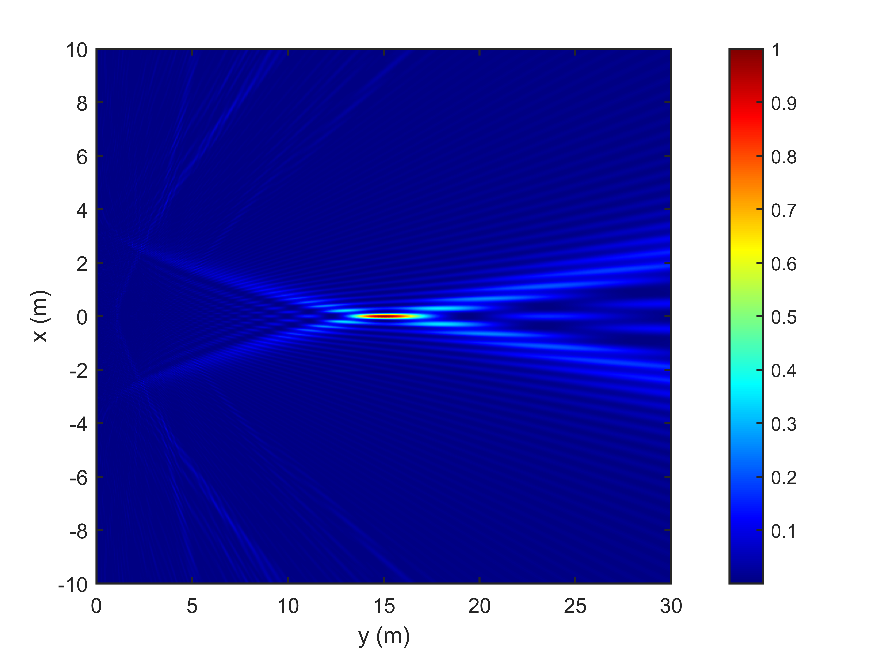}
    }
    \subfigure[ULA, $30^\circ$]{
        \includegraphics[width=0.3\textwidth]{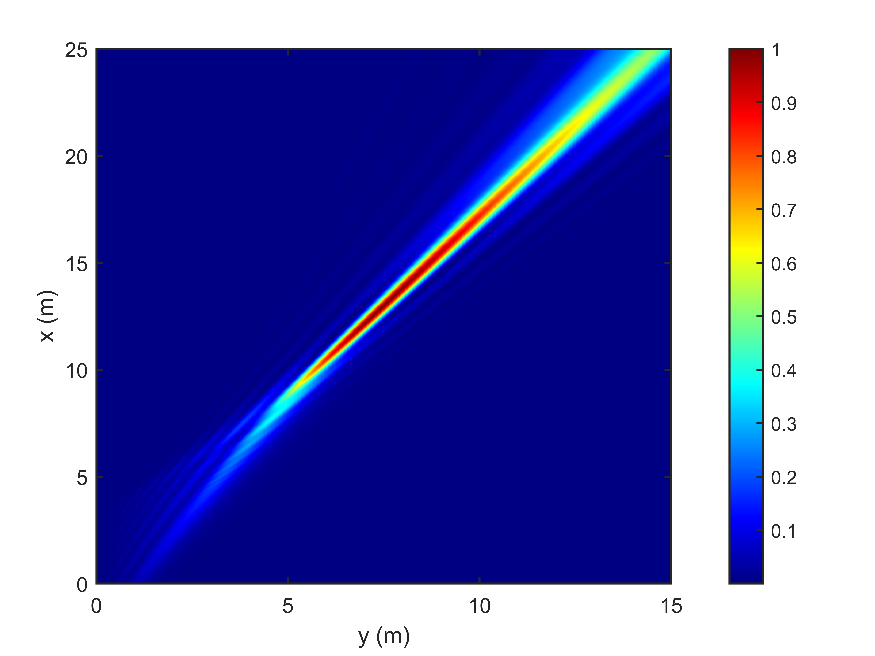}
    }
    \subfigure[MLA, $30^\circ$]{
        \includegraphics[width=0.3\textwidth]{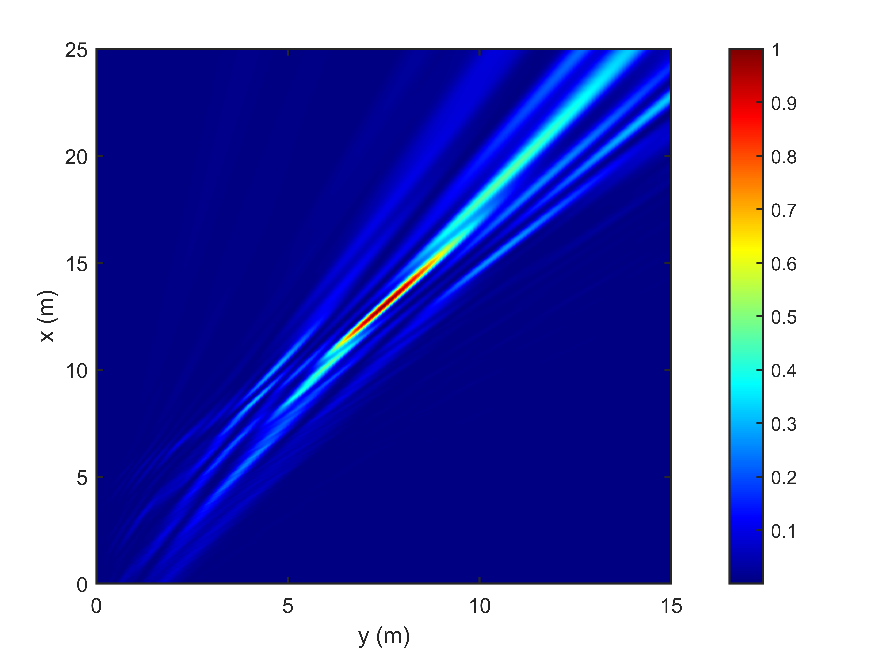}
    }
    \subfigure[UCA, $30^\circ$]{
        \includegraphics[width=0.3\textwidth]{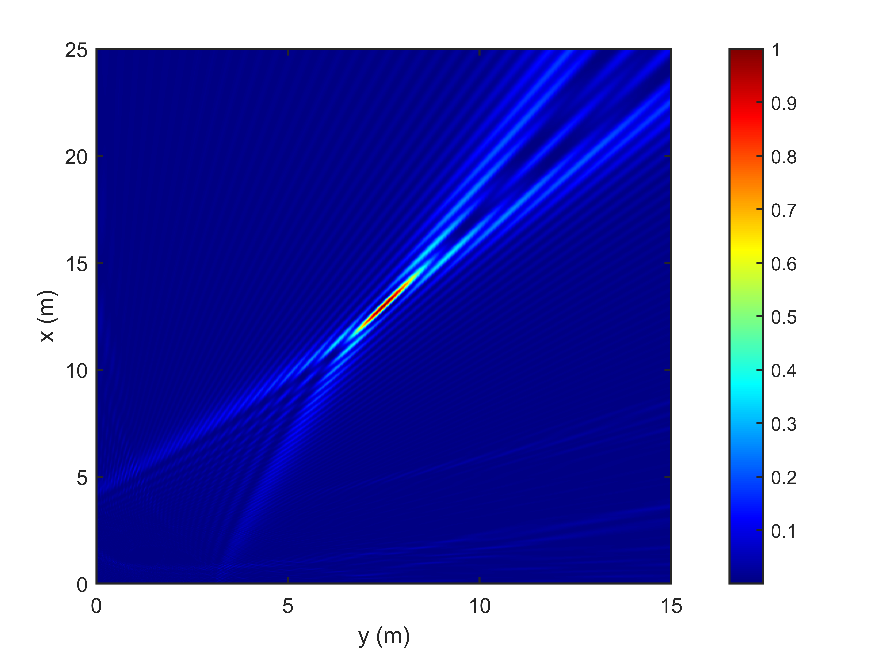}
    }
    \caption{The beamfocusing performance of different array geometries at $15$ m and directions $90^\circ$ or $30^\circ$ with 3 GHz carrier frequency and $128$ antennas. The ULA antenna spacing is set to $\lambda/2$. The MLA antennas are divided into $4$ modules separated by $10 \lambda$, and $\lambda/2$ spacing within each module. The UCA has the same aperture as the ULA. }
    \label{fig:beamfocusing}
\end{figure*}

\emph{First}, as discussed in previous sections, near-field MIMO channels exhibit distance-dependent DoFs. Specifically, the DoFs gradually decrease as the distance between the transmitter and receiver increases, resulting in a reduced number of supported data streams. To address this distance-dependent behaviour in NFC, the authors of \cite{wu2022distance} proposed a dynamic-RF-based hybrid control architecture shown in {\figurename} \ref{fig:control}(b). In this architecture, each RF chain can be configured as active or inactive via a switch based on the distance between the transmitter and receiver. Theoretically, to achieve optimal performance, in this architecture the number of active RF chains should be at least twice the number of data streams, streams, i.e., DoFs \cite{zhang2005variable, sohrabi2016hybrid}. As a more energy-efficient solution, a novel double-PS-based hybrid control architecture was developed in \cite{bogale2016number}, in which each RF chain and antenna element are connected by two PSs. Despite the doubled number of PSs, this architecture only requires the same number of active RF chains as the number of data streams, improving energy efficiency since PSs consume significantly less energy than RF chains.

\emph{Second}, the near-field effect is stronger in high-frequency bands, where greater bandwidth resources are available. In wideband systems, an ELAA may encounter the spatial-wideband effect due to non-negligible propagation delays across the antenna array and the frequency-dependent wavenumber \cite{wang2018spatial}. However, it is important to note that PSs provide the same phase shift for signals at different frequencies. Consequently, relying solely on PSs within the analog control network gives rise to an inherent problem in NFC, known as near-field beam splitting. This issue manifests itself as a misalignment between the beams at most frequencies and the user's location \cite{cui2022rainbow}. To address this challenge and enable effective wideband control, alternative solutions have been explored. One such solution is the utilization of true-time delays (TTDs), which have long been considered as a possible alternative for PSs for wideband beamforming \cite{hashemi2008integrated}. In contrast to PSs, TTDs are capable of compensating for the propagation delay difference across the antenna array and facilitating a frequency-dependent phase shift. Substituting all PSs with TTDs results in the parallel-TTD-based hybrid control architectures in Fig. \ref{fig:control}(d) \cite{hashemi2008integrated, longbrake2012true, rotman2016true, cui2022rainbow}. Nonetheless, TTDs are usually subject to a maximum delay constraint, which can pose a challenge in scenarios where extensive delay compensation is needed, particularly with extremely large antenna arrays. To address this limitation, a novel serial configuration, as exemplified in Fig. \ref{fig:control}(e), has recently emerged as a solution for short-range TTDs \cite{zhai2020thzprism, wang2023ttd}. Compared to the conventional parallel configuration, the serial configuration can effectively bypass the maximum delay limit by accumulating delays across multiple TTDs. It is important to note that implementing TTDs presents significant challenges compared to PSs primarily due to the requirement for variable TTD blocks within a constrained chip area \cite{hashemi2008integrated}. Additionally, TTDs tend to consume more power compared to PSs, making their widespread use in practical applications less desirable. The excessive use of TTDs can be avoided by adopting a PS-TTD-based hybrid control architecture \cite{gao2021wideband, dai2022delay, cui2021near, wang2023beamfocusing}, as shown in Fig. \ref{fig:control}(f). In this architecture, although only a limited number of TTDs are inserted between the RF chains and PSs, a nearly optimal communication performance can be achieved.

Additionally, with the development of EM metamaterials \cite{cui2014coding}, some new forms of antennas have emerged, which require distinct control techniques compared to conventional antennas. DMA arrays \cite{9324910} are a new type of metamaterial-based array that exploits a large number of sub-wavelength-sized metamaterial radiators that are densely arranged to realize an approximately continuous radiating surface. Compared to conventional antenna arrays, DMA arrays can achieve superior beamforming gain and spatial resolution. DMA arrays are controlled with the aid of a signal propagating along a waveguide \cite{smith2017analysis}. The metamaterial radiators are excited by the waveguide signal and emit the signal after amplitude, phase, and polarization adjustment. However, arbitrary control of the metamaterial radiators is difficult to realize. According to the analysis in \cite{smith2017analysis}, there are three possible control mechanisms: binary control, amplitude-only control, and Lorentzian-constrained phase control. Binary control is the easiest control mechanism and simply toggles each radiator between ON and OFF states. Amplitude-only control can be realized by adjusting the damping factor of the near-resonance metamaterial radiator or the oscillator strength. Phase control is more complex than amplitude control because the amplitude will necessarily vary with the phase according to the inherent Lorentzian resonance.

\begin{table*}[!t]
    \caption{Contributions on Near-Field Beamforming Design}
    \centering
    \resizebox{0.95\textwidth}{!}{
    \begin{tabular}{|c|l|l|l|l|l|}
    \hline 
    \textbf{Category} & \textbf{Ref.} & \textbf{Design Method} & \textbf{System Bandwidth} & \textbf{Control Technique}  & \textbf{Characteristics} \\ \hline
    \multirow{9}{*}{\makecell[c]{\textbf{Analytical}}} & \cite{bjornson2021primer} & Beamfocusing & Narrowband & --- & Primer on beamfocusing by MRT \\\cline{2-6}
    & \cite{kosasih2023finite} & Beamfocusing & Narrowband & --- & Finite-resolution beamfocusing for ULA, UCA, and UPA \\\cline{2-6}
    & \cite{ding2023resolution} & Beamfocusing & Narrowband & --- & NOMA with imperfect beamfocusing \\  \cline{2-6}
    &\cite{li2022analytical} & Beamfocusing & Narrowband & --- & Controlable beamfocusing via carrier frequency offset  \\\cline{2-6}
    &\cite{cui2021near} & Beamfocusing & Wideband & PS-TTD-based & Piecewise-far-field model for designing TTD coefficients \\ \cline{2-6}
    &\cite{wang2023beamfocusing} & Beamfocusing & Wideband & PS-TTD-based & Piecewise-near-field model for designing TTD coefficients\\ \cline{2-6}
    &\cite{myers2021infocus} & Beamfocusing & Wideband & PS-based & Approximately frequency-flat beamforming gain \\ \cline{2-6}
    &\cite{singh2022wavefront} & Bessel beam & Narrowband &---  & Bessel beam for near-field beamsteering  \\ \cline{2-6}
    &\cite{singh2023utilization} &Bessel beam & Wideband &--- &  Bessel beam for mitigating beam split \\ \hline 

    \multirow{10}{*}{\makecell[c]{\textbf{Optimization}\\\textbf{based}}}   & \cite{zhang2022beam} & WMMSE, BCD  & Narrowband & DMA-based & Beamforming under Lorentzian phase-shift constraint \\\cline{2-6}
    & \cite{li2023near} & SCA, ADMM  & Narrowband & DMA-based & Beamforming under various DAM constraints \\\cline{2-6}
    & \cite{wu2022distance} &Water-filling, greedy search   & Narrowband & Dynamic-RF-based & Distance-aware beamforming for near-field MIMO  \\\cline{2-6}
    & \cite{zhang2023physical} & WMMSE, BCD & Narrowband & PS-based & Distance-domain security by ULA  \\  \cline{2-6}
    &\cite{gao2022multiple} &Rayleigh
    quotient, gradient descent & Narrowband & PS-based & Distance-domain security by modular array \\\cline{2-6}
    &\cite{xu2022near} & WMMSE, MM & Wideband & DMA-based & Wideband beamforming design for DMA  \\ \cline{2-6}
    &\cite{zhang2023deep} &Signal
    model-based learning & Wideband &PS-TTD-based  & CSI-free single-user wideband beamforming with parallel TTDs \\ \cline{2-6}
    &\cite{wang2023beamfocusing} & PDD & Wideband &PS-TTD-based & Multi-user wideband beamforming with parallel TTDs \\   \cline{2-6}
    &\cite{wang2023ttd} & PDD  & Wideband &PS-TTD-based &Multi-user wideband beamforming with serial TTDs  \\ \hline 
    \multicolumn{6}{l}{\makecell[l]{${{{}^{\ddag}}}$ WMMSE: weighted minimum mean square error; BCD: block coordinate descent; SCA: successive convex approximation; \\ ADMM: alternating direction method of multipliers; MM: majorization-minimization; PDD: Penalty Dual Decomposition.}}\\
    \end{tabular}
    }
\end{table*}

\subsubsection{Beamforming Design}
Various near-field beamforming strategies have been explored, employing various antenna geometries and control techniques. In general, these existing designs can be categorized into two main groups: analytical beamforming design and optimization-based beamforming design. Specifically, analytical beamforming design primarily seeks closed-form solutions by capitalizing on the distinctive signal characteristics within the near-field region, while optimization-based beamforming design focuses on attaining optimal or near-optimal solutions under different hardware constraints in various application scenarios. In the following, we review research works on these two approaches.

$\bullet$ \emph{Analytical Beamforming Design:} In \cite{bjornson2021primer}, it was demonstrated that near-field beamforming can control not only the beamwidth in the angular domain, as in far-field scenarios, but also the beamwidth in range. This unique characteristic opens up a new paradigm, known as \emph{beamfocusing}, leveraging the ability to precisely focus the beam on a specific location. In Fig. \ref{fig:beamfocusing}, the beamfocusing performance of different array geometries is depicted. Expanding on this concept, the authors of \cite{kosasih2023finite} conducted a detailed analysis of the finite resolution of beamfocusing. This analysis is based on the 3 dB depth of focus, providing insights into the achievable beamfocusing region for different antenna geometries. Beamfocusing resolution was also studied in \cite{ding2023resolution}, with particular attention to users near the Rayleigh distance and for a case study involving non-orthogonal multiple access (NOMA), where the near-field beam was reused to serve multiple users. As a further advance, \cite{li2022analytical} introduced an analytical beamfocusing model that offers valuable insights into the factors that influence focusing performance, including carrier frequency, array dimension, and user range. Based on this model, a pair of approaches was devised to achieve flexible beamfocusing through the manipulation of the carrier frequency offset. Wideband beamfocusing was first studied in \cite{cui2021near}, where the PS-TTD-based hybrid control architecture was harnessed to tackle the spatial-wideband effect. In particular, a piecewise-far-field model was proposed for designing the coefficients of the PSs and TTDs, ensuring that beams at all frequencies could be precisely focused at the desired location. Based on a similar idea, a piecewise-near-field model was conceived in \cite{wang2023beamfocusing}, offering a more precise and accurate approach to wideband beamfocusing design, particularly in scenarios where the number of TTDs is limited. In a separate investigation, \cite{myers2021infocus} explored wideband beamforming within the context of a PS-based hybrid control architecture applied to a circular planar array. This work introduced an innovative wideband beamforming approach named “InFocus” that capitalizes on a spatial frequency-modulated continuous waveform and achieves an almost uniformly flat beamforming gain across the entire frequency band of interest. 

\begin{figure}[!t]
    \centering
    \subfigure[]{
        \includegraphics[width=0.45\textwidth]{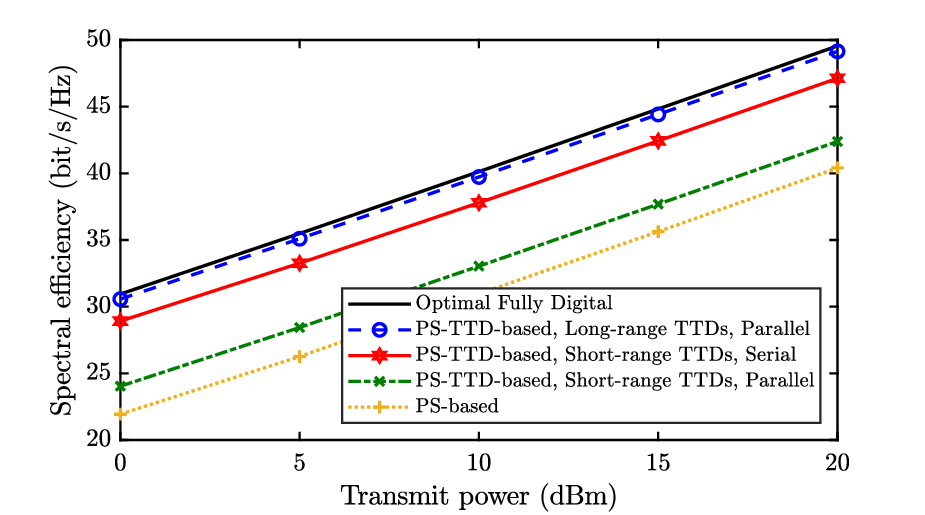}
    }        
    \subfigure[]{
        \includegraphics[width=0.45\textwidth]{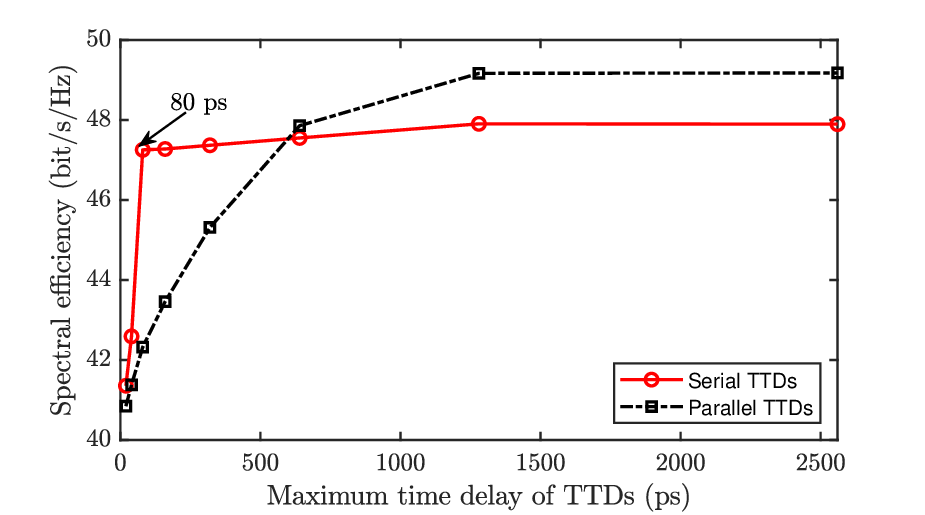}
    }
    \caption{The performance of different control architectures in near-field wideband system. (a) Spectral efficiency versus transmit power, where the maximum time delay of long-range TTDs is $1500$ picosecond (ps) and that of short-range TTDs is $80$ (ps); (b) Spectral efficiency versus the maximum time delay of TTDs, where the transmit power is $20$ dBm.}
\end{figure}

Apart from beamfocusing, the spherical-wave propagation in the near-field region also presents intriguing possibilities for more sophisticated beamforming design. For instance, a near-field beamforming approach based on Bessel beams was proposed in \cite{singh2022wavefront}. In contrast to beamfocusing, the Bessel beams facilitate “beamsteering” in the near-field region. However, a true Bessel beam requires infinite power. To address this issue, the authors devised a more practical quasi-Bessel beam approach that achieves beamsteering for a finite band of ranges in the near-field region. Furthermore, the authors of \cite{singh2023utilization} investigated the performance of Bessel beams in wideband systems, demonstrating that such beams exhibit greater resilience to the spatial-wideband effect when compared to beamfocusing.

$\bullet$ \emph{Optimization-Based Beamforming Design:} In \cite{zhang2022beam}, the authors studied optimized near-field beamfocusing employing different control architectures. In addition to the conventional full-digital and PS-based hybrid control architectures, the authors explored beamfocusing with a DMA array that can realize Lorentzian-constrained phase control. \textcolor{black}{As a further advance, \cite{li2023near} compared the performance of amplitude-controlled and phase-controlled DMA arrays for near-field beamfocusing, and proposed optimized solutions for both single- and multi-user scenarios.} Shifting the focus to near-field MIMO systems with dynamic DoFs, \cite{wu2022distance} presented a distance-aware beamforming optimization framework in which the number of active RF chains is adaptively adjusted based on the estimated DoFs of the near-field MIMO channel to maximize spectral efficiency. The study of secure near-field beamforming design was explored in \cite{zhang2023physical} and \cite{gao2022multiple}, with a specific emphasis on ULAs and modular linear arrays, respectively. To optimize secrecy rates achieved by PS-based hybrid control architectures, the authors of \cite{zhang2023physical} introduced a two-stage optimization algorithm, while the authors of \cite{gao2022multiple} proposed a non-constrained optimum-approaching algorithm. The collective findings of these studies underscore the potential of near-field beamfocusing to enhance physical layer security in the range domain. 


There are also some works exploring near-field wideband beamforming optimization subject to the spatial-wideband effect \cite{xu2022near, zhang2023deep, wang2023beamfocusing, wang2023ttd}. Specifically, near-field wideband beamforming optimization for DMA arrays was studied in \cite{xu2022near}, where both amplitude-only control and the Lorentzian-constrained phase shift control were considered. The authors of \cite{zhang2023deep} concentrated on a single-user wideband system with PS-TTD-based hybrid control, and introduced a deep-learning-based method to optimize near-field wideband beamforming without CSI. For the case of a multiuser system, \cite{wang2023beamfocusing} proposed a fully digital approximation and a heuristic two-stage approach for the design of the PSs and TTDs. Finally, the authors of \cite{wang2023ttd} showed that a serial configuration of the TTDs can substantially reduce the maximum TTD delay requirements compared with the parallel case.  

\subsubsection{Extension to Near-Field Resource Management}\label{Secty}
Beamforming design falls under the category of spatial resource management. To further improve the network utility in NG wireless networks, such as spectral efficiency, energy efficiency, and network delay, it is essential to investigate the resource management in NFC networks. Due to the spherical-wave EM propagation and non-stationary channel characteristics, NFC requires the allocation of various resources, including power control, beamforming, antenna selection, and user scheduling, all of which differ significantly from FFC. For example, near-field beamforming and user scheduling design must consider not only angular information but also distance information. Additionally, the spatial non-stationarity in NFC leads to the existence of VRs, making antenna selection critical for improving the hardware efficiency. Moreover, while spectral/energy efficiency, bit error rate, user fairness, and quality of service are common metrics for assessing FFC performance, NFC not only enhances these metrics but also introduces new ones, such as the EDoF, as discussed in Section \ref{Section: Performance Analysis of NFC: DoF and EDoF}.

Beyond the available resources and metrics, NFC brings new resource management challenges. For example, NFC typically relies on novel antenna types, such as RIS, holographic MIMO, DMA, and fluid antennas, which can introduce new constraints in problem formulation due to hardware limitations; see Section \ref{sec:NFA} for more details. Furthermore, NFC often involves large-scale antenna arrays, leading to high-dimensional optimization problems that are computationally intensive and require new mathematical tools \cite{xu2024resource}. Additionally, NFC generally operates over wide bandwidths in the mmWave/THz bands, where the available bandwidth is affected by propagation conditions due to significant molecular absorption \cite{han2022molecular}. Signal quality may degrade more rapidly over long distances at higher frequencies, making adaptive resource management crucial, with higher power allocated to lower frequencies \cite{an2023toward}.

In summary, near-field resource management differs from its far-field counterpart in several aspects, including available resources and performance metrics, resulting in different optimization objectives and constraints compared to FFC. Simultaneously, the near-field effect introduces new challenges. Resource management in NFC is a promising research direction that deserves further investigation. Currently, research in this area is still in its early stages, and more efforts are needed.
\subsection{Low-Complexity Near-Field Beam Training}
In general, CSI is necessary for beamforming design. While there have been various methods proposed for channel estimation in NFC, the resultant pilot overhead can still be a limiting factor, especially in applications with stringent latency constraints. Beam training provides a promising solution to address this issue \cite{heath2016overview, ning2023beamforming}. Specifically, beam training involves an iterative exchange of information between the transmitter and receiver. The primary objective is to select the optimal beam with the highest received power from a pre-designed codebook. Beam training can achieve rapid beamforming design without the need for CSI, and also facilitates the rapid acquisition of partial CSI of the strongest signal path connecting the transmitter and the receiver. This partial CSI can then be harnessed to conduct more advanced beamforming designs. Although beam training has been extensively investigated in FFC, this prior work cannot be directly applied to NFC because of the additional range dimension in the near-field channels. Consequently, various tailored beam training protocols and codebooks have been designed for NFC. 

\subsubsection{Polar-Domain Beam Training}
In FFC, beam training involves searching for the best beam in an angular-domain codebook \cite{xiao2016hierarchical}. However, a codebook based only on angle is not suitable for near-field beam training due to the additional dimension introduced by the near-field channels. To address this challenge, a polar-domain codebook was crafted in \cite{cui2022channel} comprising non-overlapping beams that cover the entire near-field region. A straightforward approach to identifying the optimal beams is to perform an exhaustive search within this polar-domain codebook, but this method may incur an impractical training overhead. A solution to this issue was proposed in \cite{chen2023hierarchical}, which introduced a hierarchical polar-domain codebook based on analyzing the near-field features. This codebook features multiple layers of codewords, or beams, with the upper layers serving to reduce the search overhead, and the lower layers to ensure high beamforming gains. As a further advance, optimization-based designs of hierarchical polar-domain codebooks were conceived in \cite{lu2022hierarchical}, where Gerchberg-Saxton-based and alternating-optimization-based algorithms were proposed for fully digital and PS-based hybrid control architectures, respectively. In \cite{zhang2023near}, a polar-domain codebook was designed for holographic MIMO systems, which was in turn tailored to develop a multi-user beam training scheme. Additionally, the authors of \cite{xie2023near} compared the performance of ULAs and UCAs in terms of polar-domain beam training. Their findings revealed a significant advantage for UCAs, as they can substantially reduce the codebook size, consequently mitigating beam-training complexity. Reducing near-field beam training complexity is also the goal of the deep-learning-based training method developed in \cite{liu2022deep}, which leverages a neural network to determine the optimal codeword within a polar-domain codebook based on the power received from a set of wide far-field beams. The authors of \cite{shi2023spatial} projected the polar-domain beams into the slope-intercept domain and introduced a novel spatial-chirp-based hierarchical beam training algorithm that can reduce the training overhead by over 99\% compared to conventional methods. In \cite{cui2022rainbow}, the spatial-wideband effect was exploited to reduce the training complexity. In particular, a near-field beam training method referred to as the “near-field rainbow” was proposed based on the controllable beam split effect, which enables multiple beams focused on different locations at the same time but at different frequencies.

\subsubsection{Hybrid-Domain Beam Training}
Polar-domain beam training typically involves a two-dimensional search, which, in turn, results in a substantial number of search steps. To mitigate this issue, several works have explored the incorporation of other domains to reduce the required complexity. In \cite{zhang2022fast}, a two-stage near-field beam training approach was introduced. In the first stage, the coarse angle of the optimal beam is approximated in the angular domain, relying on conventional far-field beam training. Subsequently, in the second phase, polar-domain near-field beam training is conducted within a narrow range centered around the coarse angle. Building on this concept, \cite{wu2023two} integrated hierarchical beam training into the aforementioned two-stage near-field beam training process. Hierarchical far-field and near-field codebooks were designed for the respective phases, resulting in a significant reduction in the number of search steps. 3D near-field beam training for a UPA was studied in \cite{10146329}, also employing a two-stage beam training method. Initially, 2D angular-domain far-field beam training is conducted, followed by 1D range-domain near-field beam training using an iterative Lloyd-Max algorithm. In contrast, the authors of \cite{wang2023near} proposed a distinctive approach that prioritizes range information in the first stage. They designed a codebook containing omnidirectional beams in this stage, facilitating range estimation based on received power. Subsequently, in the second stage, the optimal beam angle is determined via an angular-domain search. An alternative two-stage near-field beam training algorithm was conceived that takes into consideration the concept of VR of the antenna array \cite{10242025}. Based on the fact that different sub-arrays often have different VRs in the near-field, the authors of \cite{10242025} proposed to first identify the VR of the optimal beam,  followed by a search for the optimal beam within the identified VR. Then, the codebook for each stage was designed by exploiting contrastive learning.

\subsection{Discussions and Outlook}
In this section, we have undertaken a comprehensive review of pivotal signal processing techniques relevant to NFC, encompassing channel estimation, beamforming design, and low-complexity beam training. It is evident that the near-field effect holds the potential to enable distance-aware communication with a large number of DoFs. Moreover, substantial research works in NFC underscore the essential consideration of the near-field effect, particularly when exploiting ELAAs and high-frequency bands, to avert significant performance degradation. The effective realization of NFC demands intricate channel estimation, antenna design, and beamforming algorithms. Consequently, the development of low-complexity algorithms assumes a pivotal role in this context, where machine learning emerges as a powerful tool.

To fulfill the required performance targets of future wireless networks, several open research problems from the signal processing perspective remain.
\begin{itemize}
    \item \textbf{\emph{Channel Estimation and Beamforming of CAP Arrays:}} Compared to SPD cases, CAP arrays demonstrate significantly enhanced array gain and spatial resolution, owing to their continuous radiating surface. Nonetheless, such arrays present difficulties in channel estimation and beamforming design. Channel estimation methods developed for SPD arrays cannot be directly applied to CAP arrays due to the Green's function-based channel model. Additionally, beamforming with CAP arrays differs from SPD arrays as it relies on controlling the distribution of source currents rather than discrete element coefficients. Therefore, the channel estimation and beamforming design of CAP arrays for NFC remains an open research challenge. 

    \item \textbf{\emph{Collimation Beamforming Design for NFC:}} Due to the non-linear phase of near-field channels, it is easy to realize beamfocusing. However, beamfocusing is not always preferred due to its limited coverage. In some scenarios, collimation beamforming becomes necessary to cover multiple users with a single beam, particularly when the number of RF chains at the BS is constrained. Bessel beams are a good candidate to realize collimation beamforming \cite{singh2022wavefront, singh2023utilization}, although a true Bessel beam requires infinite power. As a result, only quasi-Bessel beams can be realized in practice, which can achieve collimation beamforming but only for a finite set of ranges. Therefore, more research efforts are required to address this challenge. 
\end{itemize}


\section{Promising Next-Generation Application Scenarios in the Near Field}\label{sec:Incorporating 6G Wireless}
Having revealed the fundamental communication performance improvements by exploiting near-field characteristics, we discuss below the incorporation of near-field models in other promising NG applications, namely NISE, massive connectivity, sustainability enhancement, and information safeguarding, as well as highlight corresponding new research opportunities in near-field scenarios.

\subsection{Near-Field Sensing (NISE)}
In addition to facilitating high-capacity communications, NG wireless networks are also envisioned to enable high-precision sensing for supporting new revolutionary digital applications, such as Metaverse, digital twins, and autonomous systems \cite{liu2022integrated,lu2024integrated}. Generally speaking, wireless sensing can estimate the angle, range, and velocity information by exploiting passively reflected echoes from targets illuminated by wireless signals instead of pilot signals transmitted by targets themselves. Note that in conventional far-field sensing, increasing the size of the antenna array only benefits angle estimation. In contrast, the resolution of range and velocity depend chiefly on the signal bandwidth and the duration of the sensing operation, respectively \cite{zhang2021overview}. Compared with far-field sensing, NISE provides the following two benefits for sensing.
\subsubsection{High-Precision Narrowband Range Sensing}
The precision of far-field range estimation typically relies on the use of a wide signal bandwidth. In contrast, the spherical wavefront characteristic of near-field models enables the acquisition of precise range information even with limited bandwidth. This capability presents new avenues for high-precision range or distance estimation in narrowband systems \cite{wang2023cram, yang2023performance, wang2023near_isac}. It suggests that NISE can achieve comparable range sensing accuracy with reduced bandwidth or spectrum resources compared to far-field methods. This advantage can be further leveraged by employing additional antennas, enhancing the near-field effect in arrays with larger apertures. These findings indicate that NISE offers a more favorable balance between sensing accuracy and the use of spectrum and spatial resources compared to far-field sensing \cite{wang2024performance}.

Owing to these benefits, near-field range sensing has garnered increasing research attention. For instance, \cite{wang2023cram} derived closed-form Cramér-Rao bounds (CRBs) for range estimation in both mono-static and bi-static configurations using ULAs. The range estimation performance was further analyzed using MLAs in \cite{yang2023performance}, demonstrating the advantages of segmenting the array into multiple widely-spaced subarrays. Research on joint waveform designs for near-field ISAC in \cite{wang2023near_isac} minimized the CRBs for NISE while ensuring minimal communication requirements. These contributions were expanded to wideband scenarios to underscore the superiority of NISE in range sensing under both ULA and UCA scenarios \cite{wang2024performance,wang2024performance1}.
\subsubsection{Feasible Transverse Velocity Sensing} 
The estimation of an object's velocity generally depends on the estimation of the Doppler frequency of radar echoes. In conventional far-field systems, the Doppler frequency reflects only the object's radial velocity (the velocity along the LoS direction) due to plane-wave propagation \cite{liu2020radar, sun20214d, xiao2023integrated, chen2023multiple}. The lack of transverse velocity information (the velocity perpendicular to the LoS direction) makes it challenging to obtain the full motion status of the object. Interferometric radar is a method to facilitate transverse velocity estimation in the far field using the measurement of phase changes \cite{nanzer2012resolution, nanzer2016estimation, wang2019simultaneous}. However, this technique conventionally relies on just two antennas, making it difficult to extend to ELAAs for achieving high spatial resolution. Additionally, its compatibility with communication systems is unclear.

In NISE systems, the Doppler frequencies of the echoes involve both radial and transverse velocities due to spherical-wave propagation \cite{wang2023rethinking, wang2023near_velocity, luo20236d}. The potential for joint estimation of radial and transverse velocities using Doppler frequencies in the near field was first studied in \cite{wang2023near_velocity}. This work also designed a predictive beamforming framework based on near-field velocity sensing. As a further advance, \cite{luo20236d} investigated the estimation of both horizontal and pitch transverse velocities in a 3D ISAC system. Besides, a joint user tracking and predictive beamforming design is presented in \cite{jiang2024near}.

The above arguments imply that the near-field effect has the potential to enable high-precision sensing while consuming less bandwidth and effectively sensing transverse velocity. Incorporating near-field effects into sensing technologies is a promising avenue for future research. Moreover, NISE systems require fewer wireless resources to achieve the same level of sensing precision compared to far-field systems. This efficiency supports better integration with communication systems, as it minimizes the impact on communication performance. This has also motivated the exploration of near-field ISAC, and several studies are currently ongoing. For example, the uplink and downlink communication-sensing rate regions of a single-user near-field ISAC were characterized in \cite{zhao2024modeling}. Additionally, a framework for near-field integrated sensing, positioning, and communications, along with the associated beamforming design, was discussed in \cite{li2023nearisac}.

\subsection{Near-Field Sustainability Enhancement}
The development of smart cities and homes in NG wireless communications requires numerous battery-powered IoT/wearable/sensor devices to be deployed and connected to wireless networks. As a result, how to enhance the sustainability of such energy-limited wireless sensor networks becomes one critical issue. Against this background, WPT is a key enabling technology for conveniently charging these devices and extending their lifetime. For conventional far-field WPT, the charging beam can be only steered towards specific angles, thus leading to a waste of energy and potential interference with other wireless networks. By contrast, near-field WPT can exploit beamfocusing capability to focus the energy on specific locations and thus improve energy efficiency. However, realizing near-field WPT presents several challenges, including channel estimation, beamfocusing and waveform design for simultaneous wireless power and information transfer (SWIPT), and antenna architecture. These challenges present numerous intriguing research opportunities, as detailed in \cite{zhang2022nearwpt}.

In existing work, the design of near-field WPT systems has been studied for different antenna structures. For example, the authors of \cite{9522074} developed an implementation of dynamic near-field WPT using a 2-bit $12 \times 12$ programmable metasurface, which employs near-field beamfocusing to improve the charging efficiency even for moving energy receivers. In \cite{9833917}, the beamforming design of a DMA-based multi-user near-field WPT system was studied in which the weighted sum-harvested energy was maximized by jointly optimizing the DMA coefficients and the baseband digital precoding. The authors of \cite{9814679} further studied indoor near-field WPT, devising a novel beam diversity scheme to reduce the fading margin for initially accessing the energy harvesting devices. Transmit antenna deployment for an indoor near-field WPT system was investigated in \cite{Mayer} with the aim of maximizing the minimum harvested energy among a collection of indoor users. Other researchers have focused on near-field SWIPT design, including \cite{Zheng} which showed that no dedicated energy beam is required for hybrid beamforming-based near-field SWIPT, and \cite{Yunpu} which studied joint beam scheduling and power allocation in a mixed near-field and far-field SWIPT system.

\subsection{Near-Field Massive Connectivity}
Given the rapidly increasing number of users and devices, massive connectivity is a stringent requirement in NG wireless networks. Despite NFC providing enhanced spatial DoFs, the fulfillment of massive connectivity will require additional multiple access techniques. NOMA is one such technology for providing flexible resource allocation in multi-user networks. The new characteristics of NFC provide new near-field NOMA design for supporting massive connectivity. By exploiting near-field beamfocusing, the signal energy can be focused on users far from the BS. Unlike far-field-NOMA, where the effective channel gains of users in the same direction generally decay with distance, a higher effective channel gain can be achieved at users farther away from the array in near-field-NOMA. This implies that a `far-to-near' channel-gain-based SIC order can be achieved in near-field-NOMA. Such an operation is beneficial for the case when the far user has a higher communication requirement than the near user. Inspired by this, the authors of \cite{Jiakuo} proposed two frameworks, namely single-location- and multiple-location-beamfocusing near-field-NOMA, and studied the corresponding resource allocation problem. Numerical results showed that the proposed near-field-NOMA schemes can achieve a higher spectral efficiency than conventional far-field-NOMA. The authors of \cite{10129111} explored the employment of NOMA-based mixed near-field/far-field communications, where the spatial beam preconfigured for the legacy near-field user is exploited to serve an additional far-field user. The work in \cite{ding2023resolution} revealed that NOMA can be employed to effectively improve the connectivity and system throughput in multi-user NFC when the near-field beamfocusing resolution is imperfect. The above research results revealed the great potential of near-field NOMA for massive connectivity in NG wireless networks.

\subsection{Near-Field Information Safeguarding}
Due to the broadcast nature of wireless communications, information safeguarding is an important problem in NG wireless networks. To this end, PLS is a promising technology for anti-eavesdropping. However, in previous far-field PLS, one critical issue is that information safeguarding is challenging if the eavesdropper has a higher channel condition than the legitimate user, e.g., the eavesdropper lies in the same direction as the legitimate user but is closer to the BS. As a remedy, near-field beamfocusing can concentrate the energy of the confidential signal on the legitimate user's location and thus greatly reduce the information leakage even to in-line eavesdroppers. Motivated by this, the study of near-field PLS was conducted in \cite{zhang2023physical} and \cite{gao2022multiple}, with a specific emphasis on ULAs and modular linear arrays, respectively. To optimize secrecy rates achieved by PS-based hybrid beamforming architectures, \cite{zhang2023physical} introduced a two-stage optimization algorithm, while \cite{gao2022multiple} proposed an unconstrained optimum-approaching algorithm. The collective findings of these studies underscore the potential of near-field beamfocusing to enhance PLS in the range domain, which constitutes a promising near-field information safeguarding application in NG wireless networks.
\section{Conclusions}\label{sec:conclusion}
This paper has presented a comprehensive survey of NFC, highlighting fundamental near-field operating principles, channel modeling, performance analysis, signal processing, and the integration of NFC with emerging technologies. \romannumeral1) In terms of fundamental principles, we delineated near-field properties from both physics and communications perspectives, emphasizing their distinctiveness from FFC. \romannumeral2) For channel modeling, we reviewed current contributions tailored for both SPD and CAP arrays, leveraging the spatial non-stationarity of near-field channels. \romannumeral3) Our survey covered current research on NFC performance analysis, encompassing DoFs/EDoFs, power scaling law, and transmission rate. \romannumeral4) For signal processing, we explored near-field channel estimation, beamforming design, and beam training, emphasizing the impact of novel array architectures and corresponding designs. \romannumeral5) We discussed the incorporation of near-field models in other emerging technologies, such as NISE, massive connectivity, sustainability enhancement, and information safeguarding. 

Throughout this survey, we have identified open problems and outlined future research directions within the realm of NFC. Although NFC has been a topic of discussion for four decades, its evolution, particularly under the new paradigms of multiple-antenna technology that include extremely large-scale aperture sizes, exceptionally high frequencies, and innovative antenna types, is still in its early stages. Given the current nascent state of NFC development, this review is intended to serve as a valuable resource, empowering researchers to explore the vast potential within this cutting-edge field.

\bibliographystyle{IEEEtran}
\bibliography{mybib}
\end{document}